\documentclass[a4paper,12pt]{article}
\usepackage{jheppub} 
\usepackage{tcilatex}
\usepackage{amsfonts}
\usepackage{amssymb}
\usepackage{multicol}
\usepackage{graphicx}
\usepackage{float}
\usepackage{caption}
\usepackage{xcolor}
\usepackage[utf8]{inputenc}
\usepackage{amsmath}
\title{Asymptotic Weak Gravity Conjecture in \\
M-theory on $K3\times K3$}
\author[1,2]{M.Charkaoui}
\author[1,2]{R. Sammani}
\author[1,2]{E.H Saidi}
\author[1,2]{R. Ahl Laamara}

\affiliation[1]{ LPHE-MS, Science Faculty, Mohammed V University in Rabat, Morocco}
\affiliation[2]{Centre of Physics and Mathematics, CPM- Morocco}
\emailAdd{sammani.rajaa@gmail.com}
\abstract{ The Asymptotic WGC has been proposed as a special case of the tower WGC that
probes infinite distances in the moduli space corresponding to weakly
coupled gauge regimes. The conjecture has been studied in M-theory on
Calabi-Yau threefold (CY3) with finite volume inducing a 5D effective QFT.
In this paper, \textrm{w}e extend the scope of the previous study to
encompass lower dimensions, particularly we generalise the obtained 5D
asymptotic WGC to the effective field theory (\emph{EFT}$_{3D}$) coupled to
3D gravity that descends from M-theory compactified on Calabi-Yau fourfold
with an emphasis on $K3\times K3$. \ We find that the CY4 has three
fibration structures labelled as line Type-$\mathbb{T}^{2}$, surface Type-$%
\mathbb{S}$ and bulk Type-$\mathbb{V}$. The emergent\textrm{\ }\emph{EFT}$%
_{3D}$\textrm{\ }is shown to have\textrm{\ }2+2 towers of particles states
termed as the BPS $\mathcal{T}_{M_{\mathrm{k}}\rightarrow 0}^{\text{\textsc{%
bps}}}$ and $\mathcal{T}_{M_{\mathrm{k}}\rightarrow \infty }^{\text{\textsc{%
bps}}}$ as well as the non-BPS $\mathcal{T}_{M_{\mathrm{k}}\rightarrow 0}^{%
\text{\textsc{n-bps}}}$ and $\mathcal{T}_{M_{\mathrm{k}}\rightarrow \infty
}^{\text{\textsc{n-bps}}}$. To ensure the viability of the 3D Asymptotic
WGC, we give explicit calculations to thoroughly test the swampland
constraint for both the weakly and strongly gauge coupled regimes.
Additional aspects, \textrm{including} the gauge symmetry breaking and
duality symmetry are also investigated.}


\begin{document}
\notoc
\maketitle
\flushbottom
\newpage
\tableofcontents
\section{Introduction}

\label{sec:intro}
Indubitably, the Swampland Program represents one of the major leaps
and breakthroughs in our comprehension of quantum gravity\textrm{\ \cite{1}-%
\cite{3C}. }It provides a protocol to filter the effective field theories
admitting a UV completion based on a set of consistency constraints known as
Swampland Conjectures. Among the inaugural Swampland criteria, we cite (i)
the Weak Gravity Conjecture (WGC) \textrm{\cite{4A}-\cite{8}} expecting the
presence of at least one super-extremal particle. One of its most potent
refinements is (ii) the so-called Tower WGC predicting the existence of a
tower of massive charged states satisfying the WGC \textrm{\cite{9A}},
instead of the lone super-extremal particle required by the mild version;
and (iii) the Swampland Distance conjecture (SDC) stipulating the emergence
of an infinite tower of states that becomes asymptotically massless at large
distances in the moduli space \textrm{\cite{10}-\cite{11B}}.

Using principles of the tower WGC and SDC, a new version of the WGC was
obtained in \textrm{\cite{12}} following an investigation of the
aforementioned conjectures in M- theory on Calabi-Yau threefold (CY3) with a
weak coupling regime\textrm{\ \cite{11A,12}}; it is labeled the
Asymptotic-WGC. In fact, to study the resulting 5D effective field
theory ($\emph{EFT}_{5D}$) in the infinite distance limit, while gravity
remains dynamical, one would require that the volume $\mathcal{V}_{CY_{3}}$
of the internal complex threefold stay finite ($\mathcal{V}%
_{CY_{3}}<\infty $). This condition on the volume imposes a certain
structure on the CY3 fibration: $\mathcal{F}_{fiber}\times \mathbb{B}_{base}$%
, to be $(1)$ either of a line fiber of Type-$\mathbb{T}^{2}$; or $(2)$ a
surface fiber of Type-K3 or Type-$\mathbb{T}^{4}$. For these three
Type-forms, the volume of the fiber $\mathcal{F}_{fiber}$ shrinks to zero in
the infinite distance limit of the Kahler moduli space of the \emph{EFT}$%
_{5D}$; while the volume of the base $\mathbb{B}_{base}$ expands to infinity
in such a way that the full volume $\mathcal{V}_{CY_{3}}$ of the CY3,
naively given by $\mathcal{V}_{\mathcal{F}_{fiber}}\times \mathcal{V}_{%
\mathbb{B}_{base}},$ remains finite.

The effective gauge theory coupled to gravity (which is now dynamical by
taking a finite CY volume) has (i) a weak gauge coupling regime in
the large distance limit ($g_{YM}\rightarrow 0$); and (ii) a charge lattice $%
\mathfrak{Q}$ endowed by rays (charge directions) hosting towers of\textrm{\
}BPS and non-BPS states satisfying the WGC \textrm{\cite{12}}. From this
perspective, the Asymptotic WGC can be seen as a special case of the Tower
WGC in the sense that it only involves a particular region of the moduli
space (i.e weakly coupled directions of the charge lattice); whereas the
Tower WGC expects towers of states in all directions independently of the
weak coupling limits. The weak gauge coupling regime is particularly
interesting because it supplies properties that are not allowed otherwise;
such as the relation between the extremality bound and the Repulsive Force
Conjecture \textrm{\cite{13}} as well as the precise calculation of the
charge to mass ratio of non-BPS particle states \textrm{\cite{12}}.

Furthermore, successful tests of the asymptotic WGC have been realized in
both F-theory on elliptically fibered Calabi-Yau fourfolds \textrm{\cite{14A}%
} and M-theory on Calabi-Yau threefolds \textrm{\cite{12}}. In the latter,
the infinite distance limit fixes the fibration structure to two Type-forms:
either Type-$\mathbb{T}^{2}$ or Type-$K3$/$\mathbb{T}^{4}$. Therein, some
directions of the \emph{EFT}$_{5D}$'s charge lattice hold the condition of
the weak gauge coupling regime formulated as\textrm{\cite{12, 14A}}:%
\begin{equation}
\frac{\Lambda _{\text{\textsc{wgc}}}^{2}}{\Lambda _{\text{\textsc{qg}}}^{2}}%
\qquad \rightarrow \qquad 0  \label{WeakCoupling}
\end{equation}%
where $\Lambda _{\text{\textsc{wgc}}}$ is the magnetic WGC scale of the
\emph{EFT}$_{5D},$ and $\Lambda _{\text{\textsc{qg}}}$ is the species scale;
i.e the scale above which quantum gravity becomes strongly coupled
instigating therefore a\textrm{\ }change of its dynamics \textrm{\cite%
{14B,14C}}. The above condition indicates the presence of weakly coupled
gauge groups along particular directions in the charge lattice of the
effective theory. In the context of \emph{EFT}$_{5D},$ two types of towers
of states were shown to satisfy the conjecture, given by BPS and non-BPS
states. Other studies has focused solely on BPS states since the BPS bound
aligns with the super-extremality bound, as well as the relation between
their charge and mass fixed by supersymmetry\textrm{\ \cite{14D}-\cite{14F}.}

In this paper, we investigate the Asymptotic WGC at infinite distance limit
of the 3D effective field theory ($\emph{EFT}_{3D}$) coupled to 3D gravity.
This effective gauge theory descends from M-theory compactified on
Calabi-Yau fourfold (CY4=$X_{4}$) with finite volume ($\mathcal{V}%
_{CY_{4}}<\infty $) following the framework of the 5D EFT of\textrm{\ \cite%
{11A} }and\textrm{\ \cite{12}. }Note however that most studies in the
context of 3D gravity has focused on AdS theories\textrm{\ \cite{14G, 6} }%
particularly a study of the basic and tower WGC conjectures in 3D from the
perspective of infrared consistency was performed in\textrm{\ \cite{9A, 9B}.}

To probe the Asymptotic WGC in\textrm{\ }\emph{EFT}$_{3D}$ emerging from the
compactification of M-theory on a Calabi-Yau fourfold, we begin by outlining
the classification of the fibral structure of the CY4 in the infinite
distance limit. We find three Type-forms labeled as the line Type-$\mathbb{T}%
^{2}$, the surface Type-$\mathbb{S}$ and the bulk Type-$\mathbb{V}$. The two
first Type-forms are common with the $CY3$\ fibral classification \textrm{%
\cite{11A}}; infinite distance limits of CY3 has also been studied in\textrm{%
\ \cite{15,16}.} The new bulk Type- $\mathbb{V}$ is due to the higher
dimension of CY4 compared to CY3. This classification will allow us to keep
working in a regime where gravity is dynamical, by taking the volume $%
\mathcal{V}_{CY_{4}}$ finite in the infinite distance limit, while the fiber
$\mathcal{F}_{fiber}$ of $X_{4}$ shrinks to zero and its base $\mathbb{B}%
_{base}$ expands to infinity.

Given the richness of the fibration structure of the Calabi-Yau
manifold, it would be more informative to focus on the surface Type-$\mathbb{%
S}$ for the following reasons: First, the fibral surface $S_{\perp }$ and
the base surface $S_{\parallel }$ have the mid- dimension of $X_{4}$.
Second, this surface Type-$\mathbb{S}$ permits the implementation of
discrete symmetries like the $\mathbb{Z}_{2}$ automorphism generated by the
transposition $S_{\perp }\longleftrightarrow $ $S_{\parallel }$ leaving the
Calabi-Yau fourfold invariant. In particular in \textrm{\cite{12} }only the
base was allowed to expand while the fiber shrinks for the purpose of providing a weak
gauge coupling regime; for our 3D investigation, we will take advantage of
the automorphism $S_{\perp }\longleftrightarrow $ $S_{\parallel }$ to go
even further and allow both of them to expand or shrink as long as the total
volume remains finite. This property will generate the discrete symmetry
predicting the existence of two asymptotic limits of the WGC related by $%
\mathbb{Z}_{2}$-symmetry: the usual asymptotic WGC with weak gauge coupling
investigated in 5D in addition to the asymptotic WGC derived in this paper
predicting a dual strong gauge coupling regime. Even though we are primarily
interested in the weak gauge coupling of the \emph{EFT}$_{3D}$ in the
infinite distance limit, we show that in the Type-$\mathbb{S}$ fibration of $%
X_{4}$ with $\mathbb{S}$ taken as the K3 surface, both strong and weak
regimes arise as the gauge symmetry breaks into $\emph{weakly}$ and \emph{%
strongly} coupled gauge groups.

As for the towers of quantum states required by\textrm{\ }the conjecture, we
show that the particles of the \emph{EFT}$_{3D}$ generate four series given
by: $(\mathbf{i})$ A tower $\mathcal{T}_{M_{\mathrm{k}}\rightarrow 0}^{\text{%
\textsc{bps}}}$ of charged BPS particles with light masses that vanish in
the infinite distance limit ($M_{\mathrm{k}}$ $\rightarrow 0$); it is
labelled by the integer \textrm{k}. $(\mathbf{ii})$ A tower $\mathcal{T}_{M_{%
\mathrm{k}}\rightarrow \infty }^{\text{\textsc{bps}}}$ of charged BPS
particles with extremely heavy masses ($M_{\mathrm{k}}\rightarrow \infty $).
$(\mathbf{iii})$ A tower $\mathcal{T}_{M_{\mathrm{k}}\rightarrow 0}^{\text{%
\textsc{n-bps}}}$ of non-BPS particles having light masses. $(\mathbf{iv})$
A tower$\mathcal{T}_{M_{\mathrm{k}}\rightarrow \infty }^{\text{\textsc{n-bps}%
}}$ of non BPS particles with extremely heavy masses. These $2+2$ towers of
particle states are related amongst others by the $\mathbb{Z}_{2}$
automorphism symmetry of the $\mathbb{S}_{\perp }\times $ $\mathbb{S}%
_{\parallel }$ fibration and the Weak/Strong gauge symmetry duality of the
3D effective gauge theory. More precisely we use the $\mathbb{Z}_{2}$
duality between weak and strong coupling limits to argue that in the
strongly coupled regions in the moduli space of the \emph{EFT}$_{3D}$, we
can also expect the presence of super-extremal states, since these regions
can be seen as weakly coupled in a dual frame. In consequence, the
asymptotic WGC is supported by even more towers of states than previously
anticipated, which is in line with the tower WGC.

To sum up, these are the major results of this current investigation:

\begin{description}
\item[(1)] We classify the possible fibrations of the Calabi-Yau fourfold in
the infinite distance limit while maintaining a finite volume inducing an
\emph{EFT}$_{3D}$ coupled to 3D gravity. We show that the $X_{4}$ has three
different fibral structures labelled as Line Type-$\mathbb{T}^{2}$, Surface
Type-$\mathbb{S}$, and Bulk Type-$\mathbb{V}$. This classification
generalises the one obtained in \textrm{\cite{11A}} for the \emph{EFT}$_{5D}$
coupled to 5D gravity induced by M-theory on CY3 in the infinite distance
limit.

\item[(2)] \ Considering the Type-$\mathbb{S}$ structure of the Calabi-Yau
fourfold namely $X_{4}$ $=$ $\mathbb{S}_{\perp }\times $ $\mathbb{S}%
_{\parallel }$ with an emphasis on the case $K3_{\perp }\times K3_{\parallel
}$ , we generate a\textrm{\ }$\mathbb{Z}_{2}$- automorphism symmetry by
allowing both the fiber and the base to either shrink or expand, provided
that the overall volume is finite. This discrete symmetry is given by the
transposition $K3_{\perp }\longleftrightarrow K3_{\parallel }$ including the
permutation of their p-cycles. By taking advantage of the special properties
of K3, we work out the two asymptotic gauge regimes (weak and strong) of the
\emph{EFT}$_{3D}$ that are compatible with the tower WGC. We also give two
claims (I and II) regarding the two gauge regimes of Asymptotic WGC that
exhibit the Weak/Strong duality of this 3D effective gauge theory.

\item[(3)] \ Using the wrapping of M2 and M5 branes on K3 cycles as well as
the duality links with heterotic string \textrm{\cite{12}}, we construct the
content of the four aforementioned towers namely: the content of the $%
\mathcal{T}_{M_{\mathrm{k}}\rightarrow 0}^{\text{\textsc{bps}}}$ having
light BPS particles. The $\mathcal{T}_{M_{\mathrm{k}}\rightarrow \infty }^{%
\text{\textsc{bps}}}$ having extremely heavy BPS particles. The content of $%
\mathcal{T}_{M_{\mathrm{k}}\rightarrow 0}^{\text{\textsc{n-bps}}}$ with
light non-BPS particles. And the $\mathcal{T}_{M_{\mathrm{k}}\rightarrow
\infty }^{\text{\textsc{n-bps}}}$ made of extremely heavy non BPS particles.
These light and the heavy towers are related by the Weak/Strong gauge
symmetry duality.

\item[(4)] \ Using the link between asymptotic WGC and the Repulsive Force
Conjecture (RFC) \textrm{\cite{13}}, we give explicit verifications of the
Asymptotic WGC for both weakly gauge coupled and strongly gauge coupled
regimes in the \emph{EFT}$_{3D}$. We perform this test first for the
towers $\mathcal{T}_{M_{\mathrm{k}}\rightarrow 0}^{\text{\textsc{bps}}}$ and
$\mathcal{T}_{M_{\mathrm{k}}\rightarrow \infty }^{\text{\textsc{bps}}}$ made
of BPS particle states, then for $\mathcal{T}_{M_{\mathrm{k}}\rightarrow 0}^{%
\text{\textsc{n-bps}}}$ and $\mathcal{T}_{M_{\mathrm{k}}\rightarrow \infty
}^{\text{\textsc{n-bps}}}$ made of non-BPS particle states. And because
classical massless 3D gravity is topological, we forfeit the classical
effects for a non trivial negative contribution\textrm{\ }$\vartheta _{QG}$%
\textrm{\ }originating from quantum corrections.
\end{description}

\ \ \ \ \newline
The organisation of this paper is as follows: \textrm{In \autoref{sec2}}, we
generalize the results of \textrm{\cite{11A}} regarding \emph{EFT}$_{5D}$,
induced by $X_{3}$ in the infinite distance limit, towards \emph{EFT}$_{3D}$
descending from M-theory on Calabi-Yau fourfold $X_{4}$ with finite volume.
\textrm{In \autoref{sec3},} we investigate the physical implications of the
geometry probed by M-theory on CY4 in the moduli space of the 3D effective
theory. We give results on the gauge coupling regimes in the \emph{EFT}$%
_{3D} $ with $X_{4}=K3\times K3$. \textrm{In \autoref{sec4}}, we distinguish the
different gauge regimes before constructing the four towers of particle
states in \emph{EFT}$_{3D}$ candidates to satisfy the asymptotic WGC.
\textrm{In \autoref{sec5}}, we give explicit calculations testing the Asymptotic
WGC for both weak and strong regime. Last section is devoted to conclusions.
In \autoref{app} we report some technical details on the underlying
properties of the towers of BPS and non-BPS states.

\section{Classification of fibers of CY4}

\label{sec2} In this section, we give first results on the large distance limit for
M-theory on Calabi-Yau fourfold (CY4) with finite volume ($\mathcal{V}_{%
{\small CY}_{4}}<\infty ).$ These results will be used later on for
compactified M-theory with background%
\begin{equation}
\mathcal{M}_{11D}=X_{4}\times \mathcal{M}_{3D}\qquad ;\qquad X_{4}:=CY_{4}
\end{equation}%
where the compact $X_{4}$ is fibered like $\mathcal{F}_{n}\times \mathbb{B}%
_{4-n}$ with complex fiber dimension taking the three values $n=1,2,3.$
These results will be also used in the study of the Asymptotic Weak Gravity
Conjecture (Asymptotic-WGC) in 3D space-time. To do so, we need two main
\textrm{properties}.

\begin{description}
\item[$(\mathbf{A})$] being particularly interested in theories coupled to
3D (topological) gravity, we need to keep the volume of the complex 4d
internal manifold $X_{4}$ finite%
\begin{equation}
\mathcal{V}_{{\small CY}_{4}}:=\left( l_{11d}\right) ^{8}\mathcal{V}_{%
{\small X}_{{\small 4}}}<\infty
\end{equation}%
where $l_{11d}=1/M_{11d}$ is the usual 11D fundamental length scale of
M-theory \textrm{\cite{17A}}. Notice that the $\mathcal{V}_{{\small X}_{%
{\small 4}}}$ is dimensionless ($\left[ \mathcal{V}_{{\small X}_{{\small 4}}}%
\right] =length^{0}$); and is related to the Planck mass $M_{\mathrm{Pl}}$
in three dimensions and the 11D mass scale \textrm{via} $4\pi \mathcal{V}_{%
{\small X}_{{\small 4}}}\simeq M_{Pl}/M_{11d};$ i.e:%
\begin{equation}
M_{11d}=\frac{M_{Pl}}{4\pi \mathcal{V}_{{\small X}_{{\small 4}}}}
\end{equation}

\item[$(\mathbf{B})$] Below, we will be interested in the infinite distance
limit in the (gauge region of the) moduli space $\mathfrak{M}_{X_{4}}$ of
the effective field theory (\emph{EFT}$_{3D}$) descending from M-theory on $%
X_{4}$. In this limit, we will take the volume $\mathcal{V}_{\mathcal{F}%
_{n}} $ of the fiber to shrink to zero whereas the volume $\mathcal{V}_{%
\mathbb{B}_{4-n}}$ of base diverges; i.e:%
\begin{equation}
\mathcal{V}_{\mathcal{F}_{n}}\rightarrow 0\qquad and\qquad \mathcal{V}_{%
\mathbb{B}_{4-n}}\rightarrow \infty
\end{equation}%
while their product $\mathcal{V}_{\mathcal{F}_{n}}\times \mathcal{V}_{%
\mathbb{B}_{4-n}}$ is finite.
\end{description}

\ \newline
Thus, we need to construct CY manifolds $X_{4}$ which obey both the two
above properties: $\left( \mathbf{A}\right) $ and $\left( \mathbf{B}\right)
. $ And this is the main goal of the investigation given in this section. To
that purpose, we revisit in subsection \textbf{2.1} useful results on \emph{%
EFT}$_{5D}$ given by M-theory on Calabi-Yau threefolds CY$_{3}:=X_{3}$
\textrm{\cite{11A,12}}. In subsection \textbf{2.2}, we give our results
\textrm{on the fibration structure of the Calabi-Yau fourfold that induces
our }\emph{EFT}$_{3D}$ in infinite distance limit in the moduli space.%
\newline
As a front matter, notice that, as for finite distances in the moduli space
of\emph{\ EFT}$_{(11-2d)D}$, one can also distinguish various \textrm{%
possible fibrations }$\mathcal{F}_{n}\times \mathbb{B}_{d-n}$\textrm{\ of
complex d-dimensional }$CY_{d}:=X_{d}$ in the infinite distance limit ($%
\lambda \rightarrow \infty $). For the case $d=4$ we are particularly
interested into here, we have\textrm{\ }the following typical fibrations
\begin{equation}
\begin{tabular}{c||c|c}
internal X$_{4}$ & fiber $\mathcal{F}_{n}$ \  & \ base $\mathbb{B}_{4-n}$ \\
\hline\hline
$\left( a\right) $ & $\mathcal{F}_{1}$ & $\mathbb{B}_{3}$ \\
$\left( b\right) $ & $\mathcal{F}_{2}$ & $\mathbb{B}_{2}$ \\
$\left( c\right) $ & $\mathcal{F}_{3}$ & $\mathbb{B}_{1}$ \\ \hline\hline
\end{tabular}
\label{4}
\end{equation}%
such that the first Chern class $c_{1}\left( \mathcal{F}_{n}\times \mathbb{B}%
_{4-n}\right) $ vanishes. Note also that a fibered Calabi-Yau manifold must
have a fiber which is also Calabi-Yau \textrm{\cite{17B}}, this will limit
tremendously the number of the possibilities, especially for\textrm{\ }$%
\mathcal{F}_{1}$\textrm{\ }and\textrm{\ }$\mathcal{F}_{2}.$ On the other
hand the possibilities of\textrm{\ }$\mathcal{F}_{3}$\textrm{\ }are vastly
larger\textrm{\ \cite{18, 19}} . \newline
To build the $\mathcal{F}_{n}$ fibers and the corresponding $\mathbb{B}%
_{4-n} $ bases that can be used in Asymptotic-WGC of \emph{EFT}$_{3D}$\emph{,%
} we borrow ideas from the classification of fiber structure of on
Calabi-Yau threefolds\textrm{\ }$X_{3}$ \textrm{\cite{11A}}\ reviewed below.

\subsection{Classification of X$_{3}$ fibers in infinite distance limit}

Following \textrm{\cite{11A}}, we distinguish two kinds of fibrations in
M-theory on Calabi-Yau threefolds in the infinite distance limit with three
possible fibers (namely $\mathbb{T}^{2},$ K3 and $\mathbb{T}^{4}$ ).
Depending on the dimension of the fiber, we have either%
\begin{equation}
\begin{tabular}{ccc}
$\mathcal{F}_{1}$ & $\rightarrow $ & X$_{3}$ \\
&  & $\downarrow $ \\
&  & $\mathbb{B}_{2}$%
\end{tabular}
\label{120}
\end{equation}

with $\mathcal{F}_{1}$\ a complex curve ($\dim _{\mathbb{C}}\mathcal{F}%
_{1}=1 $); or%
\begin{equation}
\begin{tabular}{ccc}
$\mathcal{F}_{2}$ & $\rightarrow $ & X$_{3}$ \\
&  & $\downarrow $ \\
&  & $\mathbb{B}_{1}$%
\end{tabular}
\label{121}
\end{equation}%
with $\mathcal{F}_{2}$ a complex surface ($\dim _{\mathbb{C}}\mathcal{F}%
_{2}=2$). The emerging fibers for $\mathcal{F}_{1}$ and $\mathcal{F}_{2}$ in
the infinite distance limit, result from taking a finite volume $\mathcal{V}%
_{{\small X}_{{\small 3}}}$ in order for the 5D gravity to remain dynamical.
We start by parameterizing the Kahler 2-form $J$ of the X$_{3}$ like $%
J=\sum_{\mu =1}^{h_{{\small X}_{{\small 3}}}^{1,1}}\upsilon ^{\mu }J_{\mu }$%
; which for convenience we present simply as:%
\begin{equation}
J=\lambda \upsilon ^{0}J_{0}+\frac{1}{\lambda ^{\alpha }}\sum_{i=1}^{\mathrm{%
r}}\upsilon ^{i}J_{i}  \label{J0}
\end{equation}%
with $r=h_{{\small X}_{{\small 3}}}^{1,1}-1.$ In this expansion, the real $%
\lambda $ is a spectral parameter characterizing the infinite distance limit
in the moduli space of the theory; the infinite distance limit is
implemented by the asymptotic limit $\lambda \rightarrow \infty $. The $%
J_{0} $ is a representative divisor in the base $\mathbb{B}_{2}$ (that can
be imagined as the projective $\mathbb{P}^{2}$ or $\mathbb{P}^{1}\times
\mathbb{P}^{1}$); the choice of one divisor is to simplify the calculations
in solving the constraints $\left( \mathbf{A}\right) $ and $\left( \mathbf{B}%
\right) $. The other $J_{i}$'s in (\ref{J0}) concern the fiber $\mathcal{F}%
_{1}$. The real positive $\upsilon ^{0}$ and $\upsilon ^{i}$ parameters are
the volumes of the dual 2-cycles in the homology $H_{2}(X_{3})$.

Using (\ref{J0}), the classification of\textrm{\ \cite{11A}} can be shortly
presented in terms of the value of the monomials $J_{0}^{n};$ we have:

\begin{description}
\item[$\left( \mathbf{1}\right) $] \textbf{Type-}$\mathbb{T}^{2}$ \textbf{%
form: }\newline
It corresponds to $J_{0}^{3}=0$ but $J_{0}^{2}\neq 0$ \textrm{\cite{21, 22}}
Here, the fiber $\mathcal{F}_{1}$ in the fibration (\ref{120}) is given by a
2-torus $\mathbb{T}^{2}$ (elliptic curve $\mathcal{E}$).\newline
The Kahler form is given by%
\begin{equation}
J=\lambda \upsilon ^{0}J_{0}+\frac{1}{\lambda ^{2}}\sum_{i=1}^{\mathrm{r}%
}\upsilon ^{i}J_{i}  \label{t2}
\end{equation}%
Using (\ref{t2}), it results that the volume $\mathcal{V}_{\mathcal{F}_{1}}$
of the fiber and the volume $\mathcal{V}_{\mathbb{B}_{2}}$ of the base
behave like
\begin{equation}
\mathcal{V}_{\mathcal{F}_{1}}\sim \frac{1}{\lambda ^{2}}T_{finite}\qquad
,\qquad \mathcal{V}_{\mathbb{B}_{2}}\sim \lambda ^{2}B_{finite}
\end{equation}%
where $T_{finite}$ and $B_{finite}$ take finite values in the limit $\lambda
\rightarrow \infty $. Notice that while $\lim_{\lambda \rightarrow \infty }%
\mathcal{V}_{\mathcal{F}_{1}}=0$ and $\lim_{\lambda \rightarrow \infty }%
\mathcal{V}_{\mathbb{B}_{2}}=\infty $, the total volume of the $X_{3}$
remains finite even in the infinite distance limit where it takes the value
\begin{equation}
\lim_{\lambda \rightarrow \infty }\mathcal{V}_{{\small CY}_{{\small 3}%
}}\simeq T_{finite}\times B_{finite}
\end{equation}

\item[$\left( \mathbf{2}\right) $] \textbf{Type-}$\mathbb{S}$ \textbf{forms:
}\newline
It corresponds to\emph{\ }$J_{0}^{3}=0$ and $J_{0}^{2}=0$ \textrm{\cite{23A}}%
. In this Type- surface form, the fiber $\mathcal{F}_{2}$ in the second
structure in eq(\ref{121}) is a complex surface which, in the infinite
distance limit, has been shown to be $\left( \mathbf{a}\right) $ either a K3
surface; or $\left( \mathbf{b}\right) $ an abelian \textrm{Schoen} surface
having the topology of $\mathbb{T}^{4}$.\newline
For the generalization of this classification to the case of the $X_{4},$ it
is interesting to think about the K3 and the $\mathbb{T}^{4}$\ manifolds in
terms of fibrations using the complex one dimensional lines $\mathbb{P}^{1}$
and $\mathbb{T}^{2}$. The two possible $\mathcal{F}_{2}\sim \mathbb{S}$ can
be presented as follows:%
\begin{equation}
\begin{tabular}{lll}
$\left( \mathbf{a}\right) $ & : & $\mathbb{S\sim T}^{2}\times \mathbb{P}^{1}$
\\
$\left( \mathbf{b}\right) $ & : & $\mathbb{S\sim T}^{2}\times \mathbb{T}^{2}$%
\end{tabular}
\label{S}
\end{equation}%
both of hem having $c_{1}\left( T\mathbb{S}\right) =0$; thus ruling out
surfaces $\mathbb{S}$ like $\mathbb{P}^{1}\times \mathbb{P}^{1}.$\newline
Notice that in both the Type-$\mathbb{S}$ forms (\ref{S}), the Kahler form $%
J $ can be parameterised as%
\begin{equation}
J=\lambda \upsilon ^{0}J_{0}+\frac{1}{\sqrt{\lambda }}\sum_{i=1}^{\mathrm{r}%
}\upsilon ^{i}J_{i}  \label{s}
\end{equation}%
Using this expansion, the volume $\mathcal{F}_{2}$ of the fiber and the
volume $\mathcal{V}_{\mathbb{B}_{1}}$ of the base behave like
\begin{equation}
\mathcal{V}_{\mathcal{F}_{2}}\sim \frac{1}{\lambda }T_{finite}^{\prime
}\qquad ,\qquad \mathcal{V}_{\mathbb{B}_{1}}\sim \lambda B_{finite}^{\prime }
\end{equation}%
where $T_{finite}^{\prime }$ and $B_{finite}^{\prime }$ take finite values
in the limit $\lambda \rightarrow \infty $. Notice also that these two Type-$%
\mathbb{S}$ forms (\ref{S}) are distinguished by the value of the second
Chern class c$_{2}\left( T\mathbb{S}\right) $ which is equal to 24 for K3
and vanishes for $\mathbb{T}^{4}$. Notice moreover that the full volume of
the $X_{3}$ remains finite even in the infinite distance limit; we have
\begin{equation}
\lim_{\lambda \rightarrow \infty }\mathcal{V}_{{\small X}_{{\small 3}%
}}\simeq T_{finite}^{\prime }\times B_{finite}^{\prime }
\end{equation}
\end{description}

\subsection{Generalization to X$_{4}$ in the infinite distance limit}

Mimicking the above description regarding the possible fibration forms of
the Calabi-Yau threefold $X_{3}$ in the infinite distance limit, one can
deduce straightforwardly the classification of the fibers of the manifolds (%
\ref{4}) without going much into technical details.\newline
The classification of the $\mathcal{F}_{n}$ in the fibration (\ref{4}) at $%
\lambda \rightarrow \infty $ can be done according to the values of the
monomial $J_{0}^{n}$. Extending the construction done for $X_{3}$ towards $%
X_{4}$; we end up with the following result:

\begin{description}
\item[$(\mathbf{A)}$] \textbf{Type-}$\mathbb{T}^{2}$ \textbf{form of the }X%
\textbf{$_{4}$:}\emph{\ }\newline
It corresponds to the constraint relations%
\begin{equation}
J_{0}^{4}=0\qquad ,\qquad J_{0}^{3}\neq 0  \label{Kollar}
\end{equation}%
In this case, the Kahler 2-form $J$ can be parameterized by as follows%
\begin{equation}
J=\lambda \upsilon ^{0}J_{0}+\frac{1}{\lambda ^{3}}\sum_{i=1}^{\mathrm{r}%
}\upsilon ^{i}J_{i}
\end{equation}%
with $r=h^{1,1}(X_{4})-1.$ Expressing this expansion shortly as $J=\lambda
\upsilon ^{0}J_{0}+\frac{1}{\lambda ^{3}}K$ with $K=\sum_{i=1}^{\mathrm{r}%
}\upsilon ^{i}J_{i}$ and using $J_{0}^{4}=0$ and $J_{0}^{3}\neq 0$, we have
\begin{equation}
J^{4}=\left( 4\upsilon _{0}^{3}\right) J_{0}^{3}K+\frac{6\upsilon _{0}^{2}}{%
\lambda ^{4}}J_{0}^{2}K^{2}+\frac{4\upsilon ^{0}}{\lambda ^{8}}J_{0}K^{3}+%
\frac{1}{\lambda ^{12}}K^{4}
\end{equation}%
behaving in powers of $1/\lambda $ as%
\begin{equation}
J^{4}=\left( 4\upsilon _{0}^{3}\right) J_{0}^{3}K+\mathcal{O}\left( \frac{1}{%
\lambda ^{4}}\right)
\end{equation}%
Here, the $\mathcal{F}_{1}$ in eq(\ref{4}) is a complex line homotopic to $%
\mathbb{T}^{2};$ in a similar way as in the case of $X_{3}.$ In the infinite
distance limit, the volume $\mathcal{V}_{\mathcal{F}_{1}}$ of the fiber $%
\mathcal{F}_{1}$ and the volume $\mathcal{V}_{\mathbb{B}_{3}}$ of the base $%
\mathbb{B}_{3}$ behave in terms of the spectral parameter $\lambda $ like
\begin{equation}
\mathcal{V}_{\mathcal{F}_{1}}\sim \frac{1}{\lambda ^{3}}\mathrm{\tau }%
_{finite}\qquad ,\qquad \mathcal{V}_{\mathbb{B}_{3}}\sim \lambda ^{3}\mathrm{%
\beta }_{finite}
\end{equation}%
where $\mathrm{\tau }_{finite}$ and $\mathrm{\beta }_{finite}$ are functions
of $\lambda $ taken finite everywhere in the moduli space $\mathfrak{M}_{%
{\small CY}_{{\small 4}}}$ of the 3D effective field theory. The volumes $%
\mathcal{V}_{\mathcal{F}_{1}}$ and $\mathcal{V}_{\mathbb{B}_{3}}$ have rapid
singularities in the limit $\lambda \rightarrow \infty $ (shrinking $%
\mathcal{F}_{1}$ and diverging $\mathbb{B}_{3}$); but in such way that the
full volume of the X$_{4}$ remains finite
\begin{equation}
\lim_{\lambda \rightarrow \infty }\mathcal{V}_{{\small X}_{{\small 4}%
}}\simeq \mathrm{\tau }_{finite}\times \mathrm{\beta }_{finite}
\end{equation}%
Interestingly, the constraint (\ref{Kollar}) is \textrm{closely related} to
the Kollar conjecture \textrm{\cite{23B}} characterizing Calabi-Yau
manifolds with genus-one fibration, by having some (1,1)-class $D\in
H^{1,1}(X_{4},\mathbb{Q)}$ satisfying properties among which (\ref{Kollar})
is present

\item[$(\mathbf{B)}$] \textbf{Type-}$\mathbb{S}$ \textbf{forms of the }X$%
_{4} $\textbf{:}\emph{\ }\newline
It corresponds to the constraint relations%
\begin{equation}
J_{0}^{4}=0\qquad ,\qquad J_{0}^{3}=0\qquad ,\qquad J_{0}^{2}\neq 0
\label{JK3}
\end{equation}%
Here, the Kahler 2-form is parameterised as:%
\begin{equation}
J=\lambda \upsilon ^{0}J_{0}+\frac{1}{\lambda }\sum_{i=1}^{\mathrm{r}%
}\upsilon ^{i}J_{i}  \label{22}
\end{equation}%
with $r=h^{1,1}(X_{4})-1.$ Which we can write shortly as $J=\lambda \upsilon
^{0}J_{0}+\frac{1}{\lambda }K$ with $K=\sum_{i=1}^{\mathrm{r}}\upsilon
^{i}J_{i}$ and using (\ref{JK3}), we have%
\begin{equation}
J^{4}=6\upsilon _{0}^{2}J_{0}^{2}K^{2}+\frac{4\upsilon ^{0}}{\lambda ^{2}}%
J_{0}K^{3}+\frac{1}{\lambda ^{4}}K^{4}
\end{equation}%
behaving in powers of $1/\lambda $ as follows%
\begin{equation}
J^{4}=6\upsilon _{0}^{2}J_{0}^{2}K^{2}+\mathcal{O}\left( \frac{1}{\lambda
^{2}}\right)
\end{equation}%
In this family, the fiber $\mathcal{F}_{2}$ in (\ref{4}) is a complex
surface which can be $\left( \mathbf{i}\right) $ either a K3 surface; eg $%
\mathbb{S}\sim \mathbb{T}^{2}\times \mathbb{P}^{1}$; or $\left( \mathbf{ii}%
\right) $ a Schoen surface given by $\mathbb{S\sim T}^{2}\times \mathbb{T}%
^{2}$. The volume $\mathcal{V}_{\mathcal{F}_{2}}$ of the fiber and the
volume $\mathcal{V}_{\mathbb{B}_{2}}$ of the base behave in the infinite
distance limit like
\begin{equation}
\mathcal{V}_{\mathcal{F}_{2}}\sim \frac{1}{\lambda ^{2}}\mathrm{\tau }%
_{finite}^{\prime }\qquad ,\qquad \mathcal{V}_{\mathbb{B}_{2}}\sim \lambda
^{2}\mathrm{\beta }_{finite}^{\prime }  \label{ts}
\end{equation}%
where $\mathrm{\tau }_{finite}$ and $\mathrm{\beta }_{finite}$ finite
functions of $\lambda $ and where%
\begin{equation}
\lim_{\lambda \rightarrow \infty }\mathcal{V}_{{\small CY}_{{\small 4}%
}}\simeq \mathrm{\tau }_{finite}^{\prime }\times \mathrm{\beta }%
_{finite}^{\prime }
\end{equation}%
Notice that this Type-$\mathbb{S}$ forms of the X$_{4}$ has similar
properties as in the case of the X$_{3}$; in particular the values of the
first and the second Chern numbers c$_{1}\left( T\mathbb{S}\right) $ and c$%
_{2}\left( T\mathbb{S}\right) ;$ thus ruling out surfaces like the
Hirzebruch $\mathbb{S}\sim \mathbb{P}^{1}\times \mathbb{P}^{1}.$

\item[$(\mathbf{C)}$] \textbf{Type-}$\mathbb{V}$ \textbf{forms of the X$_{4}$%
:}\emph{\ }\newline
It corresponds to the constraint relations%
\begin{equation}
J_{0}^{4}=0\qquad ,\qquad J_{0}^{3}=0\qquad ,\qquad J_{0}^{2}=0
\end{equation}%
In this case, we have the typical expansion of the Kahler 2-form:%
\begin{equation}
J=\lambda \upsilon ^{0}J_{0}+\frac{1}{\lambda ^{1/3}}K
\end{equation}%
with $K=\sum_{i=1}^{\mathrm{r}}\upsilon ^{i}J_{i}$. The volume form expands
like%
\begin{equation}
J^{4}=\left( 4\upsilon ^{0}\right) J_{0}K^{3}+\frac{1}{\lambda ^{4/3}}K^{4}
\end{equation}%
it behaves in powers of $1/\lambda $ as follows%
\begin{equation}
J^{4}=\left( 4\upsilon ^{0}\right) J_{0}K^{3}+\mathcal{O}\left( \frac{1}{%
\lambda ^{4/3}}\right)
\end{equation}%
The fiber $\mathcal{F}_{3}$ in eq(\ref{4}) is then a complex 3d space while
the base is a complex curve. Here, the volume $\mathcal{V}_{\mathcal{F}_{3}}$
of the fiber and the volume $\mathcal{V}_{\mathbb{B}_{1}}$ of the base
behaves in the infinite distance as:
\begin{equation}
\mathcal{V}_{\mathcal{F}_{3}}\sim \frac{1}{\lambda }\mathrm{\tau }%
_{finite}^{\prime \prime }\qquad ,\qquad \mathcal{V}_{\mathbb{B}_{1}}\sim
\lambda \mathrm{\beta }_{finite}^{\prime \prime }
\end{equation}%
showing that $\mathcal{V}_{{\small CY}_{{\small 4}}}$ is still finite%
\begin{equation}
\lim_{\lambda \rightarrow \infty }\mathcal{V}_{{\small X}_{{\small 4}%
}}\simeq \mathrm{\tau }_{finite}^{\prime \prime }\times \mathrm{\beta }%
_{finite}^{\prime \prime }
\end{equation}%
In general, the fiber should be a Calabi-Yau threefold, and unlike the case
of type-$\mathbb{S}$\textrm{, }where the only Calabi-Yau two folds are K3 and%
\textrm{\ }$\mathbb{T}^{4},$\textrm{\ }here the possibilities are much
larger. According to Wall's theorem \textrm{\cite{23C}}, they can be
distinguished via the following numbers:\textrm{\ }$h^{1,1},$ $h^{1,2},$ $%
c_{2}$\textrm{\ }and\textrm{\ }$J_{i}.J_{j}.J_{k}.$ If we restrain our
analysis to taking only a simple class of Calabi-Yau threefolds as a fiber,
using the $\mathbb{T}^{2}$ and $\mathbb{P}^{1}$ language used in eq(\ref{S}%
), the complex 3d fiber $\mathcal{F}_{3}$ can have one of the \textrm{three}
following Type-$\mathbb{V}$ forms:%
\begin{equation}
\begin{tabular}{lll}
$\left( i\right) $ & : & $\mathbb{V}\sim \mathbb{T}^{2}\times \mathbb{T}%
^{2}\times \mathbb{T}^{2}$ \\
$\left( ii\right) $ & : & $\mathbb{V\sim T}^{2}\times \mathbb{T}^{2}\times
\mathbb{P}^{1}$ \\
$\left( iii\right) $ & : & $\mathbb{V\sim T}^{2}\times \mathbb{P}^{1}\times
\mathbb{P}^{1}$%
\end{tabular}%
\end{equation}%
These volume forms can be discriminated by the value of the Chern-numbers $%
c_{l}\left( \mathbb{V}\right) $\textrm{\ }with\textrm{\ }$l=1,2,3.$
\end{description}

\section{EFT$_{3D}$ descending from M-theory on $K3_{\bot }\times K3_{\Vert
} $}
\label{sec3}
In this section, we study physical implications in the infinite distance
limit in the moduli space of the underlying 3D effective theory descending
from 11D M-theory on the background%
\begin{equation}
\mathcal{M}_{11D}=X_{4}\times \mathcal{M}_{3D}\qquad ;\qquad X_{4}=K3_{\bot
}\times K3_{\Vert }
\end{equation}%
Because the (dimensionless) volume $\mathcal{V}_{{\small X}_{{\small 4}}}$
of the CY4 is required to be finite, it follows that the \emph{EFT}$_{3D}$,
having four unbroken supercharges, captures remarkable aspects; in
particular $\left( i\right) $ a classical topological 3D gravity with non
trival contributions induced by quantum effects; $\left( ii\right) $\ a
weak/strong gauge duality due to the transposition of the role of the two K3
factors making the Calabi-Yau fourfold $X_{4}$; and $\left( iii\right) $
\textrm{we can effectively test} of the asymptotic weak gravity conjecture i%
\textrm{n both the weak/strong dual pictures}.

In the first subsection\textbf{\ 3.1,} we recall useful aspects on the
compactification of M-theory on X$_{4}$; this compactification has been
extensively studied in literature; we cite for example \textrm{\cite%
{24,25A,25AB}}, hence we will only be interested in the notions that we will
use in our study. In the second subsection \textbf{3.2}, we study the gauge
coupling regimes in our \emph{EFT}$_{3D}$ descending from the
compactification on $K3_{\bot }\times K3_{\Vert }$. Here, we give new
results on weak and strong gauge regimes in the large and the short distance
limits.

\subsection{Compactifying 11D supergravity down to 3D}

The action $\mathcal{S}_{11D}$ of the eleven dimensional supergravity theory
has 11D gravity field $g_{MN}$, a 3-form gauge potential $\boldsymbol{C}_{3}$
and fermionic partners. Using the 4-form field strength $\boldsymbol{F}_{4}=d%
\boldsymbol{C}_{3}$ and restricting to bosonic fields, the leading bosonic
terms in the field action read as follows%
\begin{equation}
\mathcal{S}_{11D}=\frac{2\pi }{\QTR{sl}{l}_{11d}^{9}}\int_{\mathcal{M}%
_{11D}}\left( \mathcal{R}_{11}\ast \mathbf{1}\right) -\frac{\pi }{\QTR{sl}{l}%
_{11d}^{7}}\int_{\mathcal{M}_{11D}}\boldsymbol{F}_{4}\wedge \ast (%
\boldsymbol{F}_{4})-\frac{\pi }{6\QTR{sl}{l}_{11}^{6}}\int_{\mathcal{M}%
_{11D}}\boldsymbol{C}_{3}\wedge \boldsymbol{F}_{4}\wedge \boldsymbol{F}_{4}
\end{equation}%
with 2$\kappa _{11}^{2}=\left( 2\pi \right) ^{8}/M_{11}^{9}$ and $\QTR{sl}{l}%
_{11d}=\QTR{sl}{M}_{11d}^{-1}.$ Under the compactification on $X_{4},$ the
3-form potential $\boldsymbol{C}_{3}$ decomposes in terms of the 2-form
Kahler generators \{$J_{\text{\textsc{a}}}$\} of the cohomology group $%
H^{1,1}$($X_{4}$) like%
\begin{equation}
\boldsymbol{C}_{3}=\frac{1}{2\pi l_{11d}}\sum_{\text{\textsc{a}}=1}^{h_{%
{\small X_{4}}}^{1,1}}A_{1}^{\text{\textsc{a}}}\wedge J_{\text{\textsc{a}}%
}+...\qquad ,\qquad \boldsymbol{F}_{4}=\frac{1}{2\pi }\sum_{\text{\textsc{a}}%
=1}^{h_{{\small X_{4}}}^{1,1}}F_{2}^{\text{\textsc{a}}}\wedge J_{\text{%
\textsc{a}}}+...  \label{C3}
\end{equation}%
where the $A_{1}^{\text{\textsc{a}}}$s are 1-form gauge potentials in 3D
space time given ( in $2\pi l_{11d}$ units) by%
\begin{equation}
A_{1}^{\text{\textsc{a}}}=\dint\nolimits_{\mathfrak{C}^{\text{\textsc{a}}}}%
\boldsymbol{C}_{3}\qquad ,\qquad \dint\nolimits_{\mathfrak{C}^{\text{\textsc{%
a}}}}J_{\text{\textsc{a}}}=\delta _{\text{\textsc{b}}}^{\text{\textsc{a}}}
\end{equation}%
These are abelian 1-form gauge potentials of the \emph{EFT}$_{3D}$ with
gauge symmetry
\begin{equation}
G_{{\small abelian}}=\dprod\limits_{\text{\textsc{a}}=1}^{h_{{\small X_{4}}%
}^{1,1}}U\left( 1\right) ^{\text{\textsc{a}}}  \label{34}
\end{equation}%
So, under compactification of M-theory on $X_{4}$, the leading bosonic block
terms in the induced 3D effective field action $\mathcal{S}_{3D}$ is a
functional of the scalar curvature $\mathcal{R}$ of the 3D space-time $%
\mathcal{M}_{3D}$ and the 2-form gauge field strengths $F^{\text{\textsc{a}}%
}=dA^{\text{\textsc{a}}}.$ This field action has the structure%
\begin{equation}
\mathcal{S}_{3D}=\frac{M_{Pl}}{2}\int_{\mathcal{M}_{3D}}\left( \mathcal{R}%
\ast \mathbf{1-}\mathfrak{g}_{\text{\textsc{xy}}}d\phi ^{\text{\textsc{x}}%
}\wedge \ast d\phi ^{\text{\textsc{y}}}\right) -\frac{1}{\mathrm{4}g_{3D}^{2}%
}\int_{\mathcal{M}_{3D}}G_{\text{\textsc{ab}}}\left( F^{\text{\textsc{a}}%
}\wedge \ast F^{\text{\textsc{b}}}\right) +...  \label{S3D}
\end{equation}%
where the gauge coupling metric $\mathcal{G}_{\text{\textsc{ab}}}$ is a
dimensionless quantity ([$\mathcal{G}_{\text{\textsc{ab}}}]=Mass^{0}$) given
by%
\begin{equation}
\mathcal{G}_{\text{\textsc{ab}}}=\frac{1}{\mathcal{V}_{{\small CY}_{4}}}%
\int_{{\small CY}_{4}}J_{\text{\textsc{a}}}\wedge (\ast J_{\text{\textsc{b}}%
})  \label{met}
\end{equation}%
The 3D Planck mass reads in terms of the volume of the $X_{4}$ as:%
\begin{equation}
M_{\mathrm{Pl}}=\mathrm{4\pi }M_{11}^{9}\mathcal{V}_{{\small CY}_{4}}\qquad
\Leftrightarrow \qquad \mathcal{V}_{{\small CY}_{4}}=\frac{M_{\mathrm{Pl}}}{%
\mathrm{4\pi }M_{11}^{9}}\qquad \Leftrightarrow \qquad \mathcal{V}_{{\small X%
}_{4}}=\frac{M_{\mathrm{Pl}}}{\mathrm{4\pi }M_{11}}  \label{cy4}
\end{equation}%
and the 3D gauge coupling is related to the 3D Planck mass $M_{\mathrm{Pl}}$
and the $M_{11}^{2}$ like%
\begin{equation}
\frac{1}{g_{3D}^{2}}=\frac{\mathrm{M}_{Pl}}{4\pi ^{2}M_{11}^{2}}\qquad
\Leftrightarrow \qquad \frac{1}{g_{3D}^{2}}=\frac{\mathcal{V}_{{\small X}%
_{4}}}{\pi M_{11}}  \label{g3d}
\end{equation}%
it scales like mass; that is $[g_{3D}^{2}]=\mathrm{Mass}^{1}.$\newline
We also define the following useful quantities: $\left( 1\right) $ the
partial dimensional volumes
\begin{equation}
\mathcal{V}_{\text{\textsc{a}}}=\frac{1}{3!}\dint\nolimits_{\boldsymbol{D}_{%
\text{\textsc{a}}}}J^{3}\quad ,\quad \mathcal{V}_{\text{\textsc{ab}}}=\frac{1%
}{2!}\dint\nolimits_{\boldsymbol{S}_{\text{\textsc{ab}}}}J^{2}\quad ,\quad
\mathcal{V}_{\text{\textsc{abc}}}=\dint\nolimits_{\boldsymbol{C}_{\text{%
\textsc{abc}}}}J  \label{VLS}
\end{equation}%
reading also as%
\begin{equation}
\begin{tabular}{lll}
$\mathcal{V}_{\text{\textsc{a}}}$ & $=$ & $\frac{1}{6}\dint\nolimits_{%
{\small CY}_{{\small 4}}}J_{\text{\textsc{a}}}\wedge J^{3}$ \\
$\mathcal{V}_{\text{\textsc{ab}}}$ & $=$ & $\frac{1}{2}\dint\nolimits_{%
{\small CY}_{{\small 4}}}J_{\text{\textsc{a}}}\wedge J_{\text{\textsc{b}}%
}\wedge J^{2}$ \\
$\mathcal{V}_{\text{\textsc{abc}}}$ & $=$ & $\dint\nolimits_{{\small CY}_{%
{\small 4}}}J_{\text{\textsc{a}}}\wedge J_{\text{\textsc{b}}}\wedge J_{\text{%
\textsc{c}}}\wedge J$%
\end{tabular}
\label{3V}
\end{equation}%
They are related to the parameters $\upsilon ^{\text{\textsc{a}}}$ and the
coupling tensor $\kappa _{\text{\textsc{abcd}}}$ like%
\begin{equation}
\begin{tabular}{lll}
$\mathcal{V}_{\text{\textsc{a}}}$ & $=$ & $\frac{1}{6}\kappa _{\text{\textsc{%
abcd}}}\upsilon ^{\text{\textsc{b}}}\upsilon ^{\text{\textsc{c}}}\upsilon ^{%
\text{\textsc{d}}}$ \\
$\mathcal{V}_{\text{\textsc{ab}}}$ & $=$ & $\frac{1}{2}\kappa _{\text{%
\textsc{abcd}}}\upsilon ^{\text{\textsc{c}}}\upsilon ^{\text{\textsc{d}}}$
\\
$\mathcal{V}_{\text{\textsc{abc}}}$ & $=$ & $\kappa _{\text{\textsc{abcd}}%
}\upsilon ^{\text{\textsc{d}}}$%
\end{tabular}
\label{VA}
\end{equation}%
with the properties%
\begin{equation}
\begin{tabular}{lll}
$\upsilon ^{\text{\textsc{a}}}\mathcal{V}_{\text{\textsc{a}}}$ & $=$ & $4%
\mathcal{V}_{{\small CY}_{{\small 4}}}$ \\
$\upsilon ^{\text{\textsc{a}}}\upsilon ^{\text{\textsc{b}}}\mathcal{V}_{%
\text{\textsc{ab}}}$ & $=$ & $12\mathcal{V}_{{\small CY}_{{\small 4}}}$ \\
$\upsilon ^{\text{\textsc{a}}}\upsilon ^{\text{\textsc{b}}}\upsilon ^{\text{%
\textsc{c}}}\mathcal{V}_{\text{\textsc{abc}}}$ & $=$ & $24\mathcal{V}_{%
{\small CY}_{{\small 4}}}$%
\end{tabular}
\label{AV}
\end{equation}%
$\left( 2\right) $ their associated dimensionless partners are given by:
\begin{equation}
\hat{u}_{\text{\textsc{a}}}=\frac{\mathcal{V}_{\text{\textsc{a}}}}{\mathcal{V%
}^{3/4}}\quad ,\qquad \hat{u}_{\text{\textsc{ab}}}=\frac{\mathcal{V}_{\text{%
\textsc{ab}}}}{\mathcal{V}^{2/4}}\quad ,\qquad \hat{u}_{\text{\textsc{abc}}}=%
\frac{\mathcal{V}_{\text{\textsc{abc}}}}{\mathcal{V}^{1/4}}  \label{VB}
\end{equation}%
By setting $\hat{\upsilon}^{\text{\textsc{a}}}=\upsilon ^{\text{\textsc{a}}}/%
\mathcal{V}_{{\small CY}_{{\small 4}}}^{1/4},$ we have the properties%
\begin{equation}
\begin{tabular}{lll}
$\hat{\upsilon}^{\text{\textsc{a}}}\hat{u}_{\text{\textsc{a}}}$ & $=$ & $4$
\\
$\hat{\upsilon}^{\text{\textsc{a}}}\hat{\upsilon}^{\text{\textsc{b}}}\hat{u}%
_{\text{\textsc{ab}}}$ & $=$ & $12$ \\
$\hat{\upsilon}^{\text{\textsc{a}}}\hat{\upsilon}^{\text{\textsc{b}}}\hat{%
\upsilon}^{\text{\textsc{c}}}\hat{u}_{\text{\textsc{abc}}}$ & $=$ & $24$%
\end{tabular}
\label{BV}
\end{equation}%
The metric $\mathcal{G}_{\text{\textsc{ab}}}$ (\ref{met}) plays an important
role in the physical description of the 3D effective theory. Being a
dimensionless quantity, this rank 2 tensor ($h_{{\small X_{4}}}^{1,1}\times
h_{{\small X_{4}}}^{1,1}$ symmetric matrix) can be decomposed in terms of
dimensionless volumes $\hat{u}_{\text{\textsc{a}}}$ and $\hat{u}_{\text{%
\textsc{ab}}}$ of the 6- cycles $\boldsymbol{D}_{\text{\textsc{a}}}$ and the
4-cycles $\boldsymbol{S}_{\text{\textsc{ab}}}$ in the $X_{4}$ like%
\begin{equation}
\mathcal{G}_{\text{\textsc{ab}}}=\hat{u}_{\text{\textsc{a}}}\hat{u}_{\text{%
\textsc{b}}}-\hat{u}_{\text{\textsc{ab}}}  \label{gij}
\end{equation}%
Before proceeding notice the three following:

\begin{itemize}
\item The volumes in eq(\ref{VLS}) have the scale dimensions
\begin{equation}
\lbrack \mathcal{V}_{{\small CY}_{{\small 4}}}]=\QTR{sl}{L}^{8}\quad ,\quad
\lbrack \mathcal{V}_{\text{\textsc{a}}}]=\QTR{sl}{L}^{6}\quad ,\quad \lbrack
\mathcal{V}_{\text{\textsc{ab}}}]=\QTR{sl}{L}^{4}\quad ,\quad \lbrack
\mathcal{V}_{\text{\textsc{abc}}}]=\QTR{sl}{L}^{2}
\end{equation}%
but $[\mathcal{V}_{{\small X}_{{\small 4}}}]=\QTR{sl}{L}^{0};$ the letter $%
\QTR{sl}{L}\ $refers to $\emph{Length=1/Mass}$.

\item Being a real symmetric matrix, the $\mathcal{G}_{\text{\textsc{ab}}}$ (%
\ref{gij}) can be diagonalised by an orthogonal transformation matrix $%
\mathcal{O}$; thus allowing to put it into the form \textrm{\cite{25AC,25AD}}
\begin{equation}
\mathcal{G}_{\text{\textsc{ab}}}^{diag}=\xi _{\text{\textsc{a}}}\delta _{%
\text{\textsc{ab}}}\qquad ,\qquad \frac{1}{g^{2}}\mathcal{G}_{\text{\textsc{%
ab}}}^{diag}=\frac{1}{g_{\text{\textsc{a}}}^{2}}\delta _{\text{\textsc{ab}}%
}\qquad ,\qquad \frac{1}{g_{\text{\textsc{a}}}^{2}}=\frac{\xi _{\text{%
\textsc{a}}}}{g^{2}}  \label{gi}
\end{equation}%
with some positive eigenvalues $\xi _{\text{\textsc{a}}}$. In this relation,
we have\ replaced $g_{3D}$ simply by $g$ to avoid confusing notations.
\end{itemize}

\ \newline
Regarding the weak gauge regime in M-theory on $X_{4}$, the Asymptotic WGC
probes weakly coupled directions in the charge lattice of $X_{4}$. These
charge directions are closely related to the gauge kinetic matrix $\mathcal{G%
}_{\text{\textsc{ab}}}$ presented above.\ In fact, to have a weakly coupled
gauge field $A_{1}$ in the 3D effective field theory descending from
M-theory on $X_{4}$, we proceed as follows: \newline
First, we think about the 1-form gauge potential $A_{1}$ in terms of a
linear combination of individual 3D gauge potentials like $A_{1}=\sum n_{%
\text{\textsc{a}}}A_{1}^{\text{\textsc{a}}}$ with some integers $n_{\text{%
\textsc{a}}}.$ By using (\ref{C3}), in particular the expression of $A_{1}^{%
\text{\textsc{a}}}$ in terms of the 3-form potential $\boldsymbol{C}_{3}$
namely%
\begin{equation}
A_{1}^{\text{\textsc{a}}}=\dint\nolimits_{\mathcal{C}^{\text{\textsc{a}}}}%
\boldsymbol{C}_{3}\qquad ,\qquad \dint\nolimits_{\mathcal{C}^{\text{\textsc{a%
}}}}J_{\text{\textsc{b}}}=\delta _{\text{\textsc{b}}}^{\text{\textsc{a}}}
\end{equation}%
with basis curves $\mathcal{C}^{\text{\textsc{a}}}$ dual to the divisors $J_{%
\text{\textsc{a}}}$; it results that the gauge potential $A_{1}=\sum_{\text{%
\textsc{a}}}n_{\text{\textsc{a}}}A_{1}^{\text{\textsc{a}}}$ is in fact given
by the compactification of $\boldsymbol{C}_{3}$ over a complex curve $%
\mathcal{C}_{\mathbf{n}}=\sum_{\text{\textsc{a}}}n_{\text{\textsc{a}}}%
\mathcal{C}^{\text{\textsc{a}}}$ with integer vector $\mathbf{n=(}%
n_{1},...,n_{h_{{\small X}_{{\small 4}}}^{1,1}}\mathbf{)}.$ Explicitly, this
$A_{1}$ may be denoted like
\begin{equation}
A_{1}\left[ \mathcal{C}_{\mathbf{n}}\right] =\dint\nolimits_{\mathcal{C}_{%
\mathbf{n}}}\boldsymbol{C}_{3}
\end{equation}%
which by substituting $\mathcal{C}_{\mathbf{n}}=\sum n_{\text{\textsc{a}}}%
\mathcal{C}^{\text{\textsc{a}}}$, it expands as follows%
\begin{equation}
A_{1}\left[ \mathcal{C}_{\mathbf{n}}\right] =\sum_{\text{\textsc{a}}}n_{%
\text{\textsc{a}}}A_{1}\left[ \mathcal{C}^{\text{\textsc{a}}}\right] \qquad
,\qquad A_{1}\left[ \mathcal{C}^{\text{\textsc{a}}}\right] =\dint\nolimits_{%
\mathcal{C}^{\text{\textsc{a}}}}\boldsymbol{C}_{3}
\end{equation}%
Naively, the weak gauge coupling in the infinite distance limit ($\lambda
\rightarrow \infty $) corresponds to
\begin{equation}
g_{\text{\textsc{a}}}^{2}\rightarrow 0  \label{WGCL}
\end{equation}%
As is obvious from the label \textsc{a,} this would depend on the curves in
the basis that we take into account; we will see later that these curves
should lie entirely in the fiber \textrm{(resp. in the base)} in order to
obtain the weak coupling condition \textrm{(resp. strong coupling condition)}%
.\ Also this behaviour was shown to be only a necessary condition for a more
constrained condition.\ Both these aspects are studied in the next
subsection.

\subsection{Gauge coupling regimes in M-theory on $K3_{\bot }\times
K3_{\Vert }$}

In this subsection, we investigate conditions for weak (resp. strong) gauge
coupling regimes in the large distance limit $\lambda \rightarrow \infty $ (
resp. short distance $\lambda \rightarrow 0$) in the Kahler moduli space $%
\mathfrak{M}_{{\small CY}_{{\small 4}}}$ for the fibration%
\begin{equation}
{\small CY}_{{\small 4}}=K3_{\bot }\times K3_{\Vert }
\end{equation}%
In other words, we want to look for regimes $\mathrm{g}_{\mathrm{YM}}^{weak}$
and $\mathrm{g}_{\mathrm{YM}}^{strong}$ of the effective gauge coupling
constants $\mathrm{g}_{\mathrm{YM}}$ in the 3D gauge theory descending from
M-theory on $X_{4}.$ Obviously, it is the weak coupling regime $\mathrm{g}_{%
\mathrm{YM}}^{weak}$ which is important for perturbative QFT$_{3D}$; but
here we want to also explore relevant aspects on the strong regime of $%
\mathrm{g}_{\mathrm{YM}}^{strong}$ in relationship with the mapping%
\begin{equation}
\mathrm{g}_{\mathrm{YM}}^{weak}\qquad \rightarrow \qquad \frac{1}{\mathrm{g}%
_{\mathrm{YM}}^{weak}}\sim \mathrm{g}_{\mathrm{YM}}^{strong}
\end{equation}%
To that purpose, we begin by introducing useful tools on the weak gravity
conjecture (WGC) as well as on the asymptotic weak coupling regime
compatible with WGC. After that, we turn to investigate with details the
weak (resp. strong) regimes of the gauge couplings ($\mathrm{g}_{\mathrm{YM}%
}^{weak}$/$\mathrm{g}_{\mathrm{YM}}^{strong}$) by using underlying U(1)$%
_{\Upsilon }$ gauge coupling constant $\mathrm{g}_{\Upsilon }$ with $%
\Upsilon $ labelling a complex curve in the $X_{4}.$

\subsubsection{Magnetic weak gauge coupling scale $\Lambda _{\text{\textsc{%
wgc}}}$}

We start by describing the relationships defining the weak gravity
conjecture in D- dimensional space time in the framework of M-theory on
Calabi-Yau manifolds with finite volume ($\mathcal{V}_{{\small CY}_{{\small %
11-2D}}}<\infty $). In D- dimensional gravity theory coupled to a U(1) gauge
potential with gauge coupling $g_{_{U(1)}},$ the WGC stipulates that there
should exist at least one massive charged particle state $\left\vert
m,Q\right\rangle $ with mass m and electric charge $Q=qg_{_{U(1)}}$ (q
integer) satisfying the inequality%
\begin{equation}
m\leq qg_{_{U(1)}}M_{\mathrm{Pl}}^{\frac{D-2}{2}}\sqrt{\frac{D-2}{D-3}}+....
\end{equation}%
Here, the $M_{\mathrm{Pl}}$ is the D- dimensional Planck mass\textrm{\ }%
related to Newton constant like $G_{N}\sim M_{\mathrm{Pl}}^{2-D}$. Such a
particle state is often called a \textrm{super-extremal} particle due to the
formulation of the WGC using extremal black holes \textrm{\cite{7,ZX,ZY,ZXY}}%
. Together with this electric WGC condition, we also have a magnetic WGC
relations given by%
\begin{equation}
\left. \Lambda _{_{\mathrm{WGC}}}^{2}\right\vert
_{_{U(1)}}=g_{_{U(1)}}^{2}M_{\mathrm{Pl}}^{D-2}  \label{lam}
\end{equation}%
In \textrm{\cite{14A}}, it was shown that the naive condition (\ref{WGCL})
is not enough to describe the weak coupling. In fact, the true weak gauge
coupling is defined as%
\begin{equation}
\lim_{\lambda \rightarrow \infty }\left( \frac{\left. \Lambda _{_{\mathrm{WGC%
}}}^{2}\right\vert _{_{U(1)}}}{\Lambda _{_{\mathrm{QG}}}^{2}}\right) =0
\label{lan}
\end{equation}%
where $\left. \Lambda _{_{\mathrm{WGC}}}^{2}\right\vert _{_{U(1)}}$ is the
magnetic WGC scale associated with the U(1) gauge symmetry. The $\Lambda _{_{%
\mathrm{WGC}}}$ scales like a mass ($\left[ \Lambda _{_{\mathrm{WGC}}}\right]
=Mass^{1}$); and the $g_{_{U(1)}}^{2}$ scales as,
\begin{equation}
\lbrack g_{_{U(1)}}^{2}]=\left( Length\right) ^{D-4}\qquad \Leftrightarrow
\qquad \lbrack g_{_{U(1)}}^{2}]=\left( Mass\right) ^{4-D}
\end{equation}%
The energy scale $\Lambda _{_{\mathrm{QG}}}$ in eq(\ref{lan}) is a quantum
gravity cut-off given by an energy level $\mathcal{E}_{_{\mathrm{QG}}}$
above which the gravitational coupling becomes strong; and thus the dynamics
of gravity changes in such a way that the description of the quantum theory
is no more consistent. Given the magnetic WGC relations (\ref{lam}-\ref{lan}%
) that we rewrite for a given $U(1)_{\text{\textsc{a}}}$, with a fixed label
\textsc{a}$=1,...,h_{{\small CY}_{4}}^{1,1},$ as follows%
\begin{equation}
\left. \Lambda _{_{\mathrm{WGC}}}^{2}\right\vert _{_{U(1)_{\text{\textsc{a}}%
}}}=g_{_{U(1)_{\text{\textsc{a}}}}}^{2}M_{\mathrm{Pl}}^{D-2}\qquad ,\qquad
\lim_{\lambda \rightarrow \infty }\left( \frac{\left. \Lambda _{_{\mathrm{WGC%
}}}^{2}\right\vert _{_{U(1)_{\text{\textsc{a}}}}}}{\Lambda _{_{\mathrm{QG}%
}}^{2}}\right) =0
\end{equation}%
Thus, we see that the condition of weak coupling limit $g_{_{U(1)_{\text{%
\textsc{a}}}}}^{2}\rightarrow 0$ derived in the previous subsection is
merely a necessary condition. The full condition needs to include the
quantum gravity cut-off to find the true weakly coupled directions.

\paragraph{\qquad \textbf{A. Extending eqs(\protect\ref{lam}-\protect\ref%
{lan}) to }$U(1)_{\Upsilon _{\mathbf{q}}}:$\newline
}

Here, we shall think about eqs(\ref{lam}-\ref{lan}) in terms of a generic
gauge symmetry $U(1)_{\Upsilon _{\mathbf{q}}}$ [and 1-form gauge field $%
A_{\Upsilon _{\mathbf{q}}}$ associated with $U(1)_{\Upsilon _{\mathbf{q}}}$%
]. This abelian gauge symmetry $U(1)_{\Upsilon _{\mathbf{q}}}$ is a subgroup
of $H_{\mathrm{r}_{\mathrm{g}}}=U(1)^{\mathrm{r}_{\mathrm{g}}}$ with $%
\mathrm{r}_{\mathrm{g}}=h_{{\small X}_{{\small 4}}}^{1,1}$; it is
characterised by an integral charge vector $\mathbf{q}=(q_{1},...,q_{\mathrm{%
r}_{\mathrm{g}}});$ the Lie algebra $u(1)_{\Upsilon _{\mathbf{q}}}$
associated with this gauge symmetry is given by the following linear
combination
\begin{equation}
u(1)_{\Upsilon _{\mathbf{q}}}=\sum_{\text{\textsc{a}}}q_{\text{\textsc{a}}%
}u(1)_{C^{\text{\textsc{a}}}}\qquad ,\qquad A_{\Upsilon _{\mathbf{q}}}=\sum_{%
\text{\textsc{a}}}q_{\text{\textsc{a}}}\mathfrak{A}_{C^{\text{\textsc{a}}}}
\end{equation}%
where, for later use, we set
\begin{equation}
u(1)_{\Upsilon _{\mathbf{q}}}\equiv u(1)_{\mathbf{q}}\qquad ,\qquad u(1)_{C^{%
\text{\textsc{a}}}}\equiv u(1)^{\text{\textsc{a}}}\qquad ,\qquad \mathfrak{A}%
_{C^{\text{\textsc{a}}}}\equiv \mathfrak{A}^{\text{\textsc{a}}}
\end{equation}%
By using $\left( i\right) $ the commuting Cartan-like generators \{$%
\mathfrak{h}_{\text{\textsc{a}}}$\} of the abelian $H_{\mathrm{r}_{\mathrm{g}%
}},$ and $\left( ii\right) $ the factorisation
\begin{equation}
H_{\mathrm{r}_{\mathrm{g}}}=U(1)_{\mathbf{q}}\times H^{\vee }\qquad ,\qquad
H^{\vee }\equiv \frac{H_{\mathrm{r}_{\mathrm{g}}}}{U(1)_{\mathbf{q}}}
\label{cartan}
\end{equation}%
the abelian factor $U(1)_{\mathbf{q}}$ is generated by a diagonal generator $%
h_{\mathbf{t}}^{0}$ given by a linear combination of the diagonal \{$%
\mathfrak{h}_{\text{\textsc{a}}}$\} of like%
\begin{equation}
h_{\mathbf{T}_{0}}=\sum_{\text{\textsc{a}}=0}^{\mathrm{r}_{\mathrm{g}%
}-1}t_{0}^{\text{\textsc{a}}}\mathfrak{h}_{\text{\textsc{a}}}
\end{equation}%
with parameters $\mathbf{T}_{0}=(t_{0}^{0},t_{0}^{1},...,t_{0}^{\mathrm{r}_{%
\mathrm{g}}-1})$. The other $\mathrm{r}_{\mathrm{g}}-1$ diagonal generators $%
h_{\mathbf{T}_{0}}$ generating the coset group $H^{\vee }=H_{\mathrm{r}_{%
\mathrm{g}}}/U(1)_{\mathbf{q}}$ expands like
\begin{equation}
h_{\mathbf{T}_{a}}=\sum_{\text{\textsc{a}}=0}^{\mathrm{r}_{\mathrm{g}%
}}t_{a}^{\text{\textsc{a}}}\mathfrak{h}_{\text{\textsc{a}}}\qquad ,\qquad
a=1,...,\mathrm{r}_{\mathrm{g}}-1  \label{ze}
\end{equation}%
If combining $t^{\text{\textsc{a}}}$ and $t_{a}^{\text{\textsc{a}}}$ into a $%
\mathrm{r}_{\mathrm{g}}\times \mathrm{r}_{\mathrm{g}}$ charge matrix $T_{%
\text{\textsc{a}}}^{\text{\textsc{b}}},$ we have%
\begin{equation}
h_{\mathbf{t}_{\text{\textsc{a}}}}=\sum_{\text{\textsc{b}}=0}^{\mathrm{r}_{%
\mathrm{g}}-1}T_{\text{\textsc{a}}}^{\text{\textsc{b}}}\mathfrak{h}_{\text{%
\textsc{b}}}\qquad with\qquad \det (T_{\text{\textsc{a}}}^{\text{\textsc{b}}%
})\neq 0
\end{equation}%
In what follows, we will be interested by the particular abelian $U(1)_{%
\mathbf{q}}$ \ and the associated gauge field potential $A_{\mathbf{q}}$
given by the linear combination
\begin{equation}
A_{\mathbf{q}}=\sum_{\text{\textsc{a}}=0}^{\mathrm{r}_{\mathrm{g}}-1}q_{%
\text{\textsc{a}}}\mathfrak{A}^{\text{\textsc{a}}}\qquad with\qquad
\mathfrak{A}^{\text{\textsc{a}}}=\mathfrak{A}_{\mu }^{\text{\textsc{a}}%
}dx^{\mu }
\end{equation}%
For this particular $U(1)_{\mathbf{q}}$ abelian gauge symmetry, the
conditions (\ref{lam}-\ref{lan}) for the \emph{magnetic} WGC reads as
follows,%
\begin{equation}
\left. \Lambda _{_{\mathrm{WGC}}}^{2}\right\vert _{_{U(1)_{\mathbf{q}%
}}}=g_{_{U(1)_{\mathbf{q}}}}^{2}M_{\mathrm{Pl}}^{D-2}  \label{lmn}
\end{equation}%
and%
\begin{equation}
\lim_{\lambda \rightarrow \infty }\left( \frac{\left. \Lambda _{_{\mathrm{WGC%
}}}^{2}\right\vert _{_{U(1)_{\mathbf{q}}}}}{\Lambda _{_{\mathrm{QG}}}^{2}}%
\right) =0  \label{Weak}
\end{equation}%
In this relation, the squared gauge coupling $g_{_{U(1)_{\mathbf{q}}}}^{2}$
is quadratic in the partial gauge coupling $g_{\text{\textsc{a}}}$
associated with the individual gauge group factors $U(1)_{q_{\text{\textsc{a}%
}}}\equiv U(1)_{\text{\textsc{a}}};$ it reads explicitly like
\begin{equation}
g_{_{U(1)_{\mathbf{q}}}}^{2}=\mathrm{g}_{\text{\textsc{a}}}\mathcal{G}^{%
\text{\textsc{ab}}}\mathrm{g}_{\text{\textsc{b}}}
\end{equation}%
where $\mathrm{g}_{\text{\textsc{a}}}=q_{\text{\textsc{a}}}\mathrm{g}_{_{%
\mathrm{YM}}}$ and where $\mathrm{g}_{_{\mathrm{YM}}}$ is Yang-Mills
coupling; the usual non abelian gauge constant of the underlying D-
dimensional effective field theory with gauge invariance $\boldsymbol{G}_{%
\mathrm{r}_{\mathrm{g}}}$. The $\mathcal{G}^{\text{\textsc{ab}}}$ is the
inverse of the effective gauge coupling metric $\mathcal{G}_{\text{\textsc{ab%
}}}$ (\ref{met}) involved in the field action (\ref{S3D}).

\paragraph{\qquad \textbf{B. Weak gravity conjecture in 3D:}\newline
}

In three space time dimensions, the magnetic WGC relation (\ref{lam}) reads
like $\left. \Lambda _{_{\mathrm{WGC}}}^{2}\right\vert _{_{U(1)_{\mathbf{q}%
}}}=g_{_{U(1)_{\mathbf{q}}}}^{2}M_{\mathrm{Pl}}$ with the squared $%
g_{_{U(1)_{\mathbf{q}}}}^{2}$ expressed as follows
\begin{equation}
\mathrm{g}_{_{\Upsilon _{\mathbf{q}}}}^{2}=g_{3D}^{2}\left( q_{\text{\textsc{%
a}}}\mathcal{G}^{\text{\textsc{ab}}}q_{\text{\textsc{b}}}\right)  \label{gc}
\end{equation}%
the scaling dimension being $[\mathrm{g}_{\text{\textsc{a}}%
}^{2}]=[g_{3D}^{2}]=Length^{-1}$; the couplings $\mathrm{g}_{\text{\textsc{a}%
}}$ and $g_{3D}$ have a mass dimension. Notice also the two following: $%
\left( \mathbf{i}\right) $ by substituting (\ref{gc}) into $\left. \Lambda
_{_{\mathrm{WGC}}}^{2}\right\vert _{_{U(1)_{\mathbf{q}}}}=g_{_{U(1)_{\mathbf{%
q}}}}^{2}M_{\mathrm{Pl}},$ we obtain%
\begin{equation}
\left. \Lambda _{_{\mathrm{WGC}}}^{2}\right\vert _{_{u(1)_{\mathbf{q}%
}}}=g_{3D}^{2}\left( q_{\text{\textsc{a}}}\mathcal{G}^{\text{\textsc{ab}}}q_{%
\text{\textsc{b}}}\right) M_{\mathrm{Pl}}  \label{ml}
\end{equation}%
this is a function of the gauge coupling constant $g_{3D}$ but also on the
charges $q_{\text{\textsc{a}}}$ and the inverse of the metric of the moduli
space of the 3D effective gauge theory. $\left( \mathbf{ii}\right) $ Putting
into (\ref{lmn}), the weak gravity conjecture reads like%
\begin{equation}
\lim_{\lambda \rightarrow \infty }\left( \frac{g_{3D}^{2}M_{\mathrm{Pl}%
}\left( q_{\text{\textsc{a}}}\mathcal{G}^{\text{\textsc{ab}}}q_{\text{%
\textsc{b}}}\right) }{\Lambda _{_{\mathrm{QG}}}^{2}}\right) =0
\end{equation}

To illustrate the different regimes of the gauge coupling, we consider the
example of M-theory on $X_{4}$ in the infinite distance limit of Type-$%
\mathbb{S}$, with an emphasis on type K3 \textrm{\cite{25B,25C}}; that is a $%
X_{4}$ given by K3 surface fibered over a complex base surface $%
\mathbb{B}_{2}$. This surface $\mathbb{B}_{2}$ can be imagined as one of the
familiar surfaces like del Pezzo surfaces dP$_{n},$ given by the blow up of $%
\mathbb{P}^{2}$ at n points ($n\leq 8$) \textrm{\cite{25D,25DA,25DB}}, or
the Hirzebruch surfaces $\mathbb{F}_{n}$ based on the fibration $\mathbb{P}%
^{1}\times \mathbb{P}^{1}$ \textrm{\cite{25E,25EA}}.\textrm{\ }For
illustration and in order to have quite simple calculations, we give below a
toy model where the base\textrm{\ }$\mathbb{B}_{2}$\textrm{\ }is taken as
another K3 surface, this choice turns out to exhibit some interesting
features.

\subsubsection{Gauge coupling regimes in M-theory on $K3_{\bot }\times
K3_{\Vert }$}

We have seen that a necessary condition of the weak coupling $g_{_{\Upsilon
}}^{2}=g_{3D}^{2}(q_{\text{\textsc{a}}}\mathcal{G}^{\text{\textsc{ab}}}q_{%
\text{\textsc{b}}})$ in the infinite distance limit ($\lambda \rightarrow
\infty $) that is associated with some complex curve $\Upsilon _{\mathbf{q}}$
and gauge symmetry $U(1)_{\mathbf{q}}$ as%
\begin{equation}
\Upsilon _{\mathbf{q}}=\sum_{\text{\textsc{a}}=1}^{h_{X_{4}}^{1,1}}q_{\text{%
\textsc{a}}}\mathcal{C}^{\text{\textsc{a}}}\qquad \leftrightarrow \qquad
u(1)_{\mathbf{q}}=\sum_{\text{\textsc{a}}=1}^{h_{X_{4}}^{1,1}}q_{\text{%
\textsc{a}}}u(1)^{\text{\textsc{a}}}\qquad ,\qquad u(1)_{\Upsilon _{\mathbf{q%
}}}\equiv u(1)_{\mathbf{q}}
\end{equation}%
is given by%
\begin{equation}
\lim_{\lambda \rightarrow \infty }\left[ g_{3D}^{2}(q_{\text{\textsc{a}}}%
\mathcal{G}^{\text{\textsc{ab}}}q_{\text{\textsc{b}}})\right] =0
\end{equation}%
Because $\mathcal{G}_{\text{\textsc{ab}}}$ defined by eq(\ref{met}) is a
function of the Kahler form $J,$ which depends on the parameter $\lambda ,$
it follows that $\mathcal{G}_{\text{\textsc{ab}}}$ and its inverse $\mathcal{%
G}^{\text{\textsc{ab}}}$ are also functions of $\lambda .$ Moreover,
expressing (\ref{22}) giving the asymptotic expansion of the Kahler 2-form $%
J $ for the case of $X_{4}=K3_{\perp }\times K3_{\parallel }$ like%
\begin{equation}
J=\lambda J_{\parallel }+\frac{1}{\lambda }J_{\perp }  \label{ja}
\end{equation}%
with%
\begin{equation}
J_{\parallel }=\sum_{a_{\parallel }=1}^{h_{\parallel }^{1,1}}\upsilon
^{a_{\parallel }}J_{a_{\parallel }}\qquad ,\qquad J_{\perp }=\sum_{i_{\bot
}=1}^{h_{\perp }^{1,1}}\upsilon ^{i_{\bot }}J_{i_{\bot }}  \label{jp}
\end{equation}%
where we have set $h_{\parallel }^{1,1}=h^{1,1}(K3_{\parallel })$ and $%
h_{\perp }^{1,1}=h^{1,1}(K3_{\perp }),$ it results that $\mathcal{G}_{\text{%
\textsc{ab}}}$ can be splitted like%
\begin{equation}
\mathcal{G}_{\text{\textsc{ab}}}=\left(
\begin{array}{cc}
\mathcal{G}_{\parallel \parallel } & \mathcal{G}_{\parallel \perp } \\
\mathcal{G}_{\perp \parallel } & \mathcal{G}_{\perp \perp }%
\end{array}%
\right) ,\qquad \mathcal{G}^{\text{\textsc{ab}}}=\left(
\begin{array}{cc}
\mathcal{G}^{\parallel \parallel } & \mathcal{G}^{\parallel \perp } \\
\mathcal{G}^{\perp \parallel } & \mathcal{G}^{\perp \perp }%
\end{array}%
\right)  \label{tem}
\end{equation}%
with%
\begin{equation}
\begin{tabular}{lll}
$\mathcal{G}_{\parallel \parallel }$ & $=$ & $\hat{u}_{\parallel }\hat{u}%
_{\parallel }-\hat{u}_{\parallel \parallel }$ \\
$\mathcal{G}_{\perp \perp }$ & $=$ & $\hat{u}_{\perp }\hat{u}_{\perp }-\hat{u%
}_{\perp \perp }$ \\
$\mathcal{G}_{\parallel \perp }$ & $=$ & $\hat{u}_{\parallel }\hat{u}_{\perp
}-\hat{u}_{\parallel \perp }$%
\end{tabular}%
\end{equation}%
Notice the following useful features: $\left( \mathbf{1}\right) $ the
inverse $\mathcal{G}^{\text{\textsc{ab}}}$ in terms of the components of $%
\mathcal{G}_{\text{\textsc{ab}}}$ can be formally exhibited as follows%
\begin{equation}
\mathcal{G}^{\text{\textsc{ab}}}=\left(
\begin{array}{cc}
\frac{\mathcal{G}_{\perp \perp }}{\mathcal{G}_{\parallel \parallel }\mathcal{%
G}_{\perp \perp }-\mathcal{G}_{\parallel \perp }\mathcal{G}_{\perp \parallel
}} & -\frac{\mathcal{G}_{\parallel \perp }}{\mathcal{G}_{\parallel \parallel
}\mathcal{G}_{\perp \perp }-\mathcal{G}_{\parallel \perp }\mathcal{G}^{\perp
\parallel }} \\
-\frac{\mathcal{G}_{\perp \parallel }}{\mathcal{G}_{\parallel \parallel }%
\mathcal{G}_{\perp \perp }-\mathcal{G}_{\parallel \perp }\mathcal{G}_{\perp
\parallel }} & \frac{\mathcal{G}_{\parallel \parallel }}{\mathcal{G}%
_{\parallel \parallel }\mathcal{G}_{\perp \perp }-\mathcal{G}_{\parallel
\perp }\mathcal{G}_{\perp \parallel }}%
\end{array}%
\right)
\end{equation}%
from which we see that in the infinite distance limit the $\mathcal{G}%
^{\parallel \parallel }$ behaves as the one of $\mathcal{G}_{\perp \perp }$,
and the $\mathcal{G}^{\perp \perp }$ behaves as the one of $\mathcal{G}%
_{\parallel \parallel }.$ As shown below, the behaviours for $\lambda
\rightarrow \infty $ read as follows%
\begin{equation}
\begin{tabular}{lllllll}
$\mathcal{G}_{\bot \bot }$ & $\sim $ & $\lambda ^{+2}\mathring{G}_{\bot \bot
}$ & $\qquad ,\qquad $ & $\mathcal{G}^{\bot \bot }$ & $\sim $ & $\lambda
^{-2}\mathring{G}^{\bot \bot }$ \\
$\mathcal{G}_{\bot \Vert }$ & $\sim $ & $\lambda ^{0}\mathring{G}_{\bot
\Vert }$ & $\qquad ,\qquad $ & $\mathcal{G}^{\bot \Vert }$ & $\sim $ & $%
\lambda ^{0}\mathring{G}^{\bot \Vert }$ \\
&  &  &  &  &  &  \\
$\mathcal{G}_{\Vert \Vert }$ & $\sim $ & $\lambda ^{-2}\mathring{G}_{\Vert
\Vert }$ & $\qquad ,\qquad $ & $\mathcal{G}^{\Vert \Vert }$ & $\sim $ & $%
\lambda ^{+2}\mathring{G}^{\Vert \Vert }$ \\
$\mathcal{G}_{\Vert \bot }$ & $\sim $ & $\lambda ^{0}\mathring{G}_{\Vert
\bot }$ & $\qquad ,\qquad $ & $\mathcal{G}^{\Vert \bot }$ & $\sim $ & $%
\lambda ^{0}\mathring{G}^{\Vert \bot }$%
\end{tabular}
\label{beh}
\end{equation}%
$\left( \mathbf{2}\right) $ Using the parametrisation (\ref{ja}) and the
metric (\ref{met}), we can calculate the behaviour of the variables $\left(
\hat{u}_{\parallel },\hat{u}_{\perp },\hat{u}_{\bot \bot },\hat{u}%
_{\parallel \parallel }\right) $ in the infinite distance limit in terms of
the variables $\upsilon ^{\Vert }$\ and $\hat{\upsilon}^{\bot }$. We have
for $\hat{u}_{\parallel }=\frac{1}{6}\kappa _{\Vert \text{\textsc{bcd}}}\hat{%
\upsilon}^{\text{\textsc{b}}}\hat{\upsilon}^{\text{\textsc{c}}}\hat{\upsilon}%
^{\text{\textsc{d}}}$ and $\hat{u}_{\parallel \parallel }=\frac{1}{2}\kappa
_{\Vert \Vert \text{\textsc{cd}}}\hat{\upsilon}^{\text{\textsc{c}}}\hat{%
\upsilon}^{\text{\textsc{d}}}$\ as well as $\hat{u}_{\parallel \bot }=\frac{1%
}{2}\kappa _{\Vert \bot \text{\textsc{cd}}}\hat{\upsilon}^{\text{\textsc{c}}}%
\hat{\upsilon}^{\text{\textsc{d}}}$ the following,%
\begin{equation}
\begin{tabular}{lll}
$\hat{u}_{\parallel }$ & $=$ & $\frac{1}{6\lambda ^{3}}\kappa _{_{\Vert \bot
\bot \bot }}\left( \hat{\upsilon}^{\bot }\right) ^{3}+\frac{1}{2\lambda }%
\kappa _{_{\Vert \Vert \bot \bot }}\hat{\upsilon}^{\Vert }\left( \hat{%
\upsilon}^{\bot }\right) ^{2}+\frac{\lambda }{2}\kappa _{_{\Vert \Vert \Vert
\bot }}(\hat{\upsilon}^{\Vert })^{2}\left( \hat{\upsilon}^{\bot }\right) +%
\frac{\lambda ^{3}}{6}\kappa _{_{\Vert \Vert \Vert \Vert }}^{3}(\hat{\upsilon%
}^{\Vert })^{3}$ \\
$\hat{u}_{\parallel \parallel }$ & $=$ & $\frac{1}{2\lambda ^{2}}\kappa
_{_{\Vert \Vert \bot \bot }}\left( \hat{\upsilon}^{\bot }\right) ^{2}+\kappa
_{_{\Vert \Vert \Vert \bot }}\hat{\upsilon}^{\Vert }\hat{\upsilon}^{\bot }+%
\frac{\lambda ^{2}}{2}\kappa _{_{\Vert \Vert \Vert \Vert }}(\hat{\upsilon}%
^{\Vert })^{2}$ \\
$\hat{u}_{\parallel \bot }$ & $=$ & $\frac{1}{2\lambda ^{2}}\kappa _{_{\Vert
\bot \bot \bot }}\left( \hat{\upsilon}^{\bot }\right) ^{2}+\kappa _{_{\Vert
\bot \Vert \bot }}\hat{\upsilon}^{\Vert }\hat{\upsilon}^{\bot }+\frac{%
\lambda ^{2}}{2}\kappa _{_{\Vert \bot \Vert \Vert }}(\hat{\upsilon}^{\Vert
})^{2}$%
\end{tabular}
\label{52}
\end{equation}%
For the case of the variable $\hat{u}_{\perp }=\frac{1}{6}\kappa _{\perp
\text{\textsc{bcd}}}\hat{\upsilon}^{\text{\textsc{b}}}\hat{\upsilon}^{\text{%
\textsc{c}}}\hat{\upsilon}^{\text{\textsc{d}}},$ the $\hat{u}_{\bot \bot }=%
\frac{1}{2}\kappa _{\bot \bot \text{\textsc{cd}}}\hat{\upsilon}^{\text{%
\textsc{c}}}\hat{\upsilon}^{\text{\textsc{d}}}$ and the $\hat{u}_{\bot
\parallel },$ we have%
\begin{equation}
\begin{tabular}{lll}
$\hat{u}_{\perp }$ & $=$ & $\frac{1}{6\lambda ^{3}}\kappa _{_{\bot \bot \bot
\bot }}\left( \hat{\upsilon}^{\bot }\right) ^{3}+\frac{1}{2\lambda }\kappa
_{_{\bot \bot \bot \Vert }}\hat{\upsilon}^{\Vert }\left( \hat{\upsilon}%
^{\bot }\right) ^{2}+\frac{\lambda }{2}\kappa _{_{\bot \bot \Vert \Vert }}(%
\hat{\upsilon}^{\Vert })^{2}\left( \hat{\upsilon}^{\bot }\right) +\frac{%
\lambda ^{3}}{6}\kappa _{_{\bot \Vert \Vert \Vert }}(\hat{\upsilon}^{\Vert
})^{3}$ \\
$\hat{u}_{\bot \bot }$ & $=$ & $\frac{1}{2\lambda ^{2}}\kappa _{_{\bot \bot
\bot \bot }}\left( \hat{\upsilon}^{\bot }\right) ^{2}+\kappa _{_{\bot \bot
\bot \Vert }}\left( \hat{\upsilon}^{\bot }\right) \upsilon ^{\Vert }+\frac{%
\lambda ^{2}}{2}\kappa _{_{\bot \bot \Vert \Vert }}(\hat{\upsilon}^{\Vert
})^{2}$ \\
$\hat{u}_{\bot \parallel }$ & $=$ & $\frac{1}{2\lambda ^{2}}\kappa _{_{\bot
\Vert \bot \bot }}\left( \hat{\upsilon}^{\bot }\right) ^{2}+\kappa _{_{\bot
\Vert \Vert \bot }}\hat{\upsilon}^{\Vert }\hat{\upsilon}^{\bot }+\frac{%
\lambda ^{2}}{2}\kappa _{_{\bot \Vert \Vert \Vert }}(\hat{\upsilon}^{\Vert
})^{2}$%
\end{tabular}
\label{53}
\end{equation}%
Depending on the value of $\lambda ,$ we distinguish the behaviors:

\begin{itemize}
\item \emph{Infinite distance limit} ($\lambda \rightarrow \infty $):
\begin{equation}
\begin{tabular}{lll}
$\hat{u}_{\parallel }$ & $\simeq $ & $\frac{\lambda ^{3}}{6}\kappa _{_{\Vert
\Vert \Vert \Vert }}(\hat{\upsilon}^{\Vert })^{3}$ \\
$\hat{u}_{\parallel \parallel }$ & $\simeq $ & $\frac{\lambda ^{2}}{2}\kappa
_{_{\Vert \Vert \Vert \Vert }}(\hat{\upsilon}^{\Vert })^{2}$ \\
$\hat{u}_{\parallel \bot }$ & $\simeq $ & $\frac{\lambda ^{2}}{2}\kappa
_{_{\Vert \bot \Vert \Vert }}(\hat{\upsilon}^{\Vert })^{2}$%
\end{tabular}%
\qquad ,\qquad
\begin{tabular}{lll}
$\hat{u}_{\perp }$ & $\simeq $ & $\frac{\lambda ^{3}}{6}\kappa _{_{\bot
\Vert \Vert \Vert }}(\hat{\upsilon}^{\Vert })^{3}$ \\
$\hat{u}_{\bot \bot }$ & $\simeq $ & $\frac{\lambda ^{2}}{2}\kappa _{_{\bot
\bot \Vert \Vert }}(\hat{\upsilon}^{\Vert })^{2}$ \\
$\hat{u}_{\bot \parallel }$ & $\simeq $ & $\frac{\lambda ^{2}}{2}\kappa
_{_{\bot \Vert \Vert \Vert }}(\hat{\upsilon}^{\Vert })^{2}$%
\end{tabular}
\label{up}
\end{equation}

\item \emph{Short distance limit} ($\lambda \rightarrow 0$):
\begin{equation}
\begin{tabular}{lll}
$\hat{u}_{\parallel }$ & $\simeq $ & $\frac{1}{6\lambda ^{3}}\kappa
_{_{\Vert \bot \bot \bot }}\left( \hat{\upsilon}^{\bot }\right) ^{3}\ \ $ \\
$\hat{u}_{\parallel \parallel }$ & $\simeq $ & $\frac{1}{2\lambda ^{2}}%
\kappa _{_{\Vert \Vert \bot \bot }}\left( \hat{\upsilon}^{\bot }\right)
^{2}\ $ \\
$\hat{u}_{\parallel \bot }$ & $\simeq $ & $\frac{1}{2\lambda ^{2}}\kappa
_{_{\Vert \bot \bot \bot }}\left( \hat{\upsilon}^{\bot }\right) ^{2}\ $%
\end{tabular}%
\qquad ,\qquad
\begin{tabular}{lll}
$\hat{u}_{\perp }$ & $\simeq $ & $\frac{1}{6\lambda ^{3}}\kappa _{_{\bot
\bot \bot \bot }}\left( \hat{\upsilon}^{\bot }\right) ^{3}$ \\
$\hat{u}_{\bot \bot }$ & $\simeq $ & $\frac{1}{2\lambda ^{2}}\kappa _{_{\bot
\bot \bot \bot }}\left( \hat{\upsilon}^{\bot }\right) ^{2}$ \\
$\hat{u}_{\bot \parallel }$ & $\simeq $ & $\frac{1}{2\lambda ^{2}}\kappa
_{_{\bot \Vert \bot \bot }}\left( \hat{\upsilon}^{\bot }\right) ^{2}$%
\end{tabular}
\label{pu}
\end{equation}
\end{itemize}

Notice that the above expressions (\ref{up}) and (\ref{pu}) are related
under $\lambda \rightarrow 1/\lambda .$ Notice also that by this duality,
the basis of curves in $X_{4}$ is given by the union of two sets like%
\begin{equation}
\{\mathcal{C}^{\text{\textsc{a}}}\}_{X_{4}}=\{\mathcal{C}^{\text{\textsc{a}}%
}\}_{K3_{\parallel }}\quad \dbigcup \quad \{\mathcal{C}^{\text{\textsc{a}}%
}\}_{K3_{\perp }}
\end{equation}%
with%
\begin{equation}
\begin{tabular}{lll}
$\{\mathcal{C}^{\text{\textsc{a}}}\}_{K3_{\parallel }}$ & $=$ & $\{\mathcal{C%
}^{a_{\parallel }}\}_{a_{\parallel }=1,...,h_{\parallel }^{1,1}}$ \\
$\{\mathcal{C}^{\text{\textsc{a}}}\}_{K3_{\perp }}$ & $=$ & $\{\mathcal{C}%
^{i_{\perp }}\}_{i_{\perp }=1,...,h_{\perp }^{1,1}}$%
\end{tabular}%
\end{equation}%
This notation allows to express the expansion of the complex curve $\Upsilon
_{\mathbf{q}}$ formally like $\Upsilon _{\mathbf{q}_{\parallel }}+\Upsilon _{%
\mathbf{q}_{\perp }}$ with%
\begin{equation}
\Upsilon _{\mathbf{q}_{\parallel }}=\sum_{a_{\parallel }=1_{\parallel
}}^{h_{\parallel }^{1,1}}q_{a_{\parallel }}\mathcal{C}^{a_{\parallel
}}\qquad ,\qquad \Upsilon _{\mathbf{q}_{\perp }}=\sum_{i_{\perp }=1_{\perp
}}^{h_{\perp }^{1,1}}q_{i_{\perp }}\mathcal{C}^{i_{\perp }}  \label{bf}
\end{equation}%
By focusing on the abelian gauge coupling term $\int_{M_{3D}}\mathcal{G}_{%
\text{\textsc{ab}}}\left( F^{\text{\textsc{a}}}\wedge \ast F^{\text{\textsc{b%
}}}\right) $ that we expand like%
\begin{equation}
\begin{tabular}{lll}
$\dint\nolimits_{M_{3D}}\mathcal{G}_{\text{\textsc{ab}}}\left( F^{\text{%
\textsc{a}}}\wedge \ast F^{\text{\textsc{b}}}\right) $ & $=$ & $%
\dint\nolimits_{M_{3D}}\mathcal{G}_{\bot \bot }\left( F^{\bot }\wedge \ast
F^{\bot }\right) +\dint\nolimits_{M_{3D}}\mathcal{G}_{\bot \Vert }\left(
F^{\bot }\wedge \ast F^{\Vert }\right) +$ \\
&  &  \\
&  & $\dint\nolimits_{M_{3D}}\mathcal{G}_{\Vert \Vert }\left( F^{\Vert
}\wedge \ast F^{\Vert }\right) +\dint\nolimits_{M_{3D}}\mathcal{G}_{\Vert
\bot }\left( F^{\Vert }\wedge \ast F^{\bot }\right) $%
\end{tabular}
\label{st}
\end{equation}%
\begin{equation*}
\end{equation*}%
and using the infinite distance limit of the coupling metric (\ref{beh})
namely $\mathcal{G}_{\bot \bot }\sim \mathcal{O}\left( \lambda ^{+2}\right) $
and $\mathcal{G}_{\bot \Vert }\sim \mathcal{O}\left( 1\right) $ as well as $%
\mathcal{G}_{\Vert \Vert }\sim \mathcal{O}\left( \lambda ^{-2}\right) $, we
see that (\ref{st}) involves different gauge regimes. Moreover, setting $q_{%
\text{\textsc{a}}}=(q_{a_{\parallel }},q_{i_{\perp }})$ and substituting $%
\mathcal{G}^{\text{\textsc{ab}}}$ as in eq(\ref{tem}), we learn that the
effective gauge coupling $g_{_{\Upsilon }}^{2}=g_{3D}^{2}(q_{\text{\textsc{a}%
}}\mathcal{G}^{\text{\textsc{ab}}}q_{\text{\textsc{b}}})$ splits into three
blocs as follows%
\begin{equation}
g_{_{\Upsilon }}^{2}=\left( g_{_{\Upsilon }}^{2}\right) _{\parallel
\parallel }+2\left( g_{_{\Upsilon }}^{2}\right) _{\parallel \perp }+\left(
g_{_{\Upsilon }}^{2}\right) _{\perp \perp }  \label{gamma}
\end{equation}%
with infinite distance limit behaviours like%
\begin{eqnarray}
\left( g_{_{\Upsilon }}^{2}\right) _{\parallel \parallel } &=&\lambda
^{2}g_{3D}^{2}(q_{a_{\parallel }}\mathring{G}^{a_{\parallel }b_{\parallel
}}q_{b_{\parallel }})  \label{pp1} \\
\left( g_{_{\Upsilon }}^{2}\right) _{\parallel \perp } &=&\lambda
^{0}g_{3D}^{2}(q_{a_{\parallel }}\mathring{G}^{a_{\parallel }i_{\perp
}}q_{i_{\perp }})  \label{pp2} \\
\left( g_{_{\Upsilon }}^{2}\right) _{\perp \perp } &=&\frac{1}{\lambda ^{2}}%
g_{3D}^{2}(q_{i_{\perp }}\mathring{G}^{i_{\perp }j_{\perp }}q_{j_{\perp }})
\label{pp3}
\end{eqnarray}%
showing that for $\lambda \rightarrow \infty ,$ we have $\left(
g_{_{\Upsilon }}^{2}\right) _{\parallel \parallel }\sim \mathcal{O}\left(
\lambda ^{2}\right) \rightarrow \infty $ while $\left( g_{_{\Upsilon
}}^{2}\right) _{\perp \perp }\sim \mathcal{O}\left( 1/\lambda ^{2}\right)
\rightarrow 0.$

From these behaviours, we see that for fourfold $X_{4}=K3_{\perp }\times
K3_{\parallel },$ there are two dual gauge regimes.

\begin{description}
\item[$\left( \mathbf{i}\right) $] A weakly coupled direction [$\left(
g_{_{\Upsilon }}^{2}\right) _{\perp \perp }\rightarrow 0$] associated with
fibral curve ($\Upsilon _{\mathbf{q}_{\parallel }}\rightarrow 0$). So, the
complex curve $\Upsilon _{\mathbf{q}}$ which in general is given by the
expansion (\ref{bf}) reduces down to%
\begin{equation}
q_{a_{\parallel }}=0\qquad ,\qquad \Upsilon _{\mathbf{q}_{\perp }}=\sum
q_{i_{\perp }}\mathcal{C}^{i_{\perp }}
\end{equation}
For this curve, the gauge coupling $g_{_{\Upsilon _{weak}}}^{2}$ vanishes in
the infinite distance limit as shown below%
\begin{equation}
\mathrm{g}_{_{\Upsilon _{weak}}}^{2}=g_{3D}^{2}(q_{i_{\perp }}\mathcal{G}%
^{i_{\perp }j_{\perp }}q_{j_{\perp }})\qquad ,\qquad \mathrm{g}_{_{\Upsilon
_{weak}}}^{2}\sim \mathcal{O}\left( \frac{1}{\lambda ^{2}}\right)
\rightarrow 0
\end{equation}%
For this U$\left( 1\right) _{weak}$ gauge symmetry, the corresponding 3D
Maxwell-like field action reads in terms of the field strength $%
F^{weak}=dA^{weak}$ and the coupling constant $g_{_{\Upsilon _{weak}}}^{2}$
as follows%
\begin{eqnarray}
\mathcal{S}_{\Upsilon _{weak}} &=&-\frac{1}{4\mathrm{g}_{_{\Upsilon
_{weak}}}^{2}}\dint\nolimits_{M_{3D}}F^{weak}\wedge \ast F^{weak}+  \notag \\
&&\frac{M_{\mathrm{Pl}}}{2}\int_{\mathcal{M}_{3D}}\mathcal{R}-\frac{1}{2}%
\int_{\mathcal{M}_{3D}}\mathfrak{g}_{\text{\textsc{xy}}}d\phi ^{\text{%
\textsc{x}}}\wedge \ast d\phi ^{\text{\textsc{y}}}
\end{eqnarray}%
Notice that here the 1-form gauge potential $A^{weak}$ is related to the
3-form potential $\boldsymbol{C}_{3}$ of the M-theory as follows%
\begin{equation}
A^{weak}=\dint\nolimits_{\Upsilon _{weak}}\boldsymbol{C}_{3}  \label{WEA}
\end{equation}

\item[$\left( \mathbf{ii}\right) $] A strongly coupled direction [$\left(
g_{_{\Upsilon }}^{2}\right) _{\parallel \parallel }\rightarrow \infty $]
associated with base curve ($\Upsilon _{\mathbf{q}_{\perp }}\rightarrow 0$).
Here, the curve $\Upsilon _{\mathbf{q}}$ reduces to
\begin{equation}
q_{i_{\perp }}=0\qquad ,\qquad \Upsilon _{\mathbf{q}_{\parallel }}=\sum
q_{a_{\parallel }}\mathcal{C}^{a_{\parallel }}
\end{equation}%
and the gauge coupling $g_{_{\Upsilon _{strong}}}^{2}$ is given by%
\begin{equation}
\mathrm{g}_{_{\Upsilon _{strong}}}^{2}=g_{3D}^{2}(q_{a_{\parallel }}\mathcal{%
G}^{a_{\parallel }b_{\parallel }}q_{b_{\parallel }})\qquad ,\qquad \mathrm{g}%
_{_{\Upsilon _{strong}}}^{2}\sim \mathcal{O}\left( \lambda ^{2}\right)
\rightarrow \infty
\end{equation}%
Similarly for the $\mathcal{S}_{\Upsilon ^{weak}}$, here also the
corresponding 3D Maxwell-like field action invariant under the U$\left(
1\right) _{strong}$ gauge symmetry reads in terms of the field strength $%
F^{strong}=dA^{strong}$ and the coupling constant $g_{_{\Upsilon
^{strong}}}^{2}$ as follows%
\begin{eqnarray}
\mathcal{S}_{\Upsilon _{strong}} &=&-\frac{1}{4\mathrm{g}_{_{\Upsilon
_{strong}}}^{2}}\dint\nolimits_{M_{3D}}F^{strong}\wedge \ast F^{strong}+
\notag \\
&&\frac{M_{\mathrm{Pl}}}{2}\int_{\mathcal{M}_{3D}}\mathcal{R}-\frac{1}{2}%
\int_{\mathcal{M}_{3D}}\mathfrak{g}_{\text{\textsc{xy}}}d\phi ^{\text{%
\textsc{x}}}\wedge \ast d\phi ^{\text{\textsc{y}}}
\end{eqnarray}%
with%
\begin{equation}
A^{strong}=\dint\nolimits_{\Upsilon _{strong}}\boldsymbol{C}_{3}  \label{STR}
\end{equation}
\end{description}

Notice that for the fourfold $X_{4}=K3_{\perp }\times K3_{\parallel },$ the
Hodge number $h_{\parallel }^{1,1}$ in the base $K3_{\parallel }$ and the
Hodge number $h_{\perp }^{1,1}$ of the fiber $K3_{\perp }$ are equal; so we
can take%
\begin{equation}
h_{\parallel }^{1,1}=h_{\perp }^{1,1}=\nu \qquad ,\qquad h_{X_{4}}^{1,1}=2\nu
\end{equation}%
Then, the gauge symmetry (\ref{34}) of the \emph{EFT}$_{{\small 3D}}$ is
given by $U(1)^{2\nu }$; it factorises like
\begin{equation}
G_{abelian}=U(1)_{fiber}^{\mathrm{\nu }}\times U(1)_{base}^{\mathrm{\nu }}
\label{SymBreak}
\end{equation}%
Notice also that because of the automorphism symmetry of $X_{4}$\ generated
by the permutation,%
\begin{equation}
K3_{\perp }\qquad \leftrightarrow \qquad K3_{\parallel }  \label{aut}
\end{equation}%
characteristic properties in the fiber $K3_{\perp }$ (like weakly coupled
directions) gets mapped to dual properties in the base $K3_{\parallel }$
(strongly coupled directions). This feature can be manifestly exhibited in
various ways; for instance by using the mapping $\lambda \rightarrow
1/\lambda $ exchanging $\left( g_{_{\Upsilon }}^{2}\right) _{\perp \perp }$
and $\left( g_{_{\Upsilon }}^{2}\right) _{\parallel \parallel }$ in eq(\ref%
{pp1}-\ref{pp3}). It reads also from the particular factorisation of the
gauge symmetry (\ref{SymBreak}). This phenomenon, to which we refer below to
as \emph{Weak/Strong gauge duality}, is further investigated in the next
section.

\section{Weak/Strong gauge duality in 3D}
\label{sec4}
In this section, we deepen the study of the Weak/Strong gauge duality in
M-theory on $X_{4}=K3_{\perp }\times K3_{\parallel }$ in connection with the
Asymptotic WGC and the Repulsive Force conjecture (RFC). We first develop
the Type-$\mathbb{S}$ form for this $X_{4}$ and its Weak/Strong gauge
duality properties induced by the exchange of the role of the fiber and the
base surfaces in $X_{4}$. Then, we investigate the implications of the $%
X_{4} $ automorphism (\ref{aut}) generated by the transposition $K3_{\perp
}\leftrightarrow K3_{\parallel }$ on the Asymptotic WGC. This feature
translates into Weak/Strong gauge duality in the $\emph{EFT}_{{\small 3D}}.$%
\textrm{\ }After that, we study the RFC in the $\emph{EFT}_{{\small 3D}}$
and give its verification for both towers of BPS and non BPS states.

\subsection{Type- surface form for $K3_{\perp }\times K3_{\parallel }$}

\begin{figure}[h]
\begin{center}
\includegraphics[width=12cm]{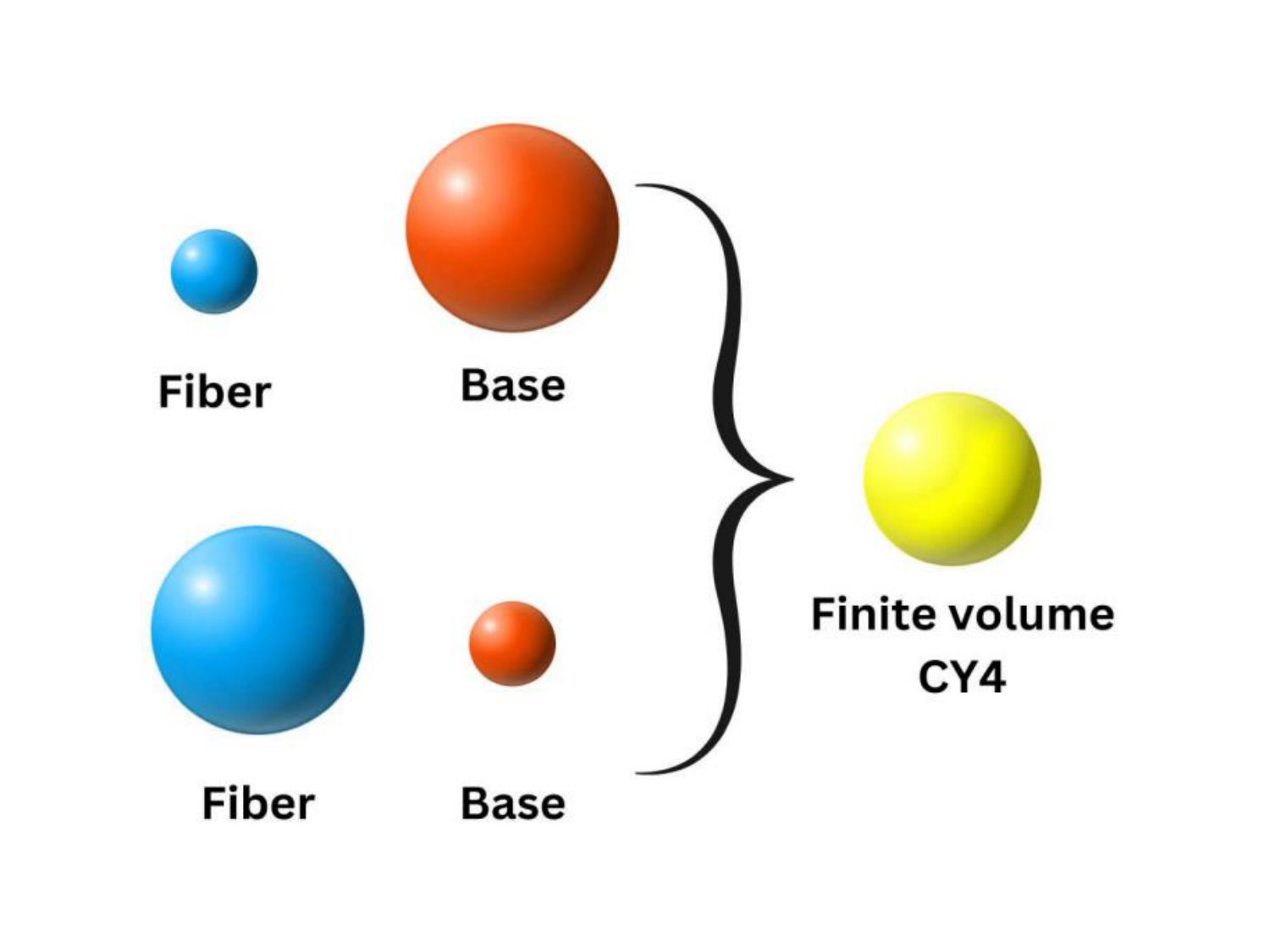}
\end{center}
\vspace{-0.5cm}
\caption{Graphic representation of the volumes of the fiber and the base of
the two K3 surfaces making the Calabi-Yau fourfold.}
\label{02K}
\end{figure}

We start by recalling that the non vanishing Hodge numbers $h_{K3_{x}}^{p,q}$
of the fibral $K3_{\perp }$ and the base $K3_{\parallel }$ surfaces namely,
\begin{equation}
h_{K3_{x}}^{0,0}=h_{K3_{x}}^{2,2}=1,\qquad h_{K3_{x}}^{1,1}=20,\qquad
h_{K3_{x}}^{2,1}=h_{K3_{x}}^{1,2}=1
\end{equation}%
This indicates that the number ${\large \nu }_{\mathcal{C}_{2}}$ of 2-cycles
$\mathcal{C}_{2}$ in the fibered CY4 is less than or equal to 40 ($\nu _{%
\mathcal{C}_{2}}\leq h_{X_{4}}^{1,1}=40).$ So, the parametrisation of the
Kahler 2-form (\ref{ja}-\ref{jp}) is generated at most by the 20 two-forms $%
J_{a_{\parallel }}$ of the $K3_{\parallel }$ (${\large \nu }_{\mathcal{C}%
_{2}^{\Vert }}\leq h_{K3_{\Vert }}^{1,1}$) as well as the 20 two-forms $%
J_{a_{\bot }}$ of the $K3_{\bot }$ (${\large \nu }_{\mathcal{C}_{2}^{\bot
}}\leq h_{K3_{\bot }}^{1,1}$). Below, we use the following shortened
representation
\begin{equation}
J=\lambda \left( \upsilon ^{\parallel }\mathcal{J}_{\parallel }\right) +%
\frac{1}{\lambda }\left( \upsilon ^{\perp }\mathcal{J}_{\perp }\right)
\label{xe}
\end{equation}%
in order to avoid cumbersome expressions that are not necessary for our
analysis. More explicit calculations are obtained by replacing $\upsilon
^{\parallel }\mathcal{J}_{\parallel }$ by the expansion $\sum \upsilon
^{a_{\parallel }}\mathcal{J}_{a_{\parallel }}$ and $\upsilon ^{\perp }%
\mathcal{J}_{\perp }$ by $\sum \upsilon ^{a_{\perp }}\mathcal{J}_{a_{\perp
}}.$ From this expansion of the Kahler 2-form on $K3_{\perp }\times
K3_{\parallel }$, we notice a manifest symmetry property; it is invariant
under the discrete transformation%
\begin{equation}
\lambda \leftrightarrow \frac{1}{\lambda }\qquad ,\qquad \upsilon
^{\parallel }\leftrightarrow \upsilon ^{\perp }\qquad ,\qquad \mathcal{J}%
_{\parallel }\leftrightarrow \mathcal{J}_{\perp }  \label{sd}
\end{equation}%
Using the expansion (\ref{xe}) of the Kahler 2-form, we can compute the
monomial 2n-form $J^{n};$ in particular volume 8-form $J^{4}$ expanding as
follows
\begin{equation}
\begin{tabular}{lll}
$J^{4}$ & $=$ & $\lambda ^{4}\left( \upsilon ^{\parallel }\right) ^{4}%
\mathcal{J}_{\parallel }^{4}+4\lambda ^{2}\upsilon ^{\perp }\left( \upsilon
^{\parallel }\right) ^{3}\left( \mathcal{J}_{\parallel }^{3}\mathcal{J}%
_{\perp }\right) +$ \\
&  & $\frac{1}{\lambda ^{4}}\left( \upsilon ^{\perp }\right) ^{4}\mathcal{J}%
_{\perp }^{4}+\frac{4}{\lambda ^{2}}\upsilon ^{\parallel }\left( \upsilon
^{\perp }\right) ^{3}\mathcal{J}_{\parallel }\mathcal{J}_{\perp }^{3}+$ \\
&  & $6\left( \upsilon ^{\parallel }\upsilon ^{\perp }\right) ^{2}\mathcal{J}%
_{\parallel }^{2}\mathcal{J}_{\perp }^{2}$%
\end{tabular}
\label{J4}
\end{equation}%
This volume 8-form is also manifestly invariant under (\ref{sd}); just
because $J$ does. Nevertheless, it is interesting to zoom on the action of
this symmetry transformation at the level of the five monomials $\mathcal{J}%
_{\parallel }^{n}\mathcal{J}_{\perp }^{4-n}$ generating $J^{4}$. The two
first lines in (\ref{J4}) gets exchanged; while the third term $6\left(
\upsilon ^{\parallel }\upsilon ^{\perp }\right) ^{2}\mathcal{J}_{\parallel
}^{2}\mathcal{J}_{\perp }^{2}$ is exactly invariant; and is remarkably
independent of the parameter $\lambda $. So, a volume 8-form $J^{4}$
independent of the parameter $\lambda $ must be restricted to%
\begin{equation}
J^{4}=6\left( \upsilon ^{\parallel }\upsilon ^{\perp }\right) ^{2}\mathcal{J}%
_{\parallel }^{2}\mathcal{J}_{\perp }^{2}
\end{equation}%
and so should obey the constraint relations $\mathcal{J}_{\parallel }^{4}=%
\mathcal{J}_{\parallel }^{3}=0$ but $\mathcal{J}_{\parallel }^{2}\neq 0;$
otherwise the $J^{4}$ diverges in the infinite distance limit $\lambda
\rightarrow \infty .$ By duality $\lambda \leftrightarrow \left( 1/\lambda
\right) ,$ it is the monomials $\mathcal{J}_{\perp }^{4}$ and $\mathcal{J}%
_{\perp }^{3}$ that should vanish in the "short distance" limit $\left(
1/\lambda \right) \rightarrow \infty .$ So, we have
\begin{equation}
\begin{tabular}{lllllllll}
$\lambda \rightarrow \infty $ & : & $\mathcal{J}_{\parallel }^{4}=\mathcal{J}%
_{\parallel }^{3}$ & $=$ & $0$ & $\qquad ,\qquad $ & $\mathcal{J}_{\parallel
}^{2}$ & $\neq $ & $0$ \\
$\frac{1}{\lambda }\rightarrow \infty $ & : & $\mathcal{J}_{\perp }^{4}=%
\mathcal{J}_{\perp }^{3}$ & $=$ & $0$ & $\qquad ,\qquad $ & $\mathcal{J}%
_{\perp }^{2}$ & $\neq $ & $0$%
\end{tabular}%
\end{equation}%
For the above specific Kahler 2-form (\ref{J4}), the volume $\mathcal{V}%
_{X_{4}}$ of the CY4 is given by%
\begin{equation}
\mathcal{V}_{CY_{4}}=\frac{1}{4!}\dint\nolimits_{CY_{4}}J^{4}  \label{vx4}
\end{equation}%
which by using
\begin{equation}
\kappa _{_{\Vert \Vert \bot \bot }}=\dint\nolimits_{CY_{4}}\mathcal{J}%
_{\parallel }^{2}\wedge \mathcal{J}_{\perp }^{2}  \label{xv4}
\end{equation}%
gives%
\begin{equation}
\mathcal{V}_{X_{4}}=\frac{1}{4}\kappa _{_{{\small \Vert \Vert \bot \bot }%
}}\left( \upsilon ^{\parallel }\right) ^{2}\left( \upsilon ^{\perp }\right)
^{2}
\end{equation}%
From these relations, we see that the total volume $\mathcal{V}_{X_{4}}$ is
given by the product $\mathcal{V}_{K3_{\parallel }}\times \mathcal{V}%
_{K3_{\perp }}$ with fiber $\mathcal{V}_{K3_{\perp }}$ and base $\mathcal{V}%
_{K3_{\parallel }}$ volumes respectively given by $\frac{1}{2}\eta _{\bot
\bot }\left( \upsilon ^{\perp }\right) ^{2}$ and $\frac{1}{2}\eta _{_{\Vert
\Vert }}\left( \upsilon ^{\parallel }\right) ^{2}.$ These expressions can be
derived by substituting $CY_{4}=K3_{\perp }\times K3_{\parallel }$ in (\ref%
{xv4}); we find that $\kappa _{_{\Vert \Vert \bot \bot }}$ factorises as the
product $\eta _{_{\Vert \Vert }}\times \eta _{_{\bot \bot }}$ with
\begin{equation}
\eta _{\Vert \Vert }=\dint\nolimits_{K3_{\parallel }}\mathcal{J}_{\parallel
}^{2}\qquad ,\qquad \eta _{\bot \bot }=\dint\nolimits_{\kappa _{\bot }}%
\mathcal{J}_{\perp }^{2}
\end{equation}%
A graphic representation of the volumes of the two\textrm{\ }K3 surfaces
making the Calabi fourfold is depicted by the Figure \textbf{\ref{02K}}
while the total volume is finite and fixed.
\newline
As a result of this description, we cite the two following:

\begin{description}
\item[$\mathbf{1)}$] \ Given the fibration $X_{4}=S_{\bot }\times S_{\Vert
}, $ the volumes of the fibral surface $S_{\bot }$ and the base $S_{\Vert }$
of the fibered Calabi-Yau fourfold scales in terms of the spectral parameter
$\lambda $ like%
\begin{equation}
vol\left( S_{\bot }\right) =\frac{1}{\lambda ^{2}}\mathcal{V}_{K3_{\perp
}}\quad ,\quad vol\left( S_{\Vert }\right) =\lambda ^{2}\mathcal{V}%
_{K3_{\parallel }}\quad ,\quad vol\left( X_{4}\right) \simeq \mathcal{V}%
_{K3_{\parallel }}\times \mathcal{V}_{K3_{\perp }}  \label{s1}
\end{equation}%
From this behavior, we see that the volumes $vol\left( S_{\bot }\right) $
and $vol\left( S_{\Vert }\right) $ take \textrm{singular values in the
limits }$\lambda \rightarrow \infty $\textrm{\ and }$\lambda \rightarrow 0$;
whilst the total volume of the Calabi-Yau fourfold $S_{\bot }\times S_{\Vert
}$ remains finite.

\item[$\mathbf{2)}$] Moreover, for $\lambda \rightarrow \infty ,$ the $%
vol\left( S_{\bot }\right) $ shrinks ($vol\left( S_{\bot }\right)
\rightarrow 0$) while the $vol\left( S_{\Vert }\right) $ diverges ($%
vol\left( S_{\Vert }\right) \rightarrow \infty $); \textrm{this gives the
Asymptotic weak coupling} [$\mathcal{G}_{\Vert \Vert
}^{weak}\int_{M_{3D}}\left( F^{\Vert }\wedge \ast F^{\Vert }\right) $] with
gauge coupling metric
\begin{equation}
\mathcal{G}_{\Vert \Vert }^{weak}=\frac{1}{\lambda ^{2}}G_{\Vert \Vert }
\label{WE}
\end{equation}%
with asymptotic behavior as $\mathcal{G}_{\Vert \Vert }^{weak}\sim \mathcal{O%
}\left( 1/\lambda ^{2}\right) \rightarrow 0$. However, for $\left( 1/\lambda
\right) \rightarrow \infty ,$ we have the inverse picture; the $vol\left(
S_{\Vert }\right) $ shrinks but the $vol\left( S_{\bot }\right) $ diverges.
The coupling [$\mathcal{G}_{\bot \bot }^{strong}\int_{M_{3D}}\left( F^{\bot
}\wedge \ast F^{\bot }\right) $] has a "strong coupling" metric%
\begin{equation}
\mathcal{G}_{\bot \bot }^{strong}=\lambda ^{2}G_{\bot \bot }  \label{ST}
\end{equation}%
with behavior at infinite distance $\mathcal{G}_{\bot \bot }^{strong}\sim
\mathcal{O}\left( \lambda ^{2}\right) \rightarrow \infty .$ Because the
change $\lambda \leftrightarrow 1/\lambda $ is a symmetry of the Kahler
2-form (\ref{J4}) and the \emph{EFT}$_{{\small 3D}}$, the asymptotic weak $%
\mathcal{G}_{\parallel \parallel }^{weak}$ and the short distance $\mathcal{G%
}_{\perp \perp }^{strong}$ are dual in M-theory on $K3_{\perp }\times
K3_{\parallel }$.\newline

\item[$\mathbf{3)}$] In the infinite distance limit ($\lambda \rightarrow
\infty $), the fibral surface $S_{\bot }$ shrinks; and there exists a non
singular curve\textrm{\ }$\Upsilon _{\mathbf{q}_{\perp }}$\textrm{\ }in the
fiber with charge $q_{\text{\textsc{a}}}=(q_{i_{\bot }};q_{a_{\Vert }}=0)$
where sits the weak gauge symmetry U$\left( 1\right) _{weak}$. The complete
curve $\Upsilon _{\mathbf{q}}$ in the CY4 is then fully located in $%
K3_{\perp }.$ By Weak/Strong gauge duality, we have a curve $\Upsilon _{%
\mathbf{q}_{\parallel }}$ and an abelian strong gauge symmetry U$\left(
1\right) _{strong}$ lying in the base$.$
\end{description}

\ \ \ \ \newline
From these results, we conjecture that in the \emph{EFT}$_{{\small 3D}}$
descending from M-theory on $K3_{\perp }\times K3_{\parallel }$ with the
symmetry (\ref{sd}), we have two dual (weak and strong) asymptotic gauge
regimes; the \emph{asymptotic strong gauge} regime is the image of the
asymptotic WGC under (\ref{sd}). Recall that asymptotic WGC stipulates that
weakly coupled directions in the fibral $K3_{\perp }$ charge lattice host
towers of BPS or non-BPS states which satisfy the weak gravity conjecture.\
By using the weak/strong duality ($K3_{\perp }\leftrightarrow K3_{\parallel
})$, we expect to also have towers of BPS or non-BPS states in the strong
gauge regime.

\subsection{BPS and non BPS towers}

As a consequence of the duality (\ref{sd}), the weakly coupled tower of BPS (%
$\mathcal{T}_{\mathrm{k}_{\mathrm{\bot }}\mathrm{,}\mathbf{q}_{\bot }}^{%
\text{\textsc{bps}}}$) or tower of non-BPS ($\mathcal{T}_{\mathrm{k}_{%
\mathrm{\bot }}\mathrm{,}\mathbf{q}_{\bot }}^{\text{\textsc{n-bps}}}$)
states in the \emph{EFT}$_{3D}.$ These towers are predicted by the
Asymptotic WGC conjecture ($\lambda \rightarrow \infty $); they gets mapped
under (\ref{sd}) into strongly coupled towers of BPS ($\mathcal{T}_{\mathrm{k%
}_{\mathrm{\Vert }}\mathrm{,}\mathbf{q}_{\Vert }}^{\text{\textsc{bps}}}$) or
tower of non-BPS ($\mathcal{T}_{\mathrm{k}_{\mathrm{\Vert }}\mathrm{,}%
\mathbf{q}_{\Vert }}^{\text{\textsc{n-bps}}}$) states for the limit $\frac{1%
}{\lambda }\rightarrow \infty .$ This correspondence between the two gauge
regimes (weak and strong) can be stated as follows%
\begin{equation}
\begin{tabular}{ccc}
weak regime: $g_{weak}^{2}$ & $\quad \leftrightarrow \quad $ & strong
regime: $g_{strong}^{2}$ \\
BPS\TEXTsymbol{\vert}$_{\bot }$ & $\leftrightarrow $ & BPS\TEXTsymbol{\vert}$%
_{\parallel }$ \\
non-BPS\TEXTsymbol{\vert}$_{\bot }$ & $\leftrightarrow $ & non-BPS%
\TEXTsymbol{\vert}$_{\parallel }$%
\end{tabular}
\label{bps}
\end{equation}%
\begin{equation*}
\end{equation*}%
with $g_{weak}^{2}=\left( g_{_{\Upsilon }}^{2}\right) _{\perp \perp }$ and $%
g_{strong}^{2}=\left( g_{_{\Upsilon }}^{2}\right) _{\parallel \parallel }$.
This result can be also motivated by the automorphism symmetry of $%
X_{4}=K3_{\parallel }\times K3_{\bot }$ exchanging the fiber $K3_{\bot }$
and the base $K3_{\parallel }.$

From the Table (\ref{bps}) and borrowing results from \cite{12}, regarding
\emph{EFT}$_{5D}$ induced from M-theory CY3 with K3 fibration, we learn that
the \emph{EFT}$_{3D}$ descending from M-theory on $K3_{\parallel }\times
K3_{\bot },$ that there are various towers of states satisfying the WGC and
its gauge dual under (\ref{sd}).

The first towers are given by $\mathcal{T}_{\mathrm{k}_{\mathrm{\bot }}%
\mathrm{,}\mathbf{q}_{\bot }}^{\text{\textsc{bps}}}$ and $\mathcal{T}_{%
\mathrm{k}_{\mathrm{\Vert }}\mathrm{,}\mathbf{q}_{\Vert }}^{\text{\textsc{bps%
}}}$ and are dual: $\left( i\right) $ The $\mathcal{T}_{\mathrm{k}_{\mathrm{%
\bot }}\mathrm{,}\mathbf{q}_{\bot }}^{\text{\textsc{bps}}}$ is labeled by
the integer $\mathrm{k}_{\mathrm{\bot }}$ and the charge vector $\mathbf{q}%
_{\bot }=\left( q_{1_{\bot }},...,q_{r_{\bot }}\right) ;$ it englobes BPS
particle states\textrm{\ }$\left\vert BPS|_{\bot }\right\rangle $
constructed from M2 brane wrapped\textrm{\ }$k_{\bot }$\textrm{\ }times
around\textrm{\ }$\Upsilon _{\mathbf{q}_{\perp }}$\textrm{\ }[BPS\TEXTsymbol{%
\vert}$_{\bot }=\text{M2/}\Upsilon _{\mathbf{q}_{\perp }}$].\ The particle
states generated have charges\textrm{\ }$\mathbf{Q}_{\perp }=\mathrm{k}%
_{\bot }\mathbf{q}_{\perp }$. By using Weak/Strong gauge duality, we get the
dual tower $\mathcal{T}_{\mathrm{k}_{\mathrm{\Vert }}\mathrm{,}\mathbf{q}%
_{\Vert }}^{\text{\textsc{bps}}}$ of BPS states $\left\vert BPS|_{\parallel
}\right\rangle $; it is given by wrapped M2 brane (BPS\TEXTsymbol{\vert}$%
_{\parallel }=\text{M2/}\Upsilon _{\mathbf{q}_{\parallel }}$) with charges$\
\mathbf{Q}_{\parallel }=\mathrm{k}_{\parallel }\mathbf{q}_{\parallel }$ as
shown by the following table,%
\begin{equation}
\begin{tabular}{ccc}
weak regime: $g_{weak}^{2}$ & $\quad \leftrightarrow \quad $ & strong
regime: $g_{strong}^{2}$ \\
BPS\TEXTsymbol{\vert}$_{\bot }=\text{M2/}\Upsilon _{\mathbf{q}_{\perp }}$ & $%
\leftrightarrow $ & BPS\TEXTsymbol{\vert}$_{\parallel }=\text{M2/}\Upsilon _{%
\mathbf{q}_{\parallel }}$%
\end{tabular}
\label{bp1}
\end{equation}%
\begin{equation*}
\end{equation*}%
Recall that the $\Upsilon _{\mathbf{q}_{\perp }}$ is purely fibral sitting
in $K3_{\bot };$ it solves the condition (\ref{Weak}) and has a positive
self-intersection
\begin{equation}
\Upsilon _{\mathbf{q}_{\perp }}\equiv \mathcal{C}_{{\small U(1)}%
_{weak}},\qquad \lbrack \mathcal{C}_{{\small U(1)}_{weak}}]^{2}>0
\label{BPS1}
\end{equation}%
whereas, the $\Upsilon _{\mathbf{q}_{\parallel }}$ is a purely base curve
sitting in $K3_{\parallel };$ it satisfies the dual of the condition (\ref%
{Weak}) and also has a positive self-intersection
\begin{equation}
\Upsilon _{\mathbf{q}_{\parallel }}\equiv \mathcal{C}_{{\small U(1)}%
_{strong}},\qquad \lbrack \mathcal{C}_{{\small U(1)}_{strong}}]^{2}>0
\label{nbp1}
\end{equation}%
The second towers are given by $\mathcal{T}_{\mathrm{k}_{\mathrm{\bot }}%
\mathrm{,}\mathbf{q}_{\bot }}^{\text{\textsc{n-bps}}}$ and $\mathcal{T}_{%
\mathrm{k}_{\mathrm{\Vert }}\mathrm{,}\mathbf{q}_{\Vert }}^{\text{\textsc{%
n-bps}}}$ contain non-BPS particle states; these two towers are also dual
under (\ref{sd}). Here, the non-BPS particles are given by excitations of
Heterotic string. This string is dual to the so called MSW string given by
M5 brane wrapping the fiber $K3$ \textrm{\cite{12,26}}. These are non-BPS
states characterised by curves $\Upsilon _{\mathbf{Q}}$ in K3 with negative
self-intersection ($\Upsilon _{\mathbf{Q}}^{2}<0)$. Depending on where we
are considering the fiber K3$_{\bot }$ or the base K3, we distinguish two
kinds of heterotic strings and then,
\begin{equation}
\begin{tabular}{lll}
$\text{M5/K3}_{\bot }$ & $\qquad \leftrightarrow \qquad $ & Het. string$%
_{\bot }$ \\
$\text{M5/K3}_{\parallel }$ & $\qquad \leftrightarrow \qquad $ & Het. string$%
_{\parallel }$%
\end{tabular}
\label{nBPS1}
\end{equation}%
\begin{equation*}
\end{equation*}%
The M-theory/heterotic string duality is explicitly formulated through the
relation between the left moving excitations $n_{L}$ of the heterotic string
and the self intersection of the curves $\Upsilon _{\mathbf{q}}$ \cite{12}.
Because $X_{4}=K3_{\parallel }\times K3_{\bot },$ we have
\begin{equation}
\begin{tabular}{lllllll}
$n_{L_{\perp }}$ & $=$ & $-\frac{1}{2}\mathbf{q}_{\perp }^{2}$ & $\qquad
,\qquad $ & $\mathbf{q}_{\perp }^{2}$ & $=$ & $q_{i_{\bot }}\eta ^{i_{\bot
}j_{\bot }}q_{j_{\bot }}$ \\
$n_{L_{\parallel }}$ & $=$ & $-\frac{1}{2}\mathbf{q}_{\parallel }^{2}$ & $%
\qquad ,\qquad $ & $\mathbf{q}_{\parallel }^{2}$ & $=$ & $q_{a_{\parallel
}}\eta ^{a_{\parallel }b_{\parallel }}q_{b_{\parallel }}$%
\end{tabular}%
\end{equation}%
Having explored the weak/strong gauge duality in the\textrm{\ }\emph{EFT}$%
_{3D}$, we turn now to investigate the asymptotic WGC in 3D.\textrm{\ }We
present the conjecture in addition to its viability constraints in claim I
and claim II; then we explicitly verify the conjecture for the different
towers of states.

\section{The Asymptotic WGC and weak/strong duality}
\label{sec5}
In this section, we give two claims (I and II) for Asymptotic WGC in the
\emph{EFT}$_{{\small 3D}}$ considered in this paper. These claims generalise
results obtained in five dimensional \emph{EFT}$_{{\small 5D}}$ descending
from M-theory on Calabi-Yau threefolds. We also give new results for the
towers of BPS and non-BPS states in \emph{EFT}$_{{\small 3D}}$.

We start by noticing that since it was first proposed, the WGC has been
formulated in different ways and refined multiple times. The basic version
of the WGC in D-dimensions can be written as%
\begin{equation}
m^{2}\leq \frac{D-2}{D-3}g^{2}q^{2}M_{\mathrm{Pl}}^{D-2}  \label{WGCd}
\end{equation}%
Clearly this condition holds for $D>3;$\ however investigations of the
conjecture has been also performed in 3D; we cite for example \textrm{\cite%
{9A, 9B}} where the WGC has been studied from the context of infrared
consistency, and the condition of the WGC in 3D was expressed as%
\begin{equation}
z^{2}\geq \frac{1}{2},\qquad z^{2}=\frac{q^{2}g^{2}}{m^{2}}M_{\mathrm{Pl}}
\end{equation}%
This constraints reads like%
\begin{equation}
m^{2}\leq 2g^{2}q^{2}M_{Pl}
\end{equation}%
and is nothing but the (\ref{WGCd}) with the factor $(D-2)/(D-3)$ replaced
by a factor of 2. Moreover, seen that 3D gravity is especially interesting
in the context of AdS$_{3}$/CFT$_{2}$, it is important to mention the study
conducted in \cite{14G} where the WGC has been formulated as follows%
\begin{equation}
\Delta =\tilde{\Delta}=\frac{\mathrm{\alpha }^{\prime }}{4}m^{2}\leq \frac{1%
}{2}\max \left( Q^{2},\tilde{Q}^{2}\right)
\end{equation}%
where%
\begin{equation}
\Delta =L_{0}-\frac{c}{24},\qquad \tilde{\Delta}=\tilde{L}_{0}-\frac{c}{24}
\end{equation}%
with $Q,$ $\tilde{Q}$ being the charges of the right and left movers
respectively. If we substitute $\mathrm{\alpha }^{\prime }$ by 1/$M_{\mathrm{%
Pl}}$ and $\max (Q^{2},\tilde{Q}^{2})$ by $q^{2},$ then the WGC\ conjecture
in 3D takes the form%
\begin{equation}
m^{2}\leq 2g^{2}q^{2}M_{\mathrm{Pl}}
\end{equation}%
Given these facts, we come now to the 3D effective gauge theory coupled to
gravity with typical field action as given by eq(\ref{S3D}). In this theory,
every direction $\Upsilon _{\mathrm{k}\mathbf{q}}$ in the \emph{even integral%
} charge lattice $\mathfrak{L}_{K3_{\parallel }\times K3_{\bot }}$, that is
dual to weakly coupled gauge group $U(1)_{weak}=U(1)_{\bot }$ [sometimes
also denoted as $U(1)_{\Upsilon _{\mathrm{k}\mathbf{q}}}$], there exists a
tower $\mathcal{T}_{\mathrm{k,}\mathbf{q}}$ of massive particle states $%
\left\vert \mathrm{k;}\mathbf{q;}M_{\mathrm{k}}\right\rangle $ that satisfy
the inequality%
\begin{equation}
M_{\mathrm{k}}^{2}\leq 2\mathrm{g}_{{\small 3D}}^{2}M_{\mathrm{Pl}}(q_{\text{%
\textsc{a}}}\mathcal{G}^{\text{\textsc{ab}}}q_{\text{\textsc{b}}})\mathrm{k}%
^{2}-\frac{1}{4}\mathfrak{g}^{\text{\textsc{xy}}}\frac{\partial M_{\mathrm{k}%
}}{\partial \phi ^{\text{\textsc{x}}}}\frac{\partial M_{\mathrm{k}}}{%
\partial \phi ^{\text{\textsc{y}}}}  \label{mcd}
\end{equation}%
In the inequality (\ref{mcd}), the $\mathbf{q}=\left( q_{1},...,q_{r}\right)
$ is integral charge vector and the other parameters have the canonical
dimensions
\begin{equation}
\left[ \mathrm{g}_{{\small 3D}}^{2}\right] =\left[ M_{\mathrm{k}}\right]
=mass^{1},\qquad \left[ \mathcal{G}^{\text{\textsc{ab}}}\right] =\left[
\mathfrak{g}^{\text{\textsc{xy}}}\right] =\left[ \phi ^{\text{\textsc{x}}}%
\right] =mass^{0}
\end{equation}%
The $M_{\mathrm{k}}=M_{\mathrm{k}}\left( \phi \right) $ is the mass of the
states in the BPS/non-BPS towers $\mathcal{T}_{\mathrm{k,}\mathbf{q}}^{\text{%
\textsc{bps}}}/\mathcal{T}_{\mathrm{k,}\mathbf{q}}^{\text{\textsc{n-bps}}};$
and the $\phi ^{\text{\textsc{x}}}$s are the scalar fields present in the
\emph{EFT}$_{3D}$ that are responsible for the Yukawa attractive force $%
\mathfrak{g}^{\text{\textsc{xy}}}(\partial _{\text{\textsc{x}}}M_{\mathrm{k}%
})(\partial _{\text{\textsc{y}}}M_{\mathrm{k}})$. Moreover, the $\mathfrak{g}%
^{\text{\textsc{xy}}}$ is the inverse of the metric $\mathfrak{g}_{\text{%
\textsc{xy}}}$ of the scalar manifold of the \emph{EFT}$_{3D}$. In what
follows, we take all the volumes of the 2-cycles to be dimensionless in
order to make the calculations easier; this is done by using the normalised
variables $\hat{\upsilon}^{\text{\textsc{a}}}=\upsilon ^{\text{\textsc{a}}}/%
\mathcal{V}_{{\small CY}_{{\small 4}}}^{1/4}$ satisfying the property%
\begin{equation}
\frac{1}{4!}\kappa _{\text{\textsc{abcd}}}\hat{\upsilon}^{\text{\textsc{a}}}%
\hat{\upsilon}^{\text{\textsc{b}}}\hat{\upsilon}^{\text{\textsc{c}}}\hat{%
\upsilon}^{\text{\textsc{d}}}=\widehat{\mathcal{V}}_{{\small CY}_{{\small 4}%
}}=1
\end{equation}%
To check the validity of the above inequality (\ref{mcd}), we proceed in
steps as follows:\newline
First, we need to express the scalar field metric $\mathfrak{g}^{\text{%
\textsc{xy}}}$ in terms of the gauge coupling metric $\mathcal{G}^{\text{%
\textsc{ab}}}$ used before. This is achieved by help of the following formula%
\begin{equation}
\begin{tabular}{lll}
$\mathfrak{g}_{\text{\textsc{xy}}}d\phi ^{\text{\textsc{x}}}\wedge \ast
d\phi ^{\text{\textsc{y}}}$ & $=$ & $\mathcal{G}_{\text{\textsc{ab}}}d\hat{%
\upsilon}^{\text{\textsc{a}}}\wedge \ast d\hat{\upsilon}^{\text{\textsc{b}}}$
\\
& $=$ & $\left[ \mathcal{G}_{\text{\textsc{ab}}}(\partial _{\text{\textsc{x}}%
}\hat{\upsilon}^{\text{\textsc{a}}})(\partial _{\text{\textsc{y}}}\hat{%
\upsilon}^{\text{\textsc{b}}})\right] d\phi ^{\text{\textsc{x}}}\wedge \ast
d\phi ^{\text{\textsc{y}}}$%
\end{tabular}%
\end{equation}%
from which we deduce the relationship $\mathfrak{g}_{\text{\textsc{xy}}}=%
\mathcal{G}_{\text{\textsc{ab}}}(\partial _{\text{\textsc{x}}}\hat{\upsilon}%
^{\text{\textsc{a}}})(\partial _{\text{\textsc{y}}}\hat{\upsilon}^{\text{%
\textsc{b}}}).$ The inverse metric $\mathcal{G}^{\text{\textsc{ab}}}$ is
related to the inverse $\mathfrak{g}^{\text{\textsc{xy}}}$ as follows%
\begin{equation}
\mathcal{G}^{\text{\textsc{ab}}}=\frac{1}{2}\mathfrak{g}^{\text{\textsc{xy}}%
}(\partial _{\text{\textsc{x}}}\hat{\upsilon}^{\text{\textsc{a}}})(\partial
_{\text{\textsc{y}}}\hat{\upsilon}^{\text{\textsc{b}}})+\varrho \hat{\upsilon%
}^{\text{\textsc{a}}}\hat{\upsilon}^{\text{\textsc{b}}}
\end{equation}%
with $\varrho =1/2.$ Using these variables, the Asymptotic WGC condition (%
\ref{mcd}) reads in terms of $\mathcal{G}^{\text{\textsc{ab}}}$ then as
follows%
\begin{equation}
M_{\mathrm{k}}^{2}\leq 2\mathrm{g}_{{\small 3D}}^{2}M_{\mathrm{Pl}}(q_{\text{%
\textsc{a}}}\mathcal{G}^{\text{\textsc{ab}}}q_{\text{\textsc{b}}})\mathrm{k}%
^{2}-\frac{1}{2}(\mathcal{G}^{\text{\textsc{ab}}}-\varrho \hat{\upsilon}^{%
\text{\textsc{a}}}\hat{\upsilon}^{\text{\textsc{b}}})\frac{\partial M_{%
\mathrm{k}}}{\partial \hat{\upsilon}^{\text{\textsc{a}}}}\frac{\partial M_{%
\mathrm{k}}}{\partial \hat{\upsilon}^{\text{\textsc{b}}}}  \label{AWGC3}
\end{equation}%
Since our discussion is focusing on the Calabi-Yau fourfold fibration $%
CY_{4}=K3_{\perp }\times K3_{\parallel },$ we obtain a generalisation of the
claims given in \textrm{\cite{12}} for the case of \emph{EFT}$_{5D}$
descending from the Calabi-Yau threefold fibration $CY_{3}=K3\times \mathbb{B%
}_{1}$.

\subsection{Two claims: \ Weak/Strong gauge regimes}

We give two claims (\textbf{I} and \textbf{II}); the claim-I concerns the
asymptotic weak gauge coupling in the WGC; the claim-\textbf{II} regards the
asymptotic strong gauge coupling in the WGC; it is the Weak/Strong gauge
dual of the claim-\textbf{I}.

\subsubsection{Claim-I: Weakly coupled regime}

In M-theory compactified on $K3_{\perp }\times K3_{\parallel }$, for every
direction in the associated even integral charge lattice $\mathfrak{L}=%
\mathfrak{L}_{\perp }\oplus \mathfrak{L}_{\parallel },$ with a primitive
integral charge $\mathbf{q}=\mathbf{q}_{\perp }\oplus \mathbf{q}_{\parallel
} $, we define two families of abelian gauge groups
\begin{equation}
U(1)_{\mathbf{q}_{\perp }}=\sum_{i_{\perp }=1}^{h_{\perp }^{1,1}}q_{i_{\perp
}}U(1)^{i_{\perp }}\qquad ,\qquad U(1)_{\mathbf{q}_{\parallel
}}=\sum_{a_{\parallel }=1}^{h_{\parallel }^{1,1}}q_{a_{\parallel
}}U(1)^{a_{\parallel }}
\end{equation}%
with the following properties:

\begin{description}
\item[(\textbf{a)}] Given the magnetic scale $\left( \Lambda _{{\small mag}%
}\right) _{U(1)_{\mathbf{q}_{\perp }}}$ and the species scale $\left(
\Lambda _{\text{\textsc{sp}}}\right) _{_{U(1)_{\mathbf{q}_{\perp }}}},$
every direction $\mathbf{q}_{\perp }$ in the fibral lattice $\mathfrak{L}%
_{\perp }$ has a $U(1)_{\mathbf{q}_{\perp }}$ gauge factor with weak
coupling limit in the infinite distance ($\lambda \rightarrow \infty )$\
\textrm{in the sense}%
\begin{equation}
\left( \frac{\Lambda _{{\small mag}}}{\Lambda _{\text{\textsc{sp}}}}\right)
_{U(1)_{\mathbf{q}_{\perp }}}\quad \rightarrow \quad 0,  \label{wcf}
\end{equation}%
admitting $\left( \mathbf{1}\right) $ either a tower $\mathcal{T}_{\mathrm{%
k_{\perp },}\mathbf{q}_{\perp }}^{\text{\textsc{bps}}}=\mathcal{T}_{M_{%
\mathrm{k_{\perp }}}\rightarrow 0}^{\text{\textsc{bps}}}$ \textit{o}f BPS
states; or $\left( \mathbf{2}\right) $ a tower $\mathcal{T}_{\mathrm{%
k_{\perp },}\mathbf{q}_{\perp }}^{\text{\textsc{n-bps}}}=\mathcal{T}_{M_{%
\mathrm{k}_{_{\perp }}}\rightarrow 0}^{\text{\textsc{n-bps}}}$ of non-BPS
states ($\mathrm{k_{\perp }}\in \mathbb{Z}$).

\item[(\textbf{b)}] \textit{T}he BPS states, represented by the typical ket $%
\left\vert \mathrm{k_{\perp }};\mathbf{q_{\perp };}M_{\mathrm{k}_{\perp
}}\right\rangle _{\text{\textsc{bps}}},$ are given M2- brane wrapping
primitive 2-cycle $\Upsilon _{\mathbf{q}_{\perp }}^{\text{\textsc{bps}}}$ of
the fiber $K3_{\perp }$ as illustrated by eqs(\ref{bp1}-\ref{BPS1}). The
complex curve $\mathbf{\Upsilon }_{\mathbf{q}_{\perp }}^{\text{\textsc{bps}}%
} $ (real 2-cycle associated with \textrm{k}$_{\perp }=1$) has positive self
intersection $\Upsilon _{\mathbf{q}_{\perp }}^{\text{\textsc{bps}}}.\Upsilon
_{\mathbf{q}_{\perp }}^{\text{\textsc{bps}}}=\mathbf{q}_{\perp }^{2}>0;$ it
sits in the even integral self dual sublattice $\mathfrak{L}_{\perp
}^{\left( +\right) }$ of the fiber lattice $\mathfrak{L}_{\perp }$ having $%
\mathbf{q}_{\perp }^{2}=2m$ belonging to $2\mathbb{Z}.$ Moreover, each state%
\textrm{\ }in the tower $\mathcal{T}_{\mathrm{k_{\perp },}\mathbf{q}_{\perp
}}^{\text{\textsc{bps}}},$ is characterised by the generic integer $\mathrm{k%
}_{\perp },$ which is given by M2- brane wrapping 2-cycles\textrm{\ }$%
\Upsilon _{\mathbf{q}_{\perp }}$\textrm{\ }lying in\textrm{\ }$\mathfrak{L}%
_{\perp }^{\left( +\right) }$\textrm{\ }multiple times.

\item[(\textbf{c)}] The non-BPS states, represented by the typical ket $%
\left\vert n_{\mathbf{q}_{\perp }}^{L}\right\rangle _{\text{\textsc{n-bps}}%
}, $ are given by heterotic string excitations as in (\ref{nbp1}-\ref{nBPS1}%
); this string follows from the usual M-theory/heterotic duality. This
heterotic string with left moving charge $n_{\mathbf{q}_{\perp }}^{L}$ is
dual to the so-called Maldacena-Strominger-Witten (MSW) string given by M5
branes wrapping the generic fiber $K3_{\perp }.$ Here, the $K3_{\perp }$
contains a complex primitive curve $\Upsilon _{\mathbf{q}_{\perp }}^{\text{%
\textsc{n-bps}}}$ sitting in the anti-self dual sublattice $\mathfrak{L}%
_{\perp }^{\left( -\right) }$ with $\left( \mathbf{\alpha }\right) $ \textit{%
s}elf intersection $\Upsilon _{\mathbf{q}_{\perp }}^{\text{\textsc{n-bps}}}.$
$\Upsilon _{\mathbf{q}_{\perp }}^{\text{\textsc{n-bps}}}=\mathbf{q}_{\perp
}^{2}<0;$ and $\left( \mathbf{\beta }\right) $ related to the string
excitation charge as $n_{\mathbf{q}_{\perp }}^{L}=-\frac{1}{2}\mathbf{q}%
_{\perp }^{2}.$ \newline
Moreover, the tower $\mathcal{T}_{\mathrm{k_{\perp },}\mathbf{q}_{\perp }}^{%
\text{\textsc{n-bps}}}$ is characterised by the generic integer $\mathrm{k}%
_{\perp };$ as for the BPS case, the non BPS tower requires $K3_{\perp }$
containing complex curve $\Upsilon _{\mathrm{k}_{\perp }\mathbf{q}_{\perp
}}= $ $\mathrm{k}_{\perp }\Upsilon _{\mathbf{q}_{\perp }}^{\text{\textsc{%
n-bps}}} $ sitting in the\textit{\ }sublattice $\mathfrak{L}_{\perp
}^{\left( -\right) }$ with self intersection $\mathbf{q}_{\perp }^{2}<0;$
and string excitation modes as
\begin{equation}
n_{\mathrm{k}_{\perp }\mathbf{q}_{\perp }}^{L}=-\frac{\mathrm{k}_{\perp }^{2}%
}{2}\mathbf{q}_{\perp }^{2}
\end{equation}
\end{description}

\ Given the claim-\textbf{I} and the $\mathbb{Z}_{2}$ symmetry (\ref{sd}),
we have the following dual claim-\textbf{II} assertion induced by
Weak/strong gauge duality.

\subsubsection{Claim-II: Strongly coupled regime}

Because of the duality symmetry of the $CY_{4}=K3_{\perp }\times
K3_{\parallel }$ acting by exchanging the two K3 surfaces ($K3_{\perp
}\leftrightarrow K3_{\parallel }$),\ the properties given in Claim-I have a
dual homologue. So, to every direction in the sublattice $\mathfrak{L}%
_{\parallel },$ dual to $\mathfrak{L}_{\perp },$ in the even integral charge
lattice $\mathfrak{L}=\mathfrak{L}_{\perp }\oplus \mathfrak{L}_{\parallel }$%
; we have:

\begin{description}
\item[\textbf{a)}] a dual gauge symmetry factor $U(1)_{\mathbf{q}%
_{_{\parallel }}}$ with vector charge $\mathbf{q}_{_{\parallel
}}=(q_{1_{\parallel }},...,q_{\nu _{\parallel }})$ with strong gauge
coupling limit in the sense,%
\begin{equation}
\left( \frac{\Lambda _{{\small mag}}}{\Lambda _{\text{\textsc{sp}}}}\right)
_{U(1)_{\mathbf{q}_{_{\parallel }}}}\quad \rightarrow \quad 0  \label{wcb}
\end{equation}%
admitting two towers of particle states in \emph{EFT}$_{{\small 3D}}:$ $%
\left( \mathbf{1}\right) $ a tower $\mathcal{T}_{\mathrm{k}_{_{\parallel }},%
\mathbf{q}_{_{\parallel }}}^{\text{\textsc{bps}}}=\mathcal{T}_{M_{\mathrm{%
k_{\parallel }}}\rightarrow 0}^{\text{\textsc{bps}}}$ of BPS particle
states; or $\left( \mathbf{2}\right) $ a tower $\mathcal{T}_{\mathrm{%
k_{\parallel },}\mathbf{q}_{\parallel }}^{\text{\textsc{n-bps}}}=\mathcal{T}%
_{M_{\mathrm{k}_{_{\parallel }}}\rightarrow 0}^{\text{\textsc{n-bps}}}$ of
non-BPS states. These two types of particles are described in the statements
(\textbf{b}) and (\textbf{c}) given below.

\item[\textbf{b)}] \textit{t}he BPS states $|\mathrm{k_{_{\parallel }};}%
\mathbf{q_{_{\parallel }};}M_{\mathrm{k}_{_{\parallel }}}>_{\text{\textsc{bps%
}}}$ are given M2 brane wrapping primitive 2-cycle $\Upsilon _{\mathbf{Q}%
_{_{\parallel }}}^{\text{\textsc{bps}}}$ of the base $K3_{_{\parallel }}$
with $\mathbf{Q}_{_{\parallel }}=\mathrm{k_{_{\parallel }}}\mathbf{q}%
_{_{\parallel }}.$\textit{\ T}he cycle $\Upsilon _{\mathbf{q}_{_{\parallel
}}}^{\text{\textsc{bps}}}$ (for the case \textrm{k}$_{_{\parallel }}=1$) has
positive self intersection with positive value $\mathbf{q}_{\parallel
}^{2}>0 $; it sits in the even integral ($\mathbf{q}_{\parallel
}^{2}=2m_{\parallel } $) \textit{s}elf dual sublattice $\mathfrak{L}%
_{\parallel }^{\left( +\right) }$ of the base lattice $\mathfrak{L}%
_{\parallel }.$ \newline
The tower $\mathcal{T}_{\mathrm{k}_{_{\parallel }},\mathbf{q}_{_{\parallel
}}}^{\text{\textsc{bps}}}$ of BPS particles is characterised by the integer $%
\mathrm{k}_{_{\parallel }};$ it is generated by M2 brane wrapping the
multi-2-cycles $\Upsilon _{\mathbf{Q}_{_{\parallel }}}=$ $\mathrm{k}%
_{\parallel }\Upsilon _{\mathbf{q}_{\parallel }}^{\text{\textsc{bps}}}$
expanding in the self dual sublattice $\mathfrak{L}_{\parallel }^{\left(
+\right) }.$

\item[\textbf{c)}] \textit{The non-BPS states} $|n_{\mathbf{q}_{\parallel
}}^{L}>_{\text{\textsc{n-bps}}}$ are given by the heterotic string
excitations. The heterotic string with left moving charge $n_{\mathbf{q}%
_{\parallel }}^{L}$ is dual to the string given by M5 brane wrapping the
base surface $K3_{\parallel }.$ Here, the $K3_{\parallel }$ contains a
complex primitive curve $\Upsilon _{\mathbf{q}_{\parallel }}^{\text{\textsc{%
n-bps}}}$ sitting in the antiself dual sublattice $\mathfrak{L}_{\parallel
}^{\left( -\right) }$ with $\left( \mathbf{\alpha }\right) $ negative self
intersection value $\mathbf{q}_{\perp }^{2}<0;$ and $\left( \mathbf{\beta }%
\right) $ related to the string excitation charge as $n_{\mathbf{q}%
_{\parallel }}^{L}=-\frac{1}{2}\mathbf{q}_{\parallel }^{2}>0.$\newline
The tower $\mathcal{T}_{\mathrm{k}_{_{\parallel }},\mathbf{q}_{_{\parallel
}}}^{\text{\textsc{n-bps}}}$ is characterised by the generic integer $%
\mathrm{k}_{\parallel };$ it requires $K3_{\parallel }$ having a complex
curve $\Upsilon _{\mathbf{Q}_{\parallel }}=$ $k_{\parallel }\Upsilon _{%
\mathbf{q}_{\parallel }}^{\text{\textsc{n-bps}}}$ sitting in the sublattice $%
\mathfrak{L}_{\perp }^{\left( -\right) }$ with $\left( i\right) $\ negative
self intersection $\mathbf{q}_{\perp }^{2}<0;$ and $\left( ii\right) $\
string excitation charge given by $n_{\mathrm{k}_{\parallel }\mathbf{q}%
_{\parallel }}^{L}=-\frac{\mathrm{k}_{\parallel }^{2}}{2}\mathbf{q}%
_{\parallel }^{2}.$
\end{description}

\ \ \ \newline
The complex curves $\Upsilon _{\perp }$ and $\Upsilon _{\parallel }$ used
above depend on the structure of $K3_{\perp }\times K3_{\parallel },$ that
is on whether they have degenerations at finite or infinite distance in the
moduli space of the \emph{EFT}$_{3D}$. Recall that in the \emph{EFT}$_{5D}$
investigated in \textrm{\cite{12}}; it was shown that in fibration of type
K3, the only curves that admit a weak coupling are of two types: $\left(
\mathbf{i}\right) $ K3 without degenerations; or $\left( \mathbf{ii}\right) $
curves that degenerate in a finite distance in the moduli space. These
correspond to a generic K3 or degenerations classified as Kulikov type I, II
or III\textrm{\ \cite{27A, 27B}. }This result is general; and so hold also
for our \emph{EFT}$_{3D}$ studied in this paper. In our claims given above,
we mentioned that in directions of the charge lattice $\mathfrak{L}_{\perp }$
(resp. $\mathfrak{L}_{\parallel }$) that allow for a weak (or the dual
strong) coupling limit, there exists towers of BPS or non-BPS states
satisfying the WGC (or its strong dual version).\newline
Notice that the BPS bound in the\textrm{\ }\emph{EFT}$_{3D}$\textrm{\ }align
with the extremality bound; and thus BPS states are super-extremal. This
feature was shown for\textrm{\ }\emph{EFT}$_{5D}$\textrm{\ }in particular in%
\textrm{\ \cite{28}}; there, the BPS states are all super-extremal, and
satisfy the WGC, not only in the infinite distance limit\textrm{\ (}$\lambda
\rightarrow \infty $\textrm{)}, but throughout the moduli space of the theory%
\textrm{\ (}$\lambda $ finite\textrm{)}; this is because charge to mass
ratio is protected by supersymmetry. However, non-BPS states do not have
this property.\newline
In what follows, we test the conjecture for BPS and non-BPS states and%
\textrm{\ }weak/strong dual\textrm{\ }characterised by the conditions (\ref%
{wcf}) and (\ref{wcb})\textrm{. }We start by determining directions in the%
\textrm{\ }even integral charge lattice $\mathfrak{L}=\mathfrak{L}_{\perp
}\oplus \mathfrak{L}_{\parallel }$ of the CY4 that admit a weak coupling
limit.\textrm{\ }Then, we\textrm{\ }construct\textrm{\ }the towers\textrm{\ }%
$\mathcal{T}_{\mathrm{k},\mathbf{q}}^{\text{\textsc{bps}}}$\textrm{\ }and%
\textrm{\ }$\mathcal{T}_{\mathrm{k},\mathbf{q}}^{\text{\textsc{n-bps}}}$%
\textrm{\ }that host BPS and non-BPS states satisfying eqs(\ref{wcf}-\ref%
{wcb}).

\subsection{Weakly/strongly coupled directions in the charge lattice}

We start by focussing on the weak coupling limit at an infinite distance in
the moduli space ($\lambda \rightarrow \infty )$ of the\textrm{\ }\emph{EFT}$%
_{3D}$; and turn after to investigating the strongly coupled region dual to
asymptotic WGC. The asymptotic weak coupling limit is defined by the
constraint relation (\ref{wcf}) namely $\left( \Lambda _{{\small mag}%
}/\Lambda _{\text{\textsc{sp}}}\right) _{U(1)_{\mathbf{q}}}\rightarrow 0$
where $\Lambda _{{\small mag}}\equiv \Lambda _{\text{\textsc{wgc}}}$ is the
magnetic WGC scale used earlier, and $\Lambda _{\text{\textsc{sp}}}\equiv
\Lambda _{\text{\textsc{qg}}}$ is the species scale below which quantum
gravity (QG) becomes weakly coupled. The reason that the weak gauge coupling
limit is defined in such a way is because when approaching the species scale
$\Lambda _{\text{\textsc{sp}}}$, the dynamics of quantum gravity change.
Given a generic complex curve\textrm{\ }$\Upsilon _{\mathbf{q}}=\sum_{\text{%
\textsc{a}}=1}^{h_{{\small X}_{{\small 4}}}^{{\small 1,1}}}q_{\text{\textsc{a%
}}}\mathcal{C}^{\text{\textsc{a}}}$ in the Calabi-Yau fourfold $K3_{\perp
}\times K3_{\parallel },$\textrm{\ }the magnetic WGC scale in terms of the
3D\ Planck mass $M_{\mathrm{Pl}}$ and the Yang Mills gauge coupling $\mathrm{%
g}_{{\small 3D}}$ is defined as%
\begin{equation}
\left. \Lambda _{{\small mag}}^{2}\right\vert _{\Upsilon _{\mathbf{q}}}=%
\mathrm{g}_{{\small 3D}}^{2}M_{\mathrm{Pl}}(q_{\text{\textsc{a}}}\mathcal{G}%
^{\text{\textsc{ab}}}q_{\text{\textsc{b}}})  \label{am}
\end{equation}%
Using the expression of the gauge coupling metric $\mathcal{G}^{\text{%
\textsc{ab}}}$ given by eq(\ref{beh}), this magnetic scale expands as follows%
\begin{eqnarray}
\left. \Lambda _{{\small mag}}^{2}\right\vert _{\Upsilon _{\mathbf{q}}}
&=&\lambda ^{2}\mathrm{g}_{{\small 3D}}^{2}M_{\mathrm{Pl}}(q_{a_{\parallel }}%
\mathring{G}^{a_{\parallel }b_{\parallel }}q_{b_{\parallel }})+  \notag \\
&&2\lambda ^{0}\mathrm{g}_{{\small 3D}}^{2}M_{\mathrm{Pl}}(q_{a_{\parallel }}%
\mathring{G}^{a_{\parallel }i_{_{\perp }}}q_{i_{_{\perp }}})+  \label{aml} \\
&&\frac{1}{\lambda ^{2}}\mathrm{g}_{{\small 3D}}^{2}M_{\mathrm{Pl}%
}(q_{i_{\perp }}\mathring{G}^{i_{\perp }j_{\perp }}q_{j_{\perp }})  \notag
\end{eqnarray}%
It has three block terms captured by diagonal coupling sub-matrices $%
\mathcal{G}^{i_{\perp }j_{\perp }}$ and $\mathcal{G}^{a_{\parallel
}b_{\parallel }}$ as well as the cross block $\mathcal{G}^{a_{\parallel
}i_{_{\perp }}}.$ From this expansion, we learn several features; in
particular the two interesting following:

$\bullet $ \emph{Weakly gauge coupled directions}\newline
For fixed values of the scale $\Lambda _{\text{\textsc{sp}}},$ the weakly
coupled directions in $K3_{\perp }\times K3_{\parallel }$ that realises the
asymptotic weak gauge coupling condition $\left( \Lambda _{{\small mag}%
}/\Lambda _{\text{\textsc{sp}}}\right) _{U(1)_{\mathbf{q}}}\rightarrow 0$ is
given by the \emph{necessary} condition $\left. \Lambda _{{\small mag}%
}^{2}\right\vert _{\Upsilon _{\mathbf{q}}}\rightarrow 0.$ This necessary
condition is solved in eq(\ref{aml}) by requiring%
\begin{equation}
q_{a_{\parallel }}=0
\end{equation}%
thus restricting the complex curve $\mathbf{\Upsilon }_{\mathbf{q}}$ to sit
completely in the fiber $K3_{\perp }$ of the CY$_{4}$; that is restricting $%
\Upsilon _{\mathbf{q}}$ to the fibral curve $\Upsilon _{\mathbf{q}_{\perp }}$
with expansion as $\sum_{i_{\perp }=1}^{h_{K3_{\perp }}^{{\small 1,1}%
}}q_{i_{\perp }}\mathcal{C}^{i_{\perp }}.$ With this restriction, eqs(\ref%
{am}-\ref{aml}) read as follows%
\begin{equation}
\left. \Lambda _{{\small mag}}^{2}\right\vert _{\Upsilon _{\mathbf{q}_{\perp
}}}=\mathrm{g}_{{\small 3D}}^{2}(q_{i_{\perp }}\mathcal{G}^{i_{\perp
}j_{\perp }}q_{j_{\perp }})=\frac{1}{\lambda ^{2}}\mathrm{g}_{{\small 3D}%
}^{2}(q_{i_{\perp }}\mathring{G}^{i_{\perp }j_{\perp }}q_{j_{\perp }})
\end{equation}%
By setting $\mathrm{\mathring{g}}_{\Upsilon _{\mathbf{q}_{\perp }}}^{2}=%
\mathrm{g}_{{\small 3D}}^{2}(q_{i_{\perp }}\mathring{G}^{i_{\perp }j_{\perp
}}q_{j_{\perp }})$ with no dependence in the scaling parameter $\lambda ,$
the above relation can be presented simply as%
\begin{equation}
\left. \Lambda _{{\small mag}}^{2}\right\vert _{\Upsilon _{\mathbf{q}_{\perp
}}}=\frac{1}{\lambda ^{2}}\mathrm{\mathring{g}}_{\Upsilon _{\mathbf{q}%
_{\perp }}}^{2}  \label{sa}
\end{equation}%
So, the magnetic scale behaves in the infinite distance limit ($\lambda
\rightarrow \infty )$ like%
\begin{equation}
\left. \Lambda _{{\small mag}}^{2}\right\vert _{\Upsilon _{\mathbf{q}_{\perp
}}}\sim \mathcal{O}\left( \frac{1}{\lambda ^{2}}\right) \rightarrow 0
\end{equation}%
and then $\left( \Lambda _{{\small mag}}/\Lambda _{\text{\textsc{sp}}%
}\right) _{U(1)_{\mathbf{q}_{\perp }}}\rightarrow 0$.

$\bullet $ \emph{Strongly gauge coupled directions}\newline
For fixed values of $\Lambda _{\text{\textsc{sp}}},$ the strongly coupled
directions in $K3_{\perp }\times K3_{\parallel },$ that are dual to the weak
ones, realises the asymptotic strong coupling condition (\textrm{\ref{scf}}%
), the inverse of (\textrm{\ref{wcf}}). This condition $\left( \Lambda _{%
{\small mag}}/\Lambda _{\text{\textsc{sp}}}\right) _{U(1)_{\mathbf{q}%
}}\rightarrow \infty $ can be realised by the \emph{necessary} condition $%
\left. \Lambda _{mag}^{2}\right\vert _{\Upsilon _{\mathbf{q}}}\rightarrow
\infty .$ From eq(\ref{am}), we see that this necessary condition is
naturally solved by demanding
\begin{equation}
q_{i_{\perp }}=0
\end{equation}%
thus restricting the complex curve $\Upsilon _{\mathbf{q}}$ to completely
sitting in the base $K3_{\parallel }$ of the CY$_{4}$; that is restricting $%
\mathcal{\Upsilon }_{\mathbf{q}}$ to base curves $\Upsilon _{\mathbf{q}%
_{_{\parallel }}}$ given by $\sum_{a_{\parallel }=1}^{h_{K3\parallel }^{%
{\small 1,1}}}q_{a_{\parallel }}\mathcal{C}^{a_{\parallel }}.$ So the
magnetic scale (\ref{am}) reads as
\begin{equation}
\left. \Lambda _{{\small mag}}^{2}\right\vert _{\Upsilon _{\mathbf{q}%
_{\parallel }}}=\mathrm{g}_{{\small 3D}}^{2}M_{\mathrm{Pl}}(q_{a_{\parallel
}}\mathcal{G}^{a_{\parallel }b_{\parallel }}q_{b_{\parallel }})=\lambda ^{2}%
\mathrm{g}_{{\small 3D}}^{2}M_{\mathrm{Pl}}(q_{a_{\parallel }}\mathring{G}%
^{a_{\parallel }b_{\parallel }}q_{b_{\parallel }})
\end{equation}%
By setting $\mathrm{\mathring{g}}_{\Upsilon _{\mathbf{n}_{\parallel }}}^{2}=%
\mathrm{g}_{{\small 3D}}^{2}(q_{a_{\parallel }}\mathring{G}^{a_{\parallel
}b_{\parallel }}q_{b_{\parallel }})$ with finite dependence in the large
limit of $\lambda ,$ this relation can be presented just like%
\begin{equation}
\left. \Lambda _{{\small mag}}^{2}\right\vert _{\Upsilon _{\mathbf{q}%
_{\parallel }}}=\lambda ^{2}\mathrm{\mathring{g}}_{\Upsilon _{\mathbf{n}%
_{\parallel }}}^{2}  \label{as}
\end{equation}%
behaving like $\left. \Lambda _{{\small mag}}^{2}\right\vert _{\mathcal{%
\Upsilon }_{\mathbf{n}_{\parallel }}}\sim \mathcal{O}\left( \lambda
^{2}\right) $ and $\left. \Lambda _{{\small mag}}^{2}\right\vert _{\mathcal{%
\Upsilon }_{\mathbf{n}_{\parallel }}}\rightarrow \infty .$ Therefore, for
fixed $\Lambda _{\text{\textsc{sp}}},$ we have $\left( \Lambda _{{\small mag}%
}/\Lambda _{\text{\textsc{sp}}}\right) _{U(1)_{\mathbf{q}_{\parallel
}}}\rightarrow \infty .$

$\bullet $ \emph{About the species scale} $\Lambda _{SP}$\newline
First, recall that the species scale $\Lambda _{SP}$ is defined as the scale
\emph{beyond} which quantum gravity becomes strongly coupled. A general
formula defining the species scale in D-dimensional space \textrm{time is
given by} \textrm{\cite{3A, 29}}%
\begin{equation}
\Lambda _{\text{\textsc{sp}}}=\frac{M_{\mathrm{Pl}}^{{\tiny [D]}}}{%
N^{1/\left( D-2\right) }}
\end{equation}%
where $M_{\mathrm{Pl}}^{{\tiny [D]}}$ is the D-dimensional Planck mass, and $%
N$ is the number of species. In our \emph{EFT}$_{3D}$ descending from
M-theory on $K3_{\perp }$-fibered Calabi-Yau fourfold on $K3_{\Vert }$, the
species scale $\Lambda _{\text{\textsc{sp}}}$ and the heterotic scale $%
\mathrm{M}_{het}$ are linked by a logarithmic behaviour via%
\begin{equation}
\Lambda _{\text{\textsc{sp}}}^{2}\sim \mathrm{M}_{{\small het}}^{2}\log (%
\frac{\mathrm{M}_{\mathrm{Pl}}}{\mathrm{M}_{{\small het}}})>0\qquad ,\qquad
\mathrm{M}_{{\small het}}<\mathrm{M}_{\mathrm{Pl}}  \label{m}
\end{equation}%
On the other hand taking the K3 volume $\mathcal{V}_{K3}$ to be
dimensionless, the heterotic string tension $T_{het}=2\pi \mathrm{M}%
_{het}^{2}$ is expressed in terms of the K3 volume as $T_{het}=2\pi \mathcal{%
V}_{K3}\mathrm{M}_{11d}^{2}$ and then $\mathrm{M}_{{\small het}}^{2}=\mathrm{%
M}_{11d}^{2}\mathcal{V}_{K3}.$ So, we distinguish%
\begin{equation}
\mathrm{M}_{{\small het}_{\perp }}^{2}=\mathrm{M}_{11d}^{2}\mathcal{V}%
_{X_{\perp }}\qquad ,\qquad \mathrm{M}_{{\small het}_{\Vert }}^{2}=\mathrm{M}%
_{11d}^{2}\mathcal{V}_{K3_{_{\Vert }}}  \label{het}
\end{equation}%
Moreover, because under infinite distance scaling ($\lambda \rightarrow
\infty $), the volume $\mathcal{V}_{K3_{\perp }}$ of the\ fiber $K3_{\perp }$
scales like $\lambda ^{-2}\mathcal{V}_{K3_{\perp }}$ while the volume $%
\mathcal{V}_{K3_{\Vert }}$ of the base $K3_{_{\Vert }}$ scales like $\lambda
^{2}\mathcal{V}_{K3_{_{\Vert }}},$ the scaling of the heterotic mass scales $%
\mathrm{M}_{het\bot }$ and $\mathrm{M}_{het\Vert }$ respectively on the
fiber and on the base of the CY4 behave as follows%
\begin{equation}
\mathrm{M}_{het\bot }\sim \frac{1}{\lambda }\mathrm{M}_{11d}\qquad ,\qquad
\mathrm{M}_{het\Vert }\sim \lambda \mathrm{M}_{11d}
\end{equation}%
As such, in the infinite distance limit, we have $\mathrm{M}_{het\bot
}\rightarrow 0$ while $\mathrm{M}_{het\Vert }\rightarrow \infty .$ By
substituting into (\ref{m}), the species scale $\Lambda _{\text{\textsc{sp}}%
_{\bot }}$ behaves for the asymptotic weak gauge coupling as%
\begin{equation}
\Lambda _{\text{\textsc{sp}}_{\bot }}^{2}\sim \frac{\mathrm{M}_{11d}^{2}}{%
\lambda ^{2}}\left[ \log \frac{\mathrm{M}_{\mathrm{Pl}}}{\mathrm{M}_{11d}}%
+\log \lambda \right]  \label{a}
\end{equation}%
For the strongly coupled regions the species scale $\Lambda _{\text{\textsc{%
sp}}_{\Vert }}$ behaves like:
\begin{equation}
\Lambda _{\text{\textsc{sp}}_{\Vert }}^{2}\sim \lambda ^{2}\mathrm{M}%
_{11d}^{2}\log \frac{\mathrm{M}_{\mathrm{Pl}}}{\lambda \mathrm{M}_{11d}}
\label{b}
\end{equation}%
indicating that we must have $\lambda \mathrm{M}_{11d}\leq \mathrm{M}_{%
\mathrm{Pl}}$ and then $\lambda $ is bounded as $\lambda \leq \mathrm{M}_{%
\mathrm{Pl}}/\mathrm{M}_{11d}.$ Consequently, the asymptotic weak gauge
coupling condition reading like
\begin{equation}
\frac{(\Lambda _{{\small mag}}^{2})_{_{\Upsilon _{\mathbf{q}_{\perp }}}}}{%
\Lambda _{\text{\textsc{sp}}_{\bot }}^{2}}=\frac{\mathrm{\mathring{g}}%
_{\Upsilon _{\mathbf{q}_{\perp }}}^{2}}{\mathrm{M}_{11d}\log \lambda }
\end{equation}%
tends indeed to zero in the infinite distance limit $\lambda \rightarrow
\infty ;$ thus showing the validity of the weak coupling limit condition (%
\ref{Weak}-\ref{wcf}). In this regards, it was shown in\textrm{\ \cite{12} t}%
hat in fibration of type K3 of the CY3 of the\textrm{\ }\emph{EFT}$_{5D}$%
\textrm{, }the curves\textrm{\ }$\mathcal{C}_{\mathbf{q}}$\textrm{\ }that
admit a weak gauge coupling symmetry\textrm{\ }$U\left( 1\right) _{\mathcal{C%
}_{\mathbf{q}}}$\textrm{\ }are either in a non degenerate fiber K3, or the
ones that degenerate in a finite distance in the moduli space. This result
is general since it only depends on the geometry of $K3$; and consequently
holds also for the curves\textrm{\ }$\Upsilon _{\mathbf{q}_{\perp }}$\textrm{%
\ }used in our\textrm{\ }\emph{EFT}$_{3D}\mathrm{.}$

As for the strongly coupled region, which is dual to the asymptotic weakly
coupled region, the check of the condition is given by $\left( \Lambda _{%
{\small mag}}/\Lambda _{\text{\textsc{sp}}}\right) _{U(1)_{\mathbf{q}%
_{\parallel }}}\rightarrow \infty $ when the scaling parameter $\lambda
\rightarrow \mathrm{M}_{\mathrm{Pl}}/\mathrm{M}_{11d}.$ By substituting $%
(\Lambda _{{\small mag}}^{2})_{\Upsilon _{\mathbf{q}_{\parallel }}}=\lambda
^{2}\mathrm{\mathring{g}}_{\Upsilon _{\mathbf{q}_{\parallel }}}^{2}$ and $%
\Lambda _{\text{\textsc{sp}}_{\Vert }}$ by its expression (\ref{b}), the
condition reads as follows%
\begin{equation}
\frac{(\Lambda _{{\small mag}}^{2})_{\Upsilon _{\mathbf{q}_{\parallel }}}}{%
\Lambda _{\text{\textsc{sp}}_{\Vert }}^{2}}=\frac{\mathrm{\mathring{g}}%
_{\Upsilon _{\mathbf{q}_{\parallel }}}^{2}}{\mathrm{M}_{11d}^{2}\log \frac{%
\mathrm{M}_{\mathrm{Pl}}}{\lambda \mathrm{M}_{11d}}}
\end{equation}%
it tends to infinity for $\lambda \rightarrow \mathrm{M}_{\mathrm{Pl}}/%
\mathrm{M}_{11d}$ in agreement with the weak/strong duality property of the
\emph{EFT}$_{3D}$ considered in this study.

In what follows we investigate the towers of states along these weak and
strong directions in the charge lattice in $K3_{\Vert }\times K3_{\bot }.$
We start with the tower $\mathcal{T}_{\mathrm{k},\mathbf{q}}^{\text{\textsc{%
bps}}}$ BPS states in the \emph{EFT}$_{3D}$, and then move on to the tower $%
\mathcal{T}_{\mathrm{k},\mathbf{q}}^{\text{\textsc{n-bps}}}$ of non-BPS
states.

\subsubsection{Towers of BPS states and asymptotic gauge couplings}

Having determined the directions in the\textrm{\ }even integral charge
lattice $\mathfrak{L}=\mathfrak{L}_{\perp }\oplus \mathfrak{L}_{\parallel }$
in the CY4 that are weakly coupled in $K3_{\bot }$ (and strongly in $%
K3_{\Vert }$ due to weak/gravity duality), we turn now to explicitly verify
if the states in these directions satisfy the tower WGC and\emph{\ its
strong dual version}.\ To that purpose, recall that the WGC tower of states
satisfies the relation (\ref{wcf}); these states include BPS states
occupying rays in the self dual sublattice $\mathfrak{L}_{\perp }^{\left(
+\right) }$; and the non BPS states living in the anti-self dual $\mathfrak{L%
}_{\perp }^{\left( -\right) }.$

For the case of the BPS tower $\mathcal{T}_{\mathrm{k}_{\bot },\mathbf{q}%
_{\bot }}^{\text{\textsc{bps}}}$, it is built out of M2 brane wrapping
complex curve $\Upsilon _{\mathrm{k}\mathbf{q}_{\bot }}$ in $\mathfrak{L}%
_{\perp }^{\left( +\right) }$ having positive self intersection ($\Upsilon _{%
\mathrm{k}_{\bot }\mathbf{q}_{\bot }}^{2}>0$). By weak/strong gauge duality,
this feature extends naturally to a dual BPS tower $\mathcal{T}_{\mathrm{k}%
_{\Vert },\mathbf{q}_{\Vert }}^{\text{\textsc{bps}}}$ based on complex
curves $\Upsilon _{\mathrm{k}_{\Vert }\mathbf{q}_{\Vert }}$ sitting in $%
\mathfrak{L}_{\Vert }^{\left( +\right) }$. For these two BPS towers, the
masses $M_{\mathrm{k}}$ of their particle states is given by
\begin{equation}
M_{\mathrm{k}}=\xi M_{11d}\dsum\limits_{\text{\textsc{a}}=1}^{h_{{\small CY4}%
}^{1,1}}\left( \mathrm{k}q_{\text{\textsc{a}}}\hat{\upsilon}^{\text{\textsc{a%
}}}\right) \qquad ,\qquad M_{11d}=\frac{M_{\mathrm{Pl}}}{\mathrm{\gamma }%
\mathcal{V}_{{\small CY}_{{\small 4}}}}  \label{km}
\end{equation}%
with integers \textrm{k}$_{x}$ and dimensionless volume $\hat{\upsilon}^{%
\text{\textsc{a}}}=\upsilon ^{\text{\textsc{a}}}/\mathcal{V}_{{\small CY}%
_{4}}^{1/4}$ and $\mathrm{\gamma =4\pi }$ and $\xi =1/\sqrt{2}.$ From this
expression of the mass, which splits like the sum of $M_{\mathrm{k}}=M_{%
\mathrm{k}_{\bot }}+M_{\mathrm{k}_{\Vert }}$ with
\begin{equation}
M_{\mathrm{k}_{\bot }}=\xi M_{11d}\sum_{i_{\bot }=1}^{h_{\bot }^{1,1}}\left(
\mathrm{k}_{\bot }q_{i_{\bot }}\hat{\upsilon}^{i_{\bot }}\right) \qquad
,\qquad M_{\mathrm{k}_{\Vert }}=\xi M_{11d}\sum_{a_{\Vert }=1}^{h_{\Vert
}^{1,1}}\left( \mathrm{k}_{\Vert }q_{a_{\Vert }}\hat{\upsilon}^{a_{\Vert
}}\right)
\end{equation}%
we can check the asymptotic WGC condition (\ref{wcf}) and its strong regime
dual (\ref{wcf}).

Now, given eq(\ref{km}), we learn that $M_{\mathrm{k}}^{2}$ is equal to $%
4\pi ^{2}\mathrm{k}^{2}\left( q_{\text{\textsc{a}}}\hat{\upsilon}^{\text{%
\textsc{a}}}\right) ^{2}M_{11d}^{2}$ , it is quadratic in the volumes $\hat{%
\upsilon}^{\text{\textsc{a}}}$ and in $\mathrm{k}$ as well as the integral
charges $q_{\text{\textsc{a}}}$; it splits into three blocks as follows
\begin{equation}
\left( M_{\mathrm{k}}\right) ^{2}=\left( M_{\mathrm{k}_{\bot }}\right)
^{2}+(M_{\mathrm{k}_{\Vert }})^{2}+2M_{\mathrm{k}_{\bot }}M_{\mathrm{k}%
_{\Vert }}
\end{equation}%
where the fibral $\left( M_{\mathrm{k}_{\bot }}\right) ^{2}$ and the base $%
(M_{\mathrm{k}_{\Vert }})^{2}$ are quadratic in ($\hat{\upsilon}^{i_{\bot }},%
\mathrm{k}_{\bot },q_{i_{\bot }})$ and ($\hat{\upsilon}^{a_{\Vert }},\mathrm{%
k}_{\Vert },q_{a_{\Vert }})$ respectively.

\paragraph{A. Repulsive Force Conjecture (RFC):\newline
}

Given a massive BPS particle in the \emph{EFT}$_{{\small 3D}}$ of mass $M_{%
\mathrm{k}}$, the repulsive force conjecture (RFC), which in the infinite
distance limit coincides with the asymptotic WGC, requires the following
inequality \textrm{\cite{3A, 6A, 6B}}
\begin{equation}
F_{Coulomb}\geq F_{Grav}+F_{Yukawa}
\end{equation}%
This inequality reads explicitly as follows%
\begin{equation}
\frac{\mathrm{g}_{{\small 3D}}^{2}}{M_{\mathrm{Pl}}}\left( \mathrm{k}^{2}q_{%
\text{\textsc{a}}}\mathcal{G}^{\text{\textsc{ab}}}q_{\text{\textsc{b}}%
}\right) \geq \left. \frac{D-3}{D-2}\right\vert _{D=3}\frac{M_{\mathrm{k}%
}^{2}}{M_{\mathrm{Pl}}^{2}}+\mathcal{Y}  \label{ne}
\end{equation}%
with Yukawa matter coupling contribution $\mathcal{Y}$ reading in terms of
the reduced mass $M_{\mathrm{k}}$ as follows%
\begin{eqnarray}
M_{\mathrm{Pl}}^{2}\mathcal{Y} &=&\frac{\mathcal{G}^{\text{\textsc{ab}}}}{%
2M_{\mathrm{k}}^{2}}\left( \frac{\partial M_{\mathrm{k}}^{2}}{\partial \hat{%
\upsilon}^{\text{\textsc{a}}}}\right) \left( \frac{\partial M_{\mathrm{k}%
}^{2}}{\partial \hat{\upsilon}^{\text{\textsc{b}}}}\right) -\frac{\varrho }{%
M_{\mathrm{k}}^{2}}\left[ \hat{\upsilon}^{\text{\textsc{a}}}\frac{\partial
M_{\mathrm{k}}^{2}}{\partial \hat{\upsilon}^{\text{\textsc{a}}}}\right] ^{2}
\\
&=&2\mathcal{G}^{\text{\textsc{ab}}}\left( \frac{\partial M_{\mathrm{k}}}{%
\partial \hat{\upsilon}^{\text{\textsc{a}}}}\right) \left( \frac{\partial M_{%
\mathrm{k}}}{\partial \hat{\upsilon}^{\text{\textsc{b}}}}\right) -4\varrho %
\left[ \hat{\upsilon}^{\text{\textsc{a}}}\frac{\partial M_{\mathrm{k}}}{%
\partial \hat{\upsilon}^{\text{\textsc{a}}}}\right] ^{2}
\end{eqnarray}%
with $\varrho =1/2.$ Notice that because here $D=3,$ the contribution from
classical gravity namely $\left( D-3\right) /\left( D-2\right) $ vanishes
identically; this is expected because massless 3D gravity is topological.
However, a non trivial contribution may come from quantum effect; we denote
it like $\mathrm{\vartheta }_{{\small QG}}$. Substituting, the RFC
inequality (\ref{ne}) becomes%
\begin{equation}
\mathrm{g}_{{\small 3D}}^{2}M_{\mathrm{Pl}}\left( \mathrm{k}^{2}q_{\text{%
\textsc{a}}}\mathcal{G}^{\text{\textsc{ab}}}q_{\text{\textsc{b}}}\right)
\geq \mathrm{\vartheta }_{{\small QG}}+\frac{\mathcal{G}^{\text{\textsc{ab}}}%
}{2M_{\mathrm{k}}^{2}}\left( \frac{\partial M_{\mathrm{k}}^{2}}{\partial
\hat{\upsilon}^{\text{\textsc{a}}}}\right) \left( \frac{\partial M_{\mathrm{k%
}}^{2}}{\partial \hat{\upsilon}^{\text{\textsc{b}}}}\right) -\frac{\varrho }{%
M_{\mathrm{k}}^{2}}\left[ \hat{\upsilon}^{\text{\textsc{a}}}\frac{\partial
M_{\mathrm{k}}^{2}}{\partial \hat{\upsilon}^{\text{\textsc{a}}}}\right] ^{2}
\label{bcfg}
\end{equation}%
with charge $q_{\text{\textsc{a}}}=\left( q_{i_{\perp }},q_{a_{\Vert
}}\right) .$ We rewrite this inequality by putting the mass contribution to
the right hand side as follows%
\begin{equation}
\frac{\mathrm{g}_{{\small 3D}}^{2}M_{\mathrm{Pl}}}{4}\left( \mathrm{k}^{2}q_{%
\text{\textsc{a}}}\mathcal{G}^{\text{\textsc{ab}}}q_{\text{\textsc{b}}%
}\right) \geq \mathrm{\vartheta }_{{\small QG}}+\frac{1}{2}\mathcal{G}^{%
\text{\textsc{ab}}}\left( \frac{\partial M_{\mathrm{k}_{x}}}{\partial \hat{%
\upsilon}^{\text{\textsc{a}}}}\right) \left( \frac{\partial M_{\mathrm{k}%
_{x}}}{\partial \hat{\upsilon}^{\text{\textsc{b}}}}\right) -\varrho \left[
\hat{\upsilon}^{\text{\textsc{a}}}\frac{\partial M_{\mathrm{k}_{x}}}{%
\partial \hat{\upsilon}^{\text{\textsc{a}}}}\right] ^{2}  \label{gf}
\end{equation}%
Because the metric inverse $\mathcal{G}^{\text{\textsc{ab}}}$ plays an
important role in this relation and its test, we first give details on its
calculation; and turn after to verify the validity of the repulsive force
conjecture for the tower of BPS states. Below, we assume\textrm{\ }$\mathrm{%
\vartheta }_{{\small QG}}<0$ for BPS particle states in vacuum \textrm{\cite%
{TPA,TPB}}; so to test\textrm{\ (\ref{bcfg}-\ref{gf})}, we can disregard%
\textrm{\ }$\mathrm{\vartheta }_{{\small QG}}$\textrm{\ }without affection
the message captured by the conjecture.

\paragraph{B. \ Deriving the inverse of gauge metric (\protect\ref{MT}):%
\newline
}

In eq(\ref{MX}), the matrix $\mathcal{G}^{\text{\textsc{ab}}}$ is the
inverse of the gauge coupling metric $\mathcal{G}_{\text{\textsc{ab}}}$;
this is a $h_{{\small CY4}}^{1,1}\times h_{{\small CY4}}^{1,1}$ matrix
decomposing into 4 bloc submatrices as follows,%
\begin{equation}
\mathcal{G}_{\text{\textsc{ab}}}=\left(
\begin{array}{cc}
\mathcal{G}_{a_{\Vert }b_{\Vert }} & \mathcal{G}_{a_{\Vert }j_{\perp }} \\
\mathcal{G}_{i_{\perp }b_{\Vert }} & \mathcal{G}_{i_{\perp }j_{\perp }}%
\end{array}%
\right) ,\qquad \mathcal{G}^{\text{\textsc{ab}}}=\left(
\begin{array}{cc}
\mathcal{G}^{a_{\Vert }b_{\Vert }} & \mathcal{G}^{a_{\Vert }j_{\perp }} \\
\mathcal{G}^{i_{\perp }b_{\Vert }} & \mathcal{G}^{i_{\perp }j_{\perp }}%
\end{array}%
\right)  \label{GM}
\end{equation}%
The sub-bloc $\mathcal{G}_{a_{\Vert }b_{\Vert }}$ is a square matrix $%
h_{\Vert }^{1,1}\times h_{\Vert }^{1,1},$ the sub-bloc $\mathcal{G}%
_{i_{\perp }j_{\perp }}$ is also a square matrix $h_{\perp }^{1,1}\times
h_{\perp }^{1,1}.$ The sub-bloc $\mathcal{G}_{a_{\Vert }j_{\perp }}$ and $%
\mathcal{G}_{i_{\perp }b_{\Vert }}$ are matrices $h_{\Vert }^{1,1}\times
h_{\perp }^{1,1}$ and $h_{\perp }^{1,1}\times h_{\Vert }^{1,1}.$ For the
case $h_{\Vert }^{1,1}=h_{\perp }^{1,1}$ required by Weak/Strong duality,
these blocs are square matrices. Below, we use le large ($\lambda
\rightarrow \infty $) and short ($\lambda \rightarrow 0$) distance limits to
compute the metric components in eq(\ref{GM}).

\begin{itemize}
\item \textbf{Infinite distance limit} ($\lambda \rightarrow \infty $):%
\newline
To determine $\mathcal{G}^{\text{\textsc{ab}}}$, we use the specific
properties (\ref{JK3}) of the $K3_{\perp }\times K3_{\Vert }$ fibration of
the CY4 showing that the coupling tensor $\kappa _{\text{\textsc{abcd}}}$ in
the infinite distance limit ($\lambda \rightarrow \infty $) obeys the
following features,
\begin{equation}
\kappa _{a_{\Vert }b_{\Vert }c_{\Vert }d_{\Vert }}=0,\qquad \kappa
_{a_{\Vert }b_{\Vert }c_{\Vert }i_{\perp }}=0,\qquad \kappa _{a_{\Vert
}b_{\Vert }i_{\perp }j_{\perp }}\neq 0
\end{equation}%
These constraint relations follow from the condition ($\mathcal{V}_{{\small %
CY}_{{\small 4}}}<\infty $) and the asymptotic behaviour of the Kahler
2-form given by $J_{\Vert }^{4}=J_{\Vert }^{3}=0$ and $J_{\Vert }^{2}\neq 0$
that translate at the level of $\kappa _{\text{\textsc{abcd}}}$ as above.%
\newline
Putting into (\ref{52}-\ref{53}), we obtain the following leading terms in $%
\lambda $,
\begin{equation}
\begin{tabular}{lll}
$\hat{u}_{\parallel }$ & $\simeq $ & $\frac{1}{2\lambda }\kappa _{_{\Vert
\Vert \bot \bot }}\hat{\upsilon}^{\Vert }\left( \hat{\upsilon}^{\bot
}\right) ^{2}$ \\
$\hat{u}_{\parallel \parallel }$ & $\simeq $ & $\frac{1}{2\lambda ^{2}}%
\kappa _{_{\Vert \Vert \bot \bot }}\left( \hat{\upsilon}^{\bot }\right) ^{2}$
\\
$\hat{u}_{\parallel \bot }$ & $\simeq $ & $\kappa _{_{\Vert \bot \Vert \bot
}}\hat{\upsilon}^{\Vert }\hat{\upsilon}^{\bot }$%
\end{tabular}%
\qquad ,\qquad
\begin{tabular}{lll}
$\hat{u}_{\perp }$ & $\simeq $ & $\frac{\lambda }{2}\kappa _{_{\bot \bot
\Vert \Vert }}(\hat{\upsilon}^{\Vert })^{2}\left( \hat{\upsilon}^{\bot
}\right) $ \\
$\hat{u}_{\bot \bot }$ & $\simeq $ & $\frac{\lambda ^{2}}{2}\kappa _{_{\bot
\bot \Vert \Vert }}(\hat{\upsilon}^{\Vert })^{2}$ \\
$\hat{u}_{\bot \parallel }$ & $\simeq $ & $\kappa _{_{\bot \Vert \Vert \bot
}}\hat{\upsilon}^{\Vert }\hat{\upsilon}^{\bot }$%
\end{tabular}%
\end{equation}%
exhibiting a manifest symmetry under the exchange $K3_{\perp
}\leftrightarrow K3_{\Vert }.$ These expressions read explicitly as follows%
\begin{equation}
\begin{tabular}{lll}
$\hat{u}_{a_{\Vert }}$ & $\simeq $ & $\frac{1}{2\lambda }\kappa _{a_{\Vert
}b_{\Vert }i_{\bot }k_{\bot }}\hat{\upsilon}^{b_{\Vert }}\left( \hat{\upsilon%
}^{i_{\bot }}\hat{\upsilon}^{k_{\bot }}\right) $ \\
$\hat{u}_{a_{\Vert }b_{\Vert }}$ & $\simeq $ & $\frac{1}{2\lambda ^{2}}%
\kappa _{a_{\Vert }b_{\Vert }i_{\bot }j_{\bot }}\left( \hat{\upsilon}%
^{i_{\bot }}\hat{\upsilon}^{j_{\bot }}\right) $ \\
$\hat{u}_{a_{\Vert }i_{\bot }}$ & $\simeq $ & $\kappa _{a_{\Vert }i_{\bot
}b_{\Vert }j_{\bot }}\hat{\upsilon}^{b_{\Vert }}\hat{\upsilon}^{j_{\bot }}$%
\end{tabular}%
\qquad ,\qquad
\begin{tabular}{lll}
$\hat{u}_{i_{\perp }}$ & $\simeq $ & $\frac{\lambda }{2}\kappa _{i_{\bot
}k_{\bot }a_{\Vert }b_{\Vert }}\hat{\upsilon}^{k_{\bot }}\hat{\upsilon}%
^{a_{\Vert }}\hat{\upsilon}^{b_{\Vert }}$ \\
$\hat{u}_{i_{\perp }j_{\perp }}$ & $\simeq $ & $\frac{\lambda ^{2}}{2}\kappa
_{i_{\perp }j_{\perp }a_{\Vert }b_{\Vert }}\hat{\upsilon}^{a_{\Vert }}\hat{%
\upsilon}^{b_{\Vert }}$ \\
$\hat{u}_{i_{\perp }a_{\Vert }}$ & $\simeq $ & $\kappa _{i_{\perp }a_{\Vert
}b_{\Vert }j_{\perp }}\hat{\upsilon}^{b_{\Vert }}\hat{\upsilon}^{j_{\bot }}$%
\end{tabular}
\label{uuu}
\end{equation}%
Moreover, using the relations
\begin{equation}
\begin{tabular}{lll}
$\kappa _{a_{\Vert }b_{\Vert }i_{\perp }j_{\perp }}$ & $=$ & $\eta
_{a_{\Vert }b_{\Vert }}\eta _{i_{\perp }j_{\perp }}$ \\
$\mathcal{\hat{V}}_{K3_{\Vert }}$ & $=$ & $\frac{1}{2}\eta _{a_{\Vert
}b_{\Vert }}\hat{\upsilon}^{a_{\Vert }}\hat{\upsilon}^{b_{\Vert }}$ \\
$\mathcal{\hat{V}}_{K3_{\perp }}$ & $=$ & $\frac{1}{2}\eta _{i_{\perp
}j_{\perp }}\hat{\upsilon}^{i_{\bot }}\hat{\upsilon}^{j_{\bot }}$%
\end{tabular}%
\end{equation}%
we can put (\ref{uuu}) in the form%
\begin{equation}
\begin{tabular}{lll}
$\hat{u}_{a_{\Vert }}$ & $\simeq $ & $\frac{\mathcal{\hat{V}}_{K3_{\perp }}}{%
\lambda }\hat{z}_{a_{\Vert }}$ \\
$\hat{u}_{a_{\Vert }b_{\Vert }}$ & $\simeq $ & $\frac{\mathcal{\hat{V}}%
_{K3_{\perp }}}{\lambda ^{2}}\eta _{a_{\Vert }b_{\Vert }}$ \\
$\hat{u}_{a_{\Vert }i_{\bot }}$ & $\simeq $ & $\hat{z}_{a_{\Vert }}\hat{z}%
_{i_{\perp }}$%
\end{tabular}%
\qquad ,\qquad
\begin{tabular}{lll}
$\hat{u}_{i_{\perp }}$ & $\simeq $ & $\left( \lambda \mathcal{\hat{V}}%
_{K3_{\Vert }}\right) \hat{z}_{i_{\perp }}$ \\
$\hat{u}_{i_{\perp }j_{\perp }}$ & $\simeq $ & $\left( \lambda ^{2}\mathcal{%
\hat{V}}_{K3_{\Vert }}\right) \eta _{i_{\perp }j_{\perp }}$ \\
$\hat{u}_{i_{\perp }a_{\Vert }}$ & $\simeq $ & $\hat{z}_{i_{\perp }}\hat{z}%
_{a_{\Vert }}$%
\end{tabular}
\label{uu}
\end{equation}%
where we set%
\begin{equation}
\begin{tabular}{lllllll}
$\hat{z}_{a_{\Vert }}$ & $=$ & $\eta _{a_{\Vert }b_{\Vert }}\hat{\upsilon}%
^{b_{\Vert }}$ & $,\qquad $ & $\hat{\upsilon}^{a_{\Vert }}\hat{z}_{a_{\Vert
}}$ & $=$ & $2\mathcal{\hat{V}}_{K3_{\Vert }}$ \\
$\hat{z}_{i_{\perp }}$ & $=$ & $\eta _{i_{\perp }j_{\perp }}\hat{\upsilon}%
^{j_{\bot }}$ & $,\qquad $ & $\hat{\upsilon}^{i_{\bot }}\hat{z}_{i_{\perp }}$
& $=$ & $2\mathcal{\hat{V}}_{K3_{\perp }}$%
\end{tabular}
\label{zz}
\end{equation}%
Substituting into $\mathcal{G}_{\text{\textsc{ab}}}=\hat{u}_{\text{\textsc{a}%
}}\hat{u}_{\text{\textsc{b}}}-\hat{u}_{\text{\textsc{ab}}},$ we get%
\begin{equation}
\begin{tabular}{lll}
$\mathcal{G}_{i_{\bot }j_{\bot }}$ & $\simeq $ & $\lambda ^{2}\mathcal{\hat{V%
}}_{K3_{\Vert }}\left[ \mathcal{\hat{V}}_{K3_{\Vert }}\hat{z}_{i_{\perp }}%
\hat{z}_{j_{\perp }}-\eta _{i_{\perp }j_{\perp }}\right] $ \\
$\mathcal{G}_{a_{\Vert }b_{\Vert }}$ & $\simeq $ & $\frac{\mathcal{\hat{V}}%
_{K3_{\perp }}}{\lambda ^{2}}\left[ \mathcal{\hat{V}}_{K3_{\perp }}\hat{z}%
_{a_{\Vert }}\hat{z}_{b_{\Vert }}-\eta _{a_{\Vert }b_{\Vert }}\right] $ \\
$\mathcal{G}_{a_{\Vert }j_{\perp }}$ & $\simeq $ & $\left( \mathcal{\hat{V}}%
_{K3_{\perp }}\mathcal{\hat{V}}_{K3_{\Vert }}\right) \hat{z}_{a_{\Vert }}%
\hat{z}_{j_{\perp }}-\hat{z}_{a_{\Vert }}\hat{z}_{j_{\perp }}$%
\end{tabular}%
\end{equation}%
Moreover, using $\mathcal{\hat{V}}_{K3_{\perp }}\mathcal{\hat{V}}_{K3_{\Vert
}}=\mathcal{\hat{V}}_{CY_{4}}=1$ and (\ref{uu}) implying $\hat{z}_{i_{\perp
}}\hat{z}_{j_{\perp }}=2\mathcal{\hat{V}}_{K3_{\perp }}\eta _{i_{\perp
}j_{\perp }}+O\left( 1/\lambda \right) $; we have $\mathcal{\hat{V}}%
_{K3_{\Vert }}\hat{z}_{i_{\perp }}\hat{z}_{j_{\perp }}=2\eta _{i_{\perp
}j_{\perp }};$ consequently we end up with $\mathcal{G}_{a_{\Vert }j_{\perp
}}\simeq 0$ as well as the following
\begin{equation}
\begin{tabular}{lll}
$\mathcal{G}_{i_{\bot }j_{\bot }}$ & $\simeq $ & $\lambda ^{2}\mathcal{\hat{V%
}}_{K3_{\Vert }}\eta _{i_{\perp }j_{\perp }}$ \\
$\mathcal{G}_{a_{\Vert }b_{\Vert }}$ & $\simeq $ & $\frac{\mathcal{\hat{V}}%
_{K3_{\perp }}}{\lambda ^{2}}\eta _{a_{\Vert }b_{\Vert }}$%
\end{tabular}%
,\qquad
\begin{tabular}{lll}
$\mathcal{G}^{i_{\bot }j_{\bot }}$ & $\simeq $ & $\frac{1}{\lambda ^{2}%
\mathcal{\hat{V}}_{K3_{\Vert }}}\eta ^{i_{\bot }j_{\bot }}$ \\
$\mathcal{G}^{a_{\Vert }b_{\Vert }}$ & $\simeq $ & $\frac{\lambda ^{2}}{%
\mathcal{\hat{V}}_{K3_{\perp }}}\eta ^{a_{\Vert }b_{\Vert }}$%
\end{tabular}
\label{TM1}
\end{equation}%
related by the Weak/Strong gauge duality.

\item \textbf{Short distance limit }($\lambda \rightarrow 0$):\newline
In the short distance limit, we have the asymptotic behavior of the Kahler
2-form: $J_{\bot }^{4}=J_{\bot }^{3}=0$ and $J_{\bot }^{2}\neq 0$. It
translates at the level of $\kappa _{\text{\textsc{abcd}}}$ as follows
\begin{equation}
\kappa _{\bot \bot \bot }=0,\qquad \kappa _{\bot \bot \bot \Vert }=0,\qquad
\kappa _{\bot \bot \Vert \Vert }\neq 0
\end{equation}%
Substituting, we get%
\begin{equation}
\begin{tabular}{lll}
$\hat{u}_{a_{\Vert }}$ & $=$ & $\frac{\mathcal{\hat{V}}_{K3_{\perp }}}{%
\lambda }\eta _{a_{\Vert }c_{\Vert }}\hat{\upsilon}^{c_{\Vert }}$ \\
$\hat{u}_{a_{\Vert }b_{\Vert }}$ & $=$ & $\frac{\mathcal{\hat{V}}_{K3_{\perp
}}}{\lambda ^{2}}\eta _{a_{\Vert }b_{\Vert }}$%
\end{tabular}%
,\qquad
\begin{tabular}{lll}
$\hat{u}_{i_{\bot }}$ & $=$ & $\lambda \mathcal{\hat{V}}_{K3_{\Vert }}\eta
_{i_{\bot }k_{\bot }}\hat{\upsilon}^{k_{\bot }}$ \\
$\hat{u}_{i_{\bot }j_{\bot }}$ & $=$ & $\lambda ^{2}\mathcal{\hat{V}}%
_{K3_{\Vert }}\eta _{i_{\bot }j_{\bot }}$%
\end{tabular}%
\end{equation}%
leading in turns to%
\begin{equation}
\begin{tabular}{lll}
$\mathcal{G}_{i_{\bot }j_{\bot }}$ & $=$ & $\lambda ^{2}\mathcal{\hat{V}}%
_{K3_{\Vert }}\eta _{i_{\bot }j_{\bot }}$ \\
$\mathcal{G}_{a_{\Vert }b_{\Vert }}$ & $=$ & $\frac{\mathcal{\hat{V}}%
_{K3_{\perp }}}{\lambda ^{2}}\eta _{a_{\Vert }b_{\Vert }}$%
\end{tabular}%
,\qquad
\begin{tabular}{lll}
$\mathcal{G}^{i_{\bot }j_{\bot }}$ & $\simeq $ & $\frac{1}{\lambda ^{2}%
\mathcal{\hat{V}}_{K3_{\Vert }}}\eta ^{i_{\bot }j_{\bot }}$ \\
$\mathcal{G}^{a_{\Vert }b_{\Vert }}$ & $\simeq $ & $\frac{\lambda ^{2}}{%
\mathcal{\hat{V}}_{K3_{\perp }}}\eta ^{a_{\Vert }b_{\Vert }}$%
\end{tabular}
\label{TM2}
\end{equation}%
manifestly related by the Weak/Strong gauge duality. We also have $\mathcal{%
\hat{V}}_{K3_{\perp }}\mathcal{\hat{V}}_{K3_{\Vert }}=1$ or equivalently $%
\mathcal{\hat{V}}_{K3_{\perp }}^{\prime }\mathcal{\hat{V}}_{K3_{\Vert
}}^{\prime }=1$ with $\mathcal{\hat{V}}_{K3_{\perp }}^{\prime }=\lambda ^{-2}%
\mathcal{\hat{V}}_{K3_{\perp }}$ and $\mathcal{\hat{V}}_{K3_{\Vert
}}^{\prime }=\lambda ^{2}\mathcal{\hat{V}}_{K3_{\Vert }}.$
\end{itemize}

\paragraph{C. \ Calculating LHS of (\protect\ref{bcfg}):\newline
}

We start from the expression of the left hand side of the Repulsive Coulomb
Conjecture namely $LHS=\frac{1}{4}\mathrm{g}_{{\small 3D}}^{2}M_{\mathrm{Pl}%
}\left( \mathrm{k}^{2}q_{\text{\textsc{a}}}\mathcal{G}^{\text{\textsc{ab}}%
}q_{\text{\textsc{b}}}\right) ;$ and use the fibration $K3_{\bot }\times
K3_{\Vert }$ of the CY4 to cast it like%
\begin{equation}
LHS=LHS_{\bot }+LHS_{\Vert }
\end{equation}%
with%
\begin{eqnarray}
LHS_{\bot } &=&\frac{\mathrm{g}_{{\small 3D}}^{2}M_{\mathrm{Pl}}}{4}\mathrm{k%
}_{\bot }^{2}\left( q_{i_{\bot }}\mathcal{G}^{i_{\bot }j_{\bot }}q_{j_{\bot
}}\right)  \label{L1} \\
LHS_{\Vert } &=&\frac{\mathrm{g}_{{\small 3D}}^{2}M_{\mathrm{Pl}}}{4}\mathrm{%
k}_{\Vert }^{2}\left( q_{a_{\Vert }}\mathcal{G}^{a_{\Vert }b_{\Vert
}}q_{b_{\Vert }}\right)  \label{L2}
\end{eqnarray}%
Then, we use the expression of the inverse metric $\mathcal{G}^{\text{%
\textsc{ab}}}$, decomposing into diagonal blocs $\mathcal{G}^{i_{\bot
}j_{\bot }}\oplus \mathcal{G}^{a_{\Vert }b_{\Vert }},$ given by
\begin{equation}
\mathcal{G}^{i_{\bot }j_{\bot }}=\mathcal{\hat{V}}_{K3_{\bot }}\eta
^{i_{\bot }j_{\bot }}\qquad ,\qquad \mathcal{G}^{a_{\Vert }b_{\Vert }}=%
\mathcal{\hat{V}}_{K3_{\Vert }}\eta ^{a_{\Vert }b_{\Vert }}  \label{GGK}
\end{equation}%
to put the $LHS_{\bot }$ and the $LHS_{\Vert }$ as follows%
\begin{equation}
\fbox{%
\begin{tabular}{lll}
$\left.
\begin{array}{c}
\text{ \ } \\
\text{ \ }%
\end{array}%
\right. $ & \ \ \ $LHS_{\bot }=\frac{1}{4}\left( \mathrm{g}_{{\small 3D}%
}^{2}M_{\mathrm{Pl}}\mathcal{\hat{V}}_{K3_{\bot }}\right) \mathrm{k}_{\bot
}^{2}\mathbf{q}_{\bot }^{2}$ \ \ \  &  \\
$\left.
\begin{array}{c}
\text{ \ } \\
\text{ \ }%
\end{array}%
\right. $ & $\ \ \ LHS_{\Vert }=\frac{1}{4}\left( \mathrm{g}_{{\small 3D}%
}^{2}M_{\mathrm{Pl}}\mathcal{\hat{V}}_{K3_{\Vert }}\right) \mathrm{k}_{\Vert
}^{2}\mathbf{q}_{\Vert }^{2}$ \ \  &
\end{tabular}%
}  \label{Q}
\end{equation}%
where we have used
\begin{equation}
\mathbf{q}_{\bot }^{2}=q_{i_{\bot }}\eta ^{i_{\bot }j_{\bot }}q_{j_{\bot
}}\qquad ,\qquad \mathbf{q}_{\Vert }^{2}=q_{a_{\Vert }}\eta ^{a_{\Vert
}b_{\Vert }}q_{b_{\Vert }}
\end{equation}

\paragraph{D. \ Calculating RHS of (\protect\ref{bcfg}):\newline
}

Here, we compute the two mass terms in the right hand side of the repulsive
force conjecture (\ref{bcfg}) separately:

\begin{description}
\item[$\mathbf{i)}$] \emph{the term} $\varrho \left[ \hat{\upsilon}^{\text{%
\textsc{a}}}\partial ^{\text{\textsc{a}}}M_{\mathrm{k}_{x}}\right] ^{2}$%
\newline
First, notice that the term $(\hat{\upsilon}^{\text{\textsc{a}}}\partial M_{%
\mathrm{k}_{x}})/\partial \hat{\upsilon}^{\text{\textsc{a}}}$ splits as the
sum of the fibral term $\hat{\upsilon}^{i_{\bot }}\partial M_{\mathrm{k}%
_{\bot }}/\partial \hat{\upsilon}^{i_{\bot }}$ and the base contribution $%
\hat{\upsilon}^{a_{\Vert }}\partial M_{\mathrm{k}_{\Vert }}/\partial \hat{%
\upsilon}^{a_{\Vert }}$. Then, we have%
\begin{eqnarray}
\left[ \hat{\upsilon}^{\text{\textsc{a}}}\frac{\partial M_{\mathrm{k}_{x}}}{%
\partial \hat{\upsilon}^{\text{\textsc{a}}}}\right] ^{2} &=&\left[ \hat{%
\upsilon}^{i_{\bot }}\frac{\partial M_{\mathrm{k}_{\bot }}}{\partial \hat{%
\upsilon}^{i_{\bot }}}\right] ^{2}+\left[ \hat{\upsilon}^{a_{\Vert }}\frac{%
\partial M_{\mathrm{k}_{\Vert }}}{\partial \hat{\upsilon}^{a_{\Vert }}}%
\right] ^{2}+  \notag \\
&&2\left( \hat{\upsilon}^{i_{\bot }}\frac{\partial M_{\mathrm{k}_{\bot }}}{%
\partial \hat{\upsilon}^{i_{\bot }}}\right) \left( \hat{\upsilon}^{a_{\Vert
}}\frac{\partial M_{\mathrm{k}_{\Vert }}}{\partial \hat{\upsilon}^{a_{\Vert
}}}\right)
\end{eqnarray}%
Using the homogenous property of the masses $M_{\mathrm{k}_{x}}$ of the
tower of BPS particle states in terms of the volumes $\hat{\upsilon}^{\text{%
\textsc{a}}}$ namely%
\begin{eqnarray}
\hat{\upsilon}^{i_{\bot }}\frac{\partial M_{\mathrm{k}_{\bot }}}{\partial
\hat{\upsilon}^{i_{\bot }}} &=&M_{\mathrm{k}_{\bot }}\qquad ,\qquad \left(
\hat{\upsilon}^{i_{\bot }}\frac{\partial M_{\mathrm{k}_{\bot }}}{\partial
\hat{\upsilon}^{i_{\bot }}}\right) ^{2}=M_{\mathrm{k}_{\bot }}^{2} \\
\hat{\upsilon}^{a_{\Vert }}\frac{\partial M_{\mathrm{k}_{\Vert }}}{\partial
\hat{\upsilon}^{a_{\Vert }}} &=&M_{\mathrm{k}_{\Vert }}\qquad ,\qquad \left(
\hat{\upsilon}^{a_{\Vert }}\frac{\partial M_{\mathrm{k}_{\Vert }}}{\partial
\hat{\upsilon}^{a_{\Vert }}}\right) ^{2}=M_{\mathrm{k}_{\Vert }}^{2}
\end{eqnarray}%
we find that $\varrho \left[ \hat{\upsilon}^{\text{\textsc{a}}}\partial M_{%
\mathrm{k}_{x}}/\partial \hat{\upsilon}^{\text{\textsc{a}}}\right] ^{2}$ ($%
\varrho =1/2$) is given by%
\begin{equation}
\varrho \left[ \hat{\upsilon}^{\text{\textsc{a}}}\frac{\partial M_{\mathrm{n}%
_{x}}}{\partial \hat{\upsilon}^{\text{\textsc{a}}}}\right] ^{2}=\varrho M_{%
\mathrm{k}_{\bot }}^{2}+\varrho M_{\mathrm{k}_{_{\Vert }}}^{2}+2\varrho M_{%
\mathrm{k}_{\bot }}M_{\mathrm{k}_{_{\Vert }}}
\end{equation}

\item[$\mathbf{ii)}$] \emph{the term} $\frac{1}{2}\mathcal{G}^{\text{\textsc{%
ab}}}\left( \partial _{\text{\textsc{a}}}M_{\mathrm{k}_{x}}\partial _{\text{%
\textsc{b}}}M_{\mathrm{k}_{x}}\right) $\newline
Because of the diagonal decomposition $\mathcal{G}^{\text{\textsc{ab}}}=%
\mathcal{G}^{i_{\bot }j_{\bot }}\oplus \mathcal{G}^{a_{\Vert }b_{\Vert }},$
the coupling mass gradients $\mathcal{G}^{\text{\textsc{ab}}}\left( \partial
_{\text{\textsc{a}}}M_{\mathrm{k}_{x}}\partial _{\text{\textsc{b}}}M_{%
\mathrm{k}_{x}}\right) $ decompose like%
\begin{eqnarray}
\frac{1}{2}\mathcal{G}^{\text{\textsc{ab}}}\left( \frac{\partial M_{\mathrm{k%
}_{x}}}{\partial \hat{\upsilon}^{\text{\textsc{a}}}}\right) \left( \frac{%
\partial M_{\mathrm{k}_{x}}}{\partial \hat{\upsilon}^{\text{\textsc{b}}}}%
\right) &=&\frac{1}{2}\mathcal{G}^{i_{\bot }j_{\bot }}\left( \frac{\partial
M_{\mathrm{k}_{x}}}{\partial \hat{\upsilon}^{i_{\bot }}}\right) \left( \frac{%
\partial M_{\mathrm{k}_{x}}}{\partial \hat{\upsilon}^{j_{\bot }}}\right) +
\notag \\
&&\frac{1}{2}\mathcal{G}^{a_{\Vert }b_{\Vert }}\left( \frac{\partial M_{%
\mathrm{k}_{x}}}{\partial \hat{\upsilon}^{a_{\Vert }}}\right) \left( \frac{%
\partial M_{\mathrm{k}_{x}}}{\partial \hat{\upsilon}^{b_{\Vert }}}\right)
\end{eqnarray}%
Using the expression $\partial _{\text{\textsc{a}}}M_{\mathrm{k}_{x}}=%
\mathrm{\xi }M_{11d}\left( \mathrm{k}_{x}q_{\text{\textsc{a}}}\right) ,$ and
$\mathcal{G}^{i_{\bot }j_{\bot }}=\mathcal{\hat{V}}_{K3_{\bot }}\eta
^{i_{\bot }j_{\bot }}$ as well as $\mathcal{G}^{a_{\Vert }b_{\Vert }}=%
\mathcal{\hat{V}}_{K3_{\Vert }}\eta ^{a_{\Vert }b_{\Vert }},$ we get%
\begin{eqnarray}
\frac{1}{2}\mathcal{G}^{i_{\bot }j_{\bot }}\left( \frac{\partial M_{\mathrm{k%
}_{x}}}{\partial \hat{\upsilon}^{i_{\bot }}}\right) \left( \frac{\partial M_{%
\mathrm{k}_{x}}}{\partial \hat{\upsilon}^{j_{\bot }}}\right) &=&\frac{%
\mathrm{\xi }^{2}M_{\mathrm{Pl}}^{2}}{2\mathrm{\gamma }^{2}\mathcal{V}_{%
{\small CY}_{{\small 4}}}^{2}}\mathrm{k}_{\bot }^{2}\mathcal{\hat{V}}%
_{K3_{\bot }}\mathbf{q}_{\bot }^{2}  \notag \\
&=&\frac{\mathrm{\xi }^{2}}{2}\left( \mathrm{g}_{{\small 3D}}^{2}M_{\mathrm{%
Pl}}\mathcal{\hat{V}}_{K3_{\bot }}\right) \mathrm{k}_{\bot }^{2}\mathbf{q}%
_{\bot }^{2}
\end{eqnarray}%
and%
\begin{eqnarray}
\frac{1}{2}\mathcal{G}^{a_{\Vert }b_{\Vert }}\left( \frac{\partial M_{%
\mathrm{k}_{x}}}{\partial \hat{\upsilon}^{a_{\Vert }}}\right) \left( \frac{%
\partial M_{\mathrm{k}_{x}}}{\partial \hat{\upsilon}^{b_{\Vert }}}\right) &=&%
\frac{\mathrm{\xi }^{2}}{2}\frac{M_{\mathrm{Pl}}^{2}}{\mathrm{\gamma }^{2}%
\mathcal{V}_{{\small CY}_{{\small 4}}}^{2}}\mathrm{k}_{\Vert }^{2}\mathcal{%
\hat{V}}_{K3_{\Vert }}\mathbf{q}_{\Vert }^{2}  \notag \\
&=&\frac{\mathrm{\xi }^{2}}{2}\left( \mathrm{g}_{{\small 3D}}^{2}M_{\mathrm{%
Pl}}\mathcal{\hat{V}}_{K3_{\Vert }}\right) \mathrm{k}_{\Vert }^{2}\mathbf{q}%
_{\Vert }^{2}
\end{eqnarray}%
where we have used $M_{11d}=(M_{\mathrm{Pl}}/\mathrm{\gamma }\mathcal{V}_{%
{\small CY}_{{\small 4}}})$ and $\mathrm{g}_{{\small 3D}}^{2}=(M_{\mathrm{Pl}%
}/\mathrm{\gamma }^{2}\mathcal{V}_{{\small CY}_{{\small 4}}}^{2}).$

\item[$\mathbf{iii)}$] \emph{Value of the right hand side}\newline
The contribution $RHS_{\bot }$ in the fiber K3$_{\bot }$ and its homologue $%
RHS_{\Vert }$ in the base K3$_{\Vert }$ are given by%
\begin{equation}
\fbox{%
\begin{tabular}{ll}
$\left.
\begin{array}{c}
\text{ \ } \\
\text{ \ }%
\end{array}%
\right. $ & $\ $ $\ \ \ RHS_{\bot }=\frac{\mathrm{\xi }^{2}}{2}\left(
\mathrm{g}_{{\small 3D}}^{2}M_{\mathrm{Pl}}\mathcal{\hat{V}}_{K3_{\bot
}}\right) \mathrm{k}_{\bot }^{2}\mathbf{q}_{\bot }^{2}-\varrho M_{\mathrm{k}%
_{\bot }}^{2}$ \ \  \\
$\left.
\begin{array}{c}
\text{ \ } \\
\text{ \ }%
\end{array}%
\right. $ & $\ \ \ RHS_{\Vert }=\frac{\mathrm{\xi }^{2}}{2}\left( \mathrm{g}%
_{{\small 3D}}^{2}M_{\mathrm{Pl}}\mathcal{\hat{V}}_{K3_{\Vert }}\right)
\mathrm{k}_{\Vert }^{2}\mathbf{q}_{\Vert }^{2}-\varrho M_{\mathrm{k}%
_{_{\Vert }}}^{2}$ \ \ \
\end{tabular}%
}  \label{QR}
\end{equation}%
Recall that we have $\xi =1/\sqrt{2}$ and $\varrho =1/2.$ Comparing with eqs(%
\ref{Q}) namely $LHS_{\bot }=\frac{1}{4}\left( \mathrm{g}_{{\small 3D}%
}^{2}M_{\mathrm{Pl}}\mathcal{\hat{V}}_{K3_{\bot }}\right) \mathrm{k}_{\bot
}^{2}\mathbf{q}_{\bot }^{2}$ and $LHS_{\Vert }=\frac{1}{4}\left( \mathrm{g}_{%
{\small 3D}}^{2}M_{\mathrm{Pl}}\mathcal{\hat{V}}_{K3_{\Vert }}\right)
\mathrm{k}_{\Vert }^{2}\mathbf{q}_{\Vert }^{2},$ we end up with the desired
test of the RFC inequalities%
\begin{equation}
\fbox{%
\begin{tabular}{ll}
$\left.
\begin{array}{c}
\text{ \ } \\
\text{ \ }%
\end{array}%
\right. $ & \ \ \ $\frac{1}{4}\left( \mathrm{g}_{{\small 3D}}^{2}M_{\mathrm{%
Pl}}\mathcal{\hat{V}}_{K3_{\bot }}\right) $\textrm{k}$_{\bot }^{2}\mathbf{q}%
_{\bot }^{2}\geq \frac{1}{4}\left( \mathrm{g}_{{\small 3D}}^{2}M_{\mathrm{Pl}%
}\mathcal{\hat{V}}_{K3_{\bot }}\right) $\textrm{k}$_{\bot }^{2}\mathbf{q}%
_{\bot }^{2}-\frac{1}{2}M_{\mathrm{k}_{\bot }}^{2}$ \ \  \\
$\left.
\begin{array}{c}
\text{ \ } \\
\text{ \ }%
\end{array}%
\right. $ & \ \ \ $\frac{1}{4}\left( \mathrm{g}_{{\small 3D}}^{2}M_{\mathrm{%
Pl}}\mathcal{\hat{V}}_{K3_{\Vert }}\right) $\textrm{k}$_{\Vert }^{2}\mathbf{q%
}_{\Vert }^{2}\geq \frac{1}{4}\left( \mathrm{g}_{{\small 3D}}^{2}M_{\mathrm{%
Pl}}\mathcal{\hat{V}}_{K3_{\Vert }}\right) $\textrm{k}$_{\Vert }^{2}\mathbf{q%
}_{\Vert }^{2}-\frac{1}{2}M_{\mathrm{k}_{\Vert }}^{2}$ \ \
\end{tabular}%
}
\end{equation}%
The two inequalities are related to each other by Weak/strong duality
exchanging the base $K3_{\Vert }$ and the fiber $K3_{\bot }$ of the CY4.
\end{description}

\subsubsection{Case of towers of non-BPS states}

The tower $\mathcal{T}_{\mathrm{k}_{x},\mathbf{q}_{x}}^{\text{\textsc{n-bps}}%
}$ of non-BPS states can be labeled like $|n_{L},Q_{i_{x}}>_{{\small het_{x}}%
}$ with $Q_{i_{x}}=\mathrm{k}_{x}q_{i_{x}}$ and $\mathrm{k}_{x}$ and $%
q_{i_{x}}$ integers ($x=\bot ,\Vert $). It has $\left( \mathbf{i}\right) $
an interpretation in terms of left moving mode excitations $\mathrm{n}_{L}$
of a heterotic string \textrm{\cite{12}} (here we have \emph{het}$_{x}$ with
$x=\bot ,\Vert $); and $\left( \mathbf{ii}\right) $ carrying charges under $%
U\left( 1\right) _{\mathbf{Q}_{x}}=\sum Q_{i_{x}}U\left( 1\right) ^{i_{x}}$
with the relationship%
\begin{equation}
\mathrm{n}_{\mathbf{Q}_{x}}^{L}=-\frac{1}{2}\mathbf{Q}_{x}^{2}>0\qquad
,\qquad \mathbf{q}_{x}=(q_{1},...,q_{h_{{\small K3}_{x}}^{1,1}})
\end{equation}%
where we have set%
\begin{equation}
\mathbf{Q}_{x}=\mathrm{k}_{x}\mathbf{q}_{x}\qquad ,\qquad \mathrm{n}_{%
\mathbf{Q}_{x}}^{L}\equiv \mathrm{n}_{x}^{L}\equiv \mathrm{n}_{x}
\end{equation}%
Recall that this heterotic string is dual to the MSW string descending M5
brane wrapping a $K3$ surface $M5/K3$ \textrm{with contribution from 2-cycle
}$\Upsilon _{\mathbf{q}_{x}}=\sum q_{i_{x}}\mathcal{C}^{i_{x}}$. But here
the $K3$ can play either the role of the fibral $K3_{\bot }$ or the base
surface $K3_{\Vert }$. The masses $M_{\mathrm{n}_{x}}$ of the excitation
modes in these towers has two contributions are given by \textrm{\cite{12}}%
\begin{equation}
M_{\mathrm{n}_{_{x}}}^{2}=8\pi (\mathrm{n}_{_{x}}-a)T_{\mathrm{s}%
_{x}}+\Delta _{{\small CB}_{x}}  \label{2M}
\end{equation}%
where $a=1$ is the zero point energy of the heterotic string \textrm{\cite%
{12, 29}}. For convenience, we disregard below the quantum shift a (assuming
$\mathrm{n}_{x}>>a$); and restrict the investigation to $\mathrm{k}_{x}=1$
implying $\mathbf{Q}_{x}=\mathbf{q}_{x}$ and then $\mathrm{n}_{x}=-\frac{1}{2%
}\mathbf{q}_{x}^{2}$. Notice that the integer $\mathrm{k}_{x}$ in (\ref{2M})
refers either to $\mathrm{k}_{\bot }$ or $\mathrm{k}_{\Vert };$ and
similarly for the other quantities like $\mathrm{n}_{x},$ $T_{\mathrm{s}%
_{x}} $\ and $\Delta _{{\small CB}_{x}}.$ This is to say that for the $%
K3_{\bot }$ fiber, we have a fibral tower $\mathcal{T}_{\mathrm{n}_{\bot }}^{%
\text{\textsc{n-bps}}}$ of particles with masse $M_{\mathrm{n}_{\bot }}^{2};$
and for the base surface $K3_{\Vert }$ we have a base tower $\mathcal{T}_{%
\mathrm{n}_{\parallel }}^{\text{\textsc{n-bps}}}$ of particles with $M_{%
\mathrm{n}_{\Vert }}^{2}.$ In sum, depending on whether M5 wraps $K3_{\bot
} $ or $K3_{\Vert },$ we distinguish two kinds of towers of states with
squared masses as follows
\begin{equation}
\begin{tabular}{lll}
$M_{\mathrm{n}_{\bot }}^{2}$ & $=$ & $8\pi \mathrm{n}_{_{\bot }}T_{\mathrm{s}%
_{\bot }}+\Delta _{{\small CB}_{\bot }}$ \\
$M_{\mathrm{n}_{\Vert }}^{2}$ & $=$ & $8\pi \mathrm{n}_{\Vert }T_{\mathrm{s}%
\Vert }+\Delta _{{\small CB}_{\Vert }}$%
\end{tabular}
\label{MM}
\end{equation}%
and values at infinite distance limit like%
\begin{equation}
\lim_{\lambda \rightarrow \infty }\left( M_{\mathrm{n}_{\bot }}^{2}\right)
=0\qquad ,\qquad \lim_{\lambda \rightarrow \infty }\left( M_{\mathrm{n}%
_{\Vert }}^{2}\right) =\infty
\end{equation}%
These towers of non BPS states fill directions $\Upsilon _{\mathbf{q}_{x}}$
in the antiself dual $\mathfrak{L}_{-}^{\left( x\right) };$ the latter is a
charge sublattice contained in the even integral lattice $\mathfrak{L}%
^{\left( x\right) }$ of K3$_{x}$. Recall that for the Calabi-Yau fourfold $%
K3_{\bot }\times K3_{\Vert }$, the charge lattice is given by $\mathfrak{L}%
^{\left( \bot \right) }\oplus \mathfrak{L}^{\left( \Vert \right) }$ with $%
\mathfrak{L}^{\left( x\right) }$ splitting into self dual and antiself dual
parts as $\mathfrak{L}^{\left( x\right) }=\mathfrak{L}_{+}^{\left( x\right)
}\oplus \mathfrak{L}_{-}^{\left( x\right) }$ with label $x=\bot ,\Vert .$
Recall also that for non BPS states, the basic curve $\Upsilon _{\mathbf{q}%
_{x}}$ has negative self intersection ($\Upsilon _{\mathbf{q}_{x}}^{2}=%
\mathbf{q}_{x}^{2}<0$); as such these non BPS states are labeled by the
positive integer $\mathrm{n}_{x}$ giving the left moving excitation mode of
the heterotic string. Moreover, the $\Delta _{CB_{x}}$ in (\ref{2M}) is%
\textrm{\ }%
\begin{equation}
\Delta _{CB_{x}}=\digamma \left( \sum_{i_{x}=1}^{h_{x}^{1,1}}q_{i_{x}}\hat{%
\upsilon}^{i_{x}}\right) ^{2}\qquad ,\qquad \digamma =4\pi ^{2}\mathrm{M}%
_{11d}^{2}  \label{D}
\end{equation}%
\textrm{\ }This $\Delta _{CB_{x}}$ is the contribution from the Coulomb
branch just as in the \emph{EFT}$_{{\small 5D}}$ \textrm{\cite{12}}; it is
quadratic in the volumes $\hat{\upsilon}^{i_{x}}$ of the 2-cycles generating%
\textrm{\ }$\Upsilon _{\mathbf{q}}=\sum q_{i_{x}}\mathcal{C}^{i_{x}}.$
Furthermore, the tension $T_{\mathrm{s}_{\bot }}$ (resp. $T_{\mathrm{s}%
_{\Vert }}$) of the string is given in terms of the volume $\mathcal{\hat{V}}%
_{K3_{\bot }}$ (resp. $\mathcal{\hat{V}}_{K3_{\parallel }}$) as follows%
\begin{equation}
T_{\mathrm{s}_{x}}=2\pi \mathcal{\hat{V}}_{K3_{x}}\mathrm{M}_{11d}^{2}\qquad
,\text{\qquad }\mathrm{M}_{11d}=\frac{M_{\mathrm{Pl}}}{\mathrm{\gamma }%
\mathcal{V}_{{\small CY}_{{\small 4}}}}  \label{T}
\end{equation}%
with $\mathcal{\hat{V}}_{{\small K3}}$ given by $\mathcal{\hat{V}}_{{\small %
K3}_{{\small \bot }}}=(\hat{\upsilon}^{i_{\perp }}\eta _{i_{\perp }j_{\perp
}}\hat{\upsilon}^{j_{\perp }})/2$\ for the fiber with scaling in the
infinite distance limit as $\lambda ^{-2}\mathcal{\hat{V}}_{{\small K3}_{%
{\small \bot }}}$; and $\mathcal{\hat{V}}_{{\small K3}_{\Vert }}=(\hat{%
\upsilon}^{a_{\Vert }}\eta _{a_{\Vert }b_{\Vert }}\hat{\upsilon}^{b_{\Vert
}})/2$\ for the base with scaling $\lambda ^{+2}\mathcal{\hat{V}}_{{\small K3%
}_{\Vert }}$. Substituting these values, we have%
\begin{equation}
\begin{tabular}{lll}
$M_{\mathrm{n}_{x}}^{2}$ & $=$ & $\varpi _{\mathrm{n}_{_{x}}}\left[ \mathcal{%
\hat{V}}_{K3_{x}}+\frac{1}{4\mathrm{n}_{_{x}}}\left( q_{i_{x}}\hat{\upsilon}%
^{i_{x}}\right) ^{2}\right] $ \\
$\mathfrak{M}_{\mathrm{n}_{x}}^{2}$ & $=$ & $\nu _{\mathrm{n}_{_{x}}}\left[
\mathcal{\hat{V}}_{K3_{x}}+\frac{1}{4\mathrm{n}_{_{x}}}\left( q_{i_{x}}\hat{%
\upsilon}^{i_{x}}\right) ^{2}\right] $%
\end{tabular}
\label{MX}
\end{equation}%
where we have set $\varpi _{\mathrm{n}_{_{x}}}=16\pi ^{2}\mathrm{M}_{11d}^{2}%
\mathrm{n}_{_{x}};$ and for later use we have introduced the reduced squared
mass $\mathfrak{M}_{\mathrm{n}_{x}}^{2}=M_{\mathrm{n}_{x}}^{2}/M_{\mathrm{Pl}%
}^{2}$ with $\varpi _{\mathrm{n}_{_{x}}}=\varpi _{\mathrm{n}_{_{x}}}/M_{%
\mathrm{Pl}}^{2}.$ By using $M_{\mathrm{Pl}}^{2}=16\pi ^{2}\mathcal{V}_{%
{\small CY}_{{\small 4}}}^{2}\mathrm{M}_{11d}^{2},$ the $\nu _{\mathrm{n}%
_{_{x}}}$ reads also as follows%
\begin{equation}
\nu _{\mathrm{n}_{_{x}}}=\frac{16\pi ^{2}\mathrm{M}_{11d}^{2}}{M_{\mathrm{Pl}%
}^{2}}\mathrm{n}_{_{x}}=\frac{1}{\mathcal{V}_{{\small X}_{{\small 4}}}^{2}}%
\mathrm{n}_{_{x}}
\end{equation}%
For later use, notice also that $M_{\mathrm{n}_{x}}^{2}$ and $\mathfrak{M}_{%
\mathrm{n}_{x}}^{2}$ are homogeneous functions in the 2-cycle volumes $\hat{%
\upsilon}^{i_{x}}$; as such they are eigenstates of the operator $L_{0}=\hat{%
\upsilon}^{i_{x}}(\partial /\partial \hat{\upsilon}^{i_{x}})$ in the sense%
\begin{equation}
L_{0}M_{\mathrm{n}_{x}}^{2}=2M_{\mathrm{n}_{x}}^{2}\qquad ,\qquad L_{0}%
\mathfrak{M}_{\mathrm{n}_{x}}^{2}=2\mathfrak{M}_{\mathrm{n}_{x}}^{2}
\end{equation}%
Notice moreover that being quadratic in the moduli $\hat{\upsilon}^{i_{x}},$
the $\mathfrak{M}_{\mathrm{n}_{x}}^{2}$ can be also nicely presented like%
\begin{equation}
\mathfrak{M}_{\mathrm{n}_{x}}^{2}=\frac{1}{2}\hat{\upsilon}^{i_{x}}\mathcal{S%
}_{i_{x}j_{x}}\hat{\upsilon}^{j_{x}}
\end{equation}%
with symmetric matrix $\mathcal{S}_{i_{x}j_{x}}$ given by%
\begin{equation}
\mathcal{S}_{i_{x}j_{x}}=\nu _{\mathrm{n}_{_{x}}}\left( \eta _{i_{x}j_{x}}+%
\frac{1}{2\mathrm{n}_{_{x}}}q_{i_{x}}q_{j_{x}}\right)
\end{equation}%
From this relation, we compute several interesting quantities; in particular%
\begin{equation}
\frac{\partial \mathfrak{M}_{\mathrm{n}_{x}}^{2}}{\partial \hat{\upsilon}%
^{i_{x}}}=\mathcal{S}_{i_{x}k_{x}}\hat{\upsilon}^{k_{x}},\qquad \frac{%
\partial \mathfrak{M}_{\mathrm{n}_{x}}^{2}}{\partial \hat{\upsilon}^{j_{x}}}=%
\mathcal{S}_{j_{x}l_{x}}\hat{\upsilon}^{l_{x}}
\end{equation}%
and%
\begin{equation}
\eta ^{k_{x}l_{x}}\left( \frac{\partial \mathfrak{M}_{\mathrm{n}_{x}}^{2}}{%
\partial \hat{\upsilon}^{i_{x}}}\right) \left( \frac{\partial \mathfrak{M}_{%
\mathrm{n}_{x}}^{2}}{\partial \hat{\upsilon}^{j_{x}}}\right) =\hat{\upsilon}%
^{k_{x}}\Delta _{i_{x}j_{x}}\hat{\upsilon}^{l_{x}}
\end{equation}%
where we have set $\Delta _{i_{x}j_{x}}=\mathcal{S}_{i_{x}k_{x}}\eta
^{k_{x}l_{x}}\mathcal{S}_{l_{x}j_{x}}$ reading explicitly as follows%
\begin{equation}
\Delta _{i_{x}j_{x}}=\nu _{\mathrm{n}_{_{x}}}^{2}\left[ \eta _{i_{x}j_{x}}+%
\frac{1}{\mathrm{n}_{_{x}}}\left( 1+\frac{\mathbf{q}_{x}^{2}}{4\mathrm{n}%
_{_{x}}}\right) q_{i_{x}}q_{j_{x}}\right]
\end{equation}%
By using the relation $\mathbf{q}_{x}^{2}=-2\mathrm{n}_{_{x}},$ it
simplifies like%
\begin{equation}
\Delta _{i_{x}j_{x}}=\nu _{\mathrm{n}_{_{x}}}^{2}\left( \eta _{i_{x}j_{x}}+%
\frac{1}{2\mathrm{n}_{_{x}}}q_{i_{x}}q_{j_{x}}\right)
\end{equation}%
By comparing this relation with the matrix $\mathcal{S}_{i_{x}j_{x}}=\nu _{%
\mathrm{n}_{_{x}}}\left( \eta _{i_{x}j_{x}}+\frac{1}{2\mathrm{n}_{_{x}}}%
q_{i_{x}}q_{j_{x}}\right) ,$ we learn that we just have
\begin{equation}
\Delta _{i_{x}j_{x}}=\nu _{\mathrm{n}_{_{x}}}\mathcal{S}_{i_{x}j_{x}}
\end{equation}%
So, we have%
\begin{equation}
\eta ^{k_{x}l_{x}}\left( \frac{\partial \mathfrak{M}_{\mathrm{n}_{x}}^{2}}{%
\partial \hat{\upsilon}^{i_{x}}}\right) \left( \frac{\partial \mathfrak{M}_{%
\mathrm{n}_{x}}^{2}}{\partial \hat{\upsilon}^{j_{x}}}\right) =\nu _{\mathrm{n%
}_{_{x}}}\hat{\upsilon}^{k_{x}}\mathcal{S}_{i_{x}j_{x}}\hat{\upsilon}^{l_{x}}
\end{equation}%
and then%
\begin{equation}
\eta ^{k_{x}l_{x}}\left( \frac{\partial \mathfrak{M}_{\mathrm{n}_{x}}^{2}}{%
\partial \hat{\upsilon}^{i_{x}}}\right) \left( \frac{\partial \mathfrak{M}_{%
\mathrm{n}_{x}}^{2}}{\partial \hat{\upsilon}^{j_{x}}}\right) =2\nu _{\mathrm{%
n}_{_{x}}}\mathfrak{M}_{\mathrm{n}_{x}}^{2}  \label{MN}
\end{equation}

\paragraph{A. Repulsive Force Conjecture (RFC):\newline
}

Given a massive particle state in the \emph{EFT}$_{{\small 3D}}$ of mass $M_{%
\mathrm{n}_{x}}$, the RFC which in the infinite distance limit coincides
with the asymptotic WGC, requires the following inequality $F_{Coulomb}\geq
F_{Grav}+F_{Yukawa}.$ This inequality reads explicitly as follows%
\begin{equation}
\frac{\mathrm{g}_{{\small 3D}}^{2}}{M_{\mathrm{Pl}}}\left( q_{\text{\textsc{a%
}}}\mathcal{G}^{\text{\textsc{ab}}}q_{\text{\textsc{b}}}\right) \geq \left.
\frac{D-3}{D-2}\right\vert _{D=3}\frac{M_{\mathrm{k}}^{2}}{M_{\mathrm{Pl}%
}^{2}}+\mathrm{\vartheta }_{{\small QG}}+\mathcal{Y}  \label{ine}
\end{equation}%
with Yukawa matter coupling contribution $\mathcal{Y}$ reading in terms of
the reduced mass $\mathfrak{M}_{\mathrm{n}_{x}}$ as follows%
\begin{equation}
\mathcal{Y}=\frac{\mathcal{G}^{\text{\textsc{ab}}}}{2\mathfrak{M}_{\mathrm{n}%
_{x}}^{2}}\left( \frac{\partial \mathfrak{M}_{\mathrm{n}_{x}}^{2}}{\partial
\hat{\upsilon}^{\text{\textsc{a}}}}\right) \left( \frac{\partial \mathfrak{M}%
_{\mathrm{n}_{x}}^{2}}{\partial \hat{\upsilon}^{\text{\textsc{b}}}}\right) -%
\frac{\varrho }{\mathfrak{M}_{\mathrm{n}_{x}}^{2}}\left[ \hat{\upsilon}^{%
\text{\textsc{a}}}\frac{\partial \mathfrak{M}_{\mathrm{n}_{x}}^{2}}{\partial
\hat{\upsilon}^{\text{\textsc{a}}}}\right] ^{2}
\end{equation}%
with $\varrho =1/2$. Notice that as for the towers of BPS states and because
here $D=3,$ the classical gravity contribution $\left( D-3\right) /\left(
D-2\right) $ to the RFC vanishes identically; however one expects a non
trivial \emph{negative} contribution $\mathrm{\vartheta }_{{\small QG}}$ of
the vacuum coming from quantum effect. \textrm{\newline
}Substituting into (\ref{ine}), the RFC inequality for non BPS states becomes%
\begin{equation}
\frac{\mathrm{g}_{{\small 3D}}^{2}}{M_{\mathrm{Pl}}}\left( q_{\text{\textsc{a%
}}}\mathcal{G}^{\text{\textsc{ab}}}q_{\text{\textsc{b}}}\right) \geq \frac{%
\mathcal{G}^{\text{\textsc{ab}}}}{2\mathfrak{M}_{\mathrm{n}_{x}}^{2}}\left(
\frac{\partial \mathfrak{M}_{\mathrm{n}_{x}}^{2}}{\partial \hat{\upsilon}^{%
\text{\textsc{a}}}}\right) \left( \frac{\partial \mathfrak{M}_{\mathrm{n}%
_{x}}^{2}}{\partial \hat{\upsilon}^{\text{\textsc{b}}}}\right) -\frac{%
\varrho }{\mathfrak{M}_{\mathrm{n}_{x}}^{2}}4\mathfrak{M}_{\mathrm{n}%
_{x}}^{4}+\mathrm{\vartheta }_{{\small QG}}^{\prime }
\end{equation}%
where we have used $L_{0}\mathfrak{M}_{\mathrm{n}_{x}}^{2}=2\mathfrak{M}_{%
\mathrm{n}_{x}}^{2}.$ Multiplying both sides by $\left( -1\right) ,$ we
bring this inequality to%
\begin{equation*}
-\frac{\mathrm{g}_{{\small 3D}}^{2}}{M_{\mathrm{Pl}}}\left( q_{\text{\textsc{%
a}}}\mathcal{G}^{\text{\textsc{ab}}}q_{\text{\textsc{b}}}\right) \leq \left(
4\varrho \mathfrak{M}_{\mathrm{n}_{x}}^{2}-\mathrm{\vartheta }_{{\small QG}%
}\right) -\frac{\mathcal{G}^{\text{\textsc{ab}}}}{2\mathfrak{M}_{\mathrm{n}%
_{x}}^{2}}\left( \frac{\partial \mathfrak{M}_{\mathrm{n}_{x}}^{2}}{\partial
\hat{\upsilon}^{\text{\textsc{a}}}}\right) \left( \frac{\partial \mathfrak{M}%
_{\mathrm{n}_{x}}^{2}}{\partial \hat{\upsilon}^{\text{\textsc{b}}}}\right)
\end{equation*}%
and then into the following stronger form%
\begin{equation}
\fbox{ $\left.
\begin{array}{c}
\text{ } \\
\text{ }%
\end{array}%
\right. $\ \ $-\frac{\mathrm{g}_{{\small 3D}}^{2}}{M_{\mathrm{Pl}}}\left( q_{%
\text{\textsc{a}}}\mathcal{G}^{\text{\textsc{ab}}}q_{\text{\textsc{b}}%
}\right) \leq 4\varrho \mathfrak{M}_{\mathrm{n}_{x}}^{2}-\frac{\mathcal{G}^{%
\text{\textsc{ab}}}}{2\mathfrak{M}_{\mathrm{n}_{x}}^{2}}\left( \frac{%
\partial \mathfrak{M}_{\mathrm{n}_{x}}^{2}}{\partial \hat{\upsilon}^{\text{%
\textsc{a}}}}\right) \left( \frac{\partial \mathfrak{M}_{\mathrm{n}_{x}}^{2}%
}{\partial \hat{\upsilon}^{\text{\textsc{b}}}}\right) $ \ \ \ \ \ }
\label{rfc}
\end{equation}%
just because $\mathrm{\vartheta }_{{\small QG}}<0.$

\paragraph{B. Computing left hand side of eq(\protect\ref{rfc}):\newline
}

The LHS term of eq(\ref{rfc}) is given by the term $-\frac{\mathrm{g}_{%
{\small 3D}}^{2}}{M_{\mathrm{Pl}}}\left( q_{\text{\textsc{a}}}\mathcal{G}^{%
\text{\textsc{ab}}}q_{\text{\textsc{b}}}\right) ;$ depending on whether the
charge $q_{\text{\textsc{a}}}$ is given by the fibral value $q_{i_{\bot
}}\neq 0$ with $q_{a_{\Vert }}=0;$ or the base one namely $q_{a_{\Vert
}}\neq 0$ with $q_{i_{\bot }}=0,$ we split this contribution in two blocs as
follows%
\begin{equation}
\begin{tabular}{lll}
fiber K3$_{\bot }$ & : & $-\frac{\mathrm{g}_{{\small 3D}}^{2}}{M_{\mathrm{Pl}%
}}\left( q_{i_{\bot }}\mathcal{G}^{i_{\bot }j_{\bot }}q_{j_{\bot }}\right) $
\\
base K3$_{\Vert }$ & : & $-\frac{\mathrm{g}_{{\small 3D}}^{2}}{M_{\mathrm{Pl}%
}}\left( q_{a_{\Vert }}\mathcal{G}^{a_{\Vert }b_{\Vert }}q_{b_{\Vert
}}\right) $%
\end{tabular}%
\end{equation}%
We do the calculation for the fiber K3$_{\bot };$ and deduce the result for
the base by using the automorphism symmetry permuting K3$_{\bot }$ and K3$%
_{\Vert }$ (Weak/Strong duality).\newline
Using eqs(\ref{TM1}-\ref{TM2}), we can solve the relation $\mathcal{G}^{%
\text{\textsc{ac}}}\mathcal{G}_{\text{\textsc{cb}}}=\delta _{\text{\textsc{b}%
}}^{\text{\textsc{a}}}$ as%
\begin{equation}
\begin{tabular}{lllll}
$\mathcal{G}_{\text{\textsc{ab}}}$ & $=$ & $\left(
\begin{array}{cc}
\frac{1}{\mathcal{\hat{V}}_{K3_{\Vert }}}\eta _{a_{\Vert }b_{\Vert }} & 0 \\
0 & \frac{1}{\mathcal{\hat{V}}_{K3_{\perp }}}\eta _{i_{\perp }j_{\perp }}%
\end{array}%
\right) $ & $=$ & $\left(
\begin{array}{cc}
\mathcal{\hat{V}}_{K3_{\perp }}\eta _{a_{\Vert }b_{\Vert }} & 0 \\
0 & \mathcal{\hat{V}}_{K3_{\Vert }}\eta _{i_{\perp }j_{\perp }}%
\end{array}%
\right) $ \\
&  &  &  &  \\
$\mathcal{G}^{\text{\textsc{ab}}}$ & $=$ & $\left(
\begin{array}{cc}
\mathcal{\hat{V}}_{K3_{\Vert }}\eta ^{a_{\Vert }b_{\Vert }} & 0 \\
0 & \mathcal{\hat{V}}_{K3_{\perp }}\eta ^{i_{\perp }j_{\perp }}%
\end{array}%
\right) $ & $=$ & $\left(
\begin{array}{cc}
\frac{1}{\mathcal{\hat{V}}_{K3_{\perp }}}\eta ^{a_{\Vert }b_{\Vert }} & 0 \\
0 & \frac{1}{\mathcal{\hat{V}}_{K3_{\Vert }}}\eta ^{i_{\perp }j_{\perp }}%
\end{array}%
\right) $%
\end{tabular}
\label{MT}
\end{equation}%
with
\begin{equation}
\begin{tabular}{lll}
$\mathcal{\hat{V}}_{{\small K3}_{{\small \bot }}}$ & $=$ & $\frac{1}{2}\eta
_{i_{\perp }j_{\perp }}\hat{\upsilon}^{i_{\perp }}\hat{\upsilon}^{j_{\perp
}} $ \\
$\mathcal{\hat{V}}_{{\small K3}_{\Vert }}$ & $=$ & $\frac{1}{2}\eta
_{a_{\Vert }b_{\Vert }}\hat{\upsilon}^{a_{\Vert }}\hat{\upsilon}^{b_{\Vert
}} $%
\end{tabular}
\label{KK}
\end{equation}%
\ and scaling behaviours in the infinite distance limit like $\lambda ^{-2}%
\mathcal{\hat{V}}_{{\small K3}_{{\small \bot }}}$; and $\lambda ^{+2}%
\mathcal{\hat{V}}_{{\small K3}_{\Vert }}.$ \newline
Using these values, the LHS in (\ref{rfc}) reads as%
\begin{eqnarray}
-\frac{\mathrm{g}_{{\small 3D}}^{2}}{M_{\mathrm{Pl}}}\left( q_{i_{\bot }}%
\mathcal{G}^{i_{\perp }j_{\perp }}q_{j_{\bot }}\right) &=&-\frac{\mathrm{g}_{%
{\small 3D}}^{2}}{M_{\mathrm{Pl}}}\mathcal{\hat{V}}_{K3_{\perp }}\mathbf{q}%
_{\bot }^{2} \\
&=&\frac{2\mathrm{g}_{{\small 3D}}^{2}}{M_{\mathrm{Pl}}}\mathcal{\hat{V}}%
_{K3_{\perp }}\mathrm{n}_{\bot }
\end{eqnarray}%
where we have used $\mathbf{q}_{\bot }^{2}=\left( q_{i_{\bot }}\eta
^{i_{\perp }j_{\perp }}q_{j_{\bot }}\right) $ and $\mathbf{q}_{\bot }^{2}=-2%
\mathrm{n}_{\bot }.$ Thus, the contribution of the LHS is given by%
\begin{equation}
\fbox{%
\begin{tabular}{ll}
$\left.
\begin{array}{c}
\text{ } \\
\text{ }%
\end{array}%
\right. $ & $\ \ LHS_{\bot }=-\frac{\mathrm{g}_{{\small 3D}}^{2}}{M_{\mathrm{%
Pl}}}\left( q_{i_{\bot }}\mathcal{G}^{i_{\perp }j_{\perp }}q_{j_{\bot
}}\right) =\frac{2\mathrm{g}_{{\small 3D}}^{2}}{M_{\mathrm{Pl}}}\mathcal{%
\hat{V}}_{{\small K3}_{{\small \bot }}}$\textrm{n}$_{\bot }$\ \ \ \ \ \ \ \
\  \\
$\left.
\begin{array}{c}
\text{ } \\
\text{ }%
\end{array}%
\right. $ & $\ \ LHS_{\Vert }=-\frac{\mathrm{g}_{{\small 3D}}^{2}}{M_{%
\mathrm{Pl}}}\left( q_{i_{\Vert }}\mathcal{G}^{a_{\Vert }b_{\Vert
}}q_{j_{\Vert }}\right) =\frac{2\mathrm{g}_{{\small 3D}}^{2}}{M_{\mathrm{Pl}}%
}\mathcal{\hat{V}}_{{\small K3}_{\Vert }}$\textrm{n}$_{\Vert }$\ \ \
\end{tabular}%
}
\end{equation}%
These $LHS_{\bot }$ and the $LHS_{\Vert }$ are manifestly related under
Weak/Strong duality generated by the transposition $K3_{\bot
}\leftrightarrow K3_{\Vert }.$

\paragraph{B. Computing right hand side of (\protect\ref{rfc})\newline
}

The contribution in the RHS of eq(\ref{rfc}) for the fiber K3$_{\perp }$ is
given by%
\begin{equation}
RHS_{\perp }=\left( 2\mathfrak{M}_{\mathrm{n}_{\perp }}^{2}+\mathrm{%
\vartheta }_{{\small QG}}\right) -\frac{\mathcal{\hat{V}}_{K3_{\perp }}}{2%
\mathfrak{M}_{\mathrm{n}_{\perp }}^{2}}\eta ^{i_{\perp }j_{\perp }}\frac{%
\partial \mathfrak{M}_{\mathrm{n}_{\perp }}^{2}}{\partial \hat{\upsilon}%
^{i_{\perp }}}\frac{\partial \mathfrak{M}_{\mathrm{n}_{\perp }}^{2}}{%
\partial \hat{\upsilon}^{j_{\perp }}}
\end{equation}%
where we have substituted $\mathcal{G}^{i_{\perp }j_{\perp }}=\mathcal{\hat{V%
}}_{K3_{\perp }}\eta ^{i_{\perp }j_{\perp }}.$ \newline
Using (\ref{MN}) namely
\begin{equation}
\frac{\partial \mathfrak{M}_{\mathrm{n}_{\perp }}^{2}}{\partial \hat{\upsilon%
}^{i_{\perp }}}\eta ^{i_{\perp }j_{\perp }}\frac{\partial \mathfrak{M}_{%
\mathrm{n}_{\perp }}^{2}}{\partial \hat{\upsilon}^{j_{x}}}=2\nu _{\mathrm{n}%
_{_{\perp }}}\mathfrak{M}_{\mathrm{n}_{\perp }}^{2}
\end{equation}%
the above expression reduces down to%
\begin{equation}
RHS_{\perp }=\left( 2\mathfrak{M}_{\mathrm{n}_{\perp }}^{2}+\mathrm{%
\vartheta }_{{\small QG}}-\nu _{\mathrm{n}_{_{\perp }}}\mathcal{\hat{V}}%
_{K3_{\perp }}\right)  \label{rhs}
\end{equation}%
By using (\ref{MX}), namely
\begin{equation}
2\mathfrak{M}_{\mathrm{n}_{\perp }}^{2}=2\nu _{\mathrm{n}_{_{\perp }}}%
\mathcal{\hat{V}}_{K3_{x}}+\frac{\nu _{\mathrm{n}_{_{\perp }}}}{2\mathrm{n}%
_{_{\perp }}}\left( q_{i_{\perp }}\hat{\upsilon}^{i_{\perp }}\right) ^{2}
\end{equation}%
and putting into (\ref{rhs}), we get
\begin{equation}
RHS_{\perp }=\nu _{\mathrm{n}_{_{_{\perp }}}}\left[ \mathcal{\hat{V}}%
_{K3_{\perp }}+\frac{1}{2\mathrm{n}_{_{_{\perp }}}}\left( q_{i_{_{\perp }}}%
\hat{\upsilon}^{i_{_{\perp }}}\right) ^{2}\right]  \label{hs}
\end{equation}%
By replacing $\nu _{\mathrm{n}_{_{_{\perp }}}}=\mathrm{n}_{_{_{\perp }}}/%
\mathcal{V}_{{\small CY}_{{\small 4}}}^{2}$ and using $\mathrm{g}_{{\small 3D%
}}^{2}/M_{\mathrm{Pl}}=1/(4\mathcal{V}_{{\small X}_{{\small 4}}}^{2})$, we
have $\nu _{\mathrm{n}_{_{_{\perp }}}}=\left( 4\mathrm{g}_{{\small 3D}%
}^{2}/M_{\mathrm{Pl}}\right) \mathrm{n}_{_{_{\perp }}}.$ \newline
Then, putting back into (\ref{hs}), we end up with the following result%
\begin{equation}
\fbox{%
\begin{tabular}{ll}
$\left.
\begin{array}{c}
\text{ \ } \\
\text{ \ }%
\end{array}%
\right. $ & \ \ \ \ $RHS_{\perp }=2\left[ \frac{2\mathrm{g}_{{\small 3D}}^{2}%
}{M_{\mathrm{Pl}}}\mathcal{\hat{V}}_{K3_{\perp }}\mathrm{n}_{_{\perp }}+%
\frac{\mathrm{g}_{{\small 3D}}^{2}}{M_{\mathrm{Pl}}}\left( q_{i_{_{_{\perp
}}}}\hat{\upsilon}^{i_{_{_{\perp }}}}\right) ^{2}\right] $ \ \ \ \ \ \ \ \ \
\ \ \  \\
$\left.
\begin{array}{c}
\text{ \ } \\
\text{ \ }%
\end{array}%
\right. $ & \ \ \ \ $RHS_{\Vert }=2\left[ \frac{2\mathrm{g}_{{\small 3D}}^{2}%
}{M_{\mathrm{Pl}}}\mathcal{\hat{V}}_{K3_{\Vert }}\mathrm{n}_{_{\Vert }}+%
\frac{\mathrm{g}_{{\small 3D}}^{2}}{M_{\mathrm{Pl}}}\left( q_{a_{_{_{\Vert
}}}}\hat{\upsilon}^{a_{_{_{\Vert }}}}\right) ^{2}\right] $ \ \ \
\end{tabular}%
}  \label{NRH}
\end{equation}%
It is indeed greater than to the contribution of left hand sides $%
LHS_{x}=\left( 2\mathrm{g}_{{\small 3D}}^{2}/M_{\mathrm{Pl}}\right) \mathcal{%
\hat{V}}_{{\small K3}_{{\small x}}}$\textrm{n}$_{x}$ ( with $x=\bot ,\Vert $%
) in agreement with the repulsive force conjecture (\ref{rfc}).

\section{Conclusion and discussions}

\qquad The WGC has been definitively one of the most studied conjectures of
the Swampland; it has undergone numerous refinements and generalisations
such as the one investigated in this paper for the\textrm{\ }\emph{EFT}$_{%
{\small 3D}}$\emph{.} Here, we took a particular interest in the Repulsive
Force Conjecture (RFC), the tower WGC and its most outstanding derivative
the Asymptotic WGC, where we studied different regimes of gauge couplings,
weak and strong. First, we deepened the study in the weak coupling regime of
the \emph{EFT}$_{{\small 3D}}$ descending from M-theory on the family of
Calabi-Yau fourfold $K3_{\bot }\times K3_{\Vert };$ and constructed the
towers of BPS or non-BPS states satisfying the Asymptotic WGC. Then, we used
the discrete symmetry of the CY4 generated by the permutation $K3_{\bot
}\leftrightarrow K3_{\Vert }$ and showed through explicit calculations the
existence of a dual version of the Asymptotic WGC. This dual description
involves strong gauge coupling dual to the weak gauge coupling investigated
in \textrm{\cite{12}} for M-theory on CY3 with K3 fibration.

In order to classify the infinite distance limits in M-theory on Calabi-Yau
fourfolds, we use the so called type-$\mathbb{T}$ and type-$\mathbb{S}$
studied in \textrm{\cite{11A}}. For generic \emph{EFT}$_{{\small 3D}},$ we
showed that there exists three possible classes termed type-$\mathbb{T}$,
type-$\mathbb{S}$ (as in \emph{EFT}$_{{\small 5D}}$) \textrm{in addition to}
type-$\mathbb{V}$ for \emph{EFT}$_{{\small 3D}}$. These represent
respectively a $X_{4}$ with an elliptic fiber, a surface fiber, and a volume
fiber; they correspond to the internal manifolds $X_{4}$ taking the form $%
\mathcal{F}_{n}\times \mathcal{B}_{n-4}$ with $n=1,2,3$ being the complex
dimension of the fiber. Thus, we constructed a Calabi-Yau fourfold with
shrinking fibers and diverging bases, keeping the total volume of the
manifold finite ($\mathcal{V}_{X_{4}}<\infty $) to insure the dynamics of
gravity; this allowed us to probe infinite distance in the moduli space
which correspond to weak coupling regime (\ref{WeakCoupling}).

In fact to make contact with this physical aspect, we compactified M-theory
on the aforementioned manifold, where we derived the 3D action from the 11D
supergravity action with M-brane. Then, we discussed the weak gauge coupling
limit that constitutes half of our main focus in this paper; the other half
regards the strong gauge regime. This has been motivated by the type-$%
\mathbb{S}$ form of the CY4 fibration. For this surface form, the Calabi-Yau
fourfold is fibered as $S_{\perp }\times S_{\parallel }$ with finite total
volume $\mathcal{V}_{X_{4}}=\mathcal{V}_{S_{\perp }}\mathcal{V}_{S_{\Vert
}}; $ \textrm{with} $\left( \mathbf{i}\right) $ a fibral surface volume $%
\mathcal{V}_{S_{\perp }}$ shrinking in the infinite distance limit as $%
\mathcal{O}\left( \lambda ^{-2}\right) ;$ and $\left( \mathbf{ii}\right) $
base surface volume $\mathcal{V}_{S_{\Vert }}$ expanding in the infinite
distance limit as $\mathcal{O}\left( \lambda ^{+2}\right) .$ Because the $%
S_{\perp }$ and the $S_{\parallel }$ have some complex dimension, the CY4
has a discrete $\mathbb{Z}_{2}$ symmetry given by transposition of the roles
of $S_{\perp }$ and $S_{\parallel };$ that is $S_{\perp }\leftrightarrow
S_{\parallel }.$ This discrete duality induces a Weak/Strong duality between
weak and strong gauge couplings generated by the $\mathbb{Z}_{2}$ symmetry $%
\lambda \rightarrow 1/\lambda .$ In the present study, we took the
interesting CY4 geometry $X_{4}=K3_{\perp }\times K3_{\parallel }$ and we
showed that the full abelian gauge symmetry of the \emph{EFT}$_{{\small 3D}}$
factorises like $U\left( 1\right) _{\bot }^{\nu }\times U\left( 1\right)
_{\Vert }^{\nu }$ as in (\ref{SymBreak}). This CY4 is of particular interest
due to the intrinsic properties of $K3$ where several calculations can be
performed explicitly. Using the $\mathbb{Z}_{2}$ duality, we can switch from
the fiber to the base using the mapping $\lambda \rightarrow 1/\lambda $;
and then switch from weak to strong gauge groups. This is an interesting
property for the study of the Asymptotic WGC in \emph{EFT}$_{{\small 3D}}$
and its test as done explicitly in section 5; \textrm{it also} predicts a
strong coupling regime with a dual Asymptotic WGC which \textrm{can be
interpreted} as a strong version of Asymptotic WGC. There, towers $\mathcal{T%
}_{\bot }^{\text{\textsc{bps}}}$ of BPS and towers $\mathcal{T}_{\bot }^{%
\text{\textsc{n-bps}}}$ of non BPS states with light masses have $\mathbb{Z}%
_{2}$ duals given by towers $\mathcal{T}_{\Vert }^{\text{\textsc{bps}}}$ of
BPS and towers $\mathcal{T}_{\Vert }^{\text{\textsc{n-bps}}}$ of non BPS
states with heavy masses. \ These towers of BPS and non BPS particle states
in the \emph{EFT}$_{{\small 3D}}$ are described in the two claims of
subsection 4.3.\newline
To insure the viability of the conjecture in 3d gravity induced by
M-theory on a Calabi-Yau fourfold of the form $X_{4}=K3_{\perp }\times
K3_{\parallel }$\textrm{\ we thoroughly tested the Swampland conjecture in
both weakly and strongly coupled regimes in section 5}$.$\textrm{\ This
successful test is motivating the use} of this internal geometry to probe
other conjectures of the Swampland. In this regard, one can mention in
particular the Distance Conjecture and the No-Global Symmetries Conjecture.
The importance of such investigations \textrm{lies in }the possibility of
\textrm{unifying} the Swampland conjectures in such a way that one could
reduce the number of conjectures to the most fundamental ones. Progress in
this direction will be reported in a future occasion.

 \appendix

\section{Appendix: Towers of BPS/non BPS states}

\label{app}
In this appendix, we give useful details regarding the BPS and the
non-BPS towers in the K3 fibered Calabi-Yau fourfold $X_{4}=K3\times K3$ as
in the Figure \textbf{\ref{02K}}. We also comment on the existence of
asymptotic light (resp. heavy) towers of particle states in the weak (resp.
strong) coupling regime associated with the infinite (resp. short ) distance
limit where the K3 fiber shrinks (resp. diverges) and the base expands
(resp. shrinks) such that the CY4 volume $\mathcal{V}_{X_{4}}$ is finite.

\subsection*{M-theory on $K3_{\bot }\times K3_{\Vert }$ in weak/strong gauge
regimes}

\qquad In the investigation of the underlying 3D effective field theory (%
\emph{EFT}$_{{\small 3D}}$) in weak (strong) regime descending from M-theory
on $K3_{\bot }\times K3_{\Vert }$ with finite volume $\mathcal{V}%
_{X_{4}}\sim \mathcal{V}_{K3_{\bot }}\mathcal{V}_{K3_{\Vert }},$ one has to
distinguish between different classes of fibers depending on whether the K3
fiber (base) is a regular surface, or it has singularities at isolated
points on the base (fiber) occurring $\left( i\right) $ either at finite
distances in the moduli space (Kulikov of type I); or $\left( ii\right) $ at
the infinite distances (Kulikov of type II and III) \textrm{\cite{27A, 27B}}%
. Below, we focus on regular K3s at infinite distances with finite volume of
the CY4.

Given a basis set of basic holomorphic curves $\mathcal{I}_{\mathrm{\rho }%
}=\{C^{\text{\textsc{a}}}\}_{1\leq \text{\textsc{a}}\leq \mathrm{\rho }\leq
h_{{\small X}_{{\small 4}}}^{{\small 1,1}}}$ generating a subgroup of H$%
_{2}\left( X_{4}\right) ,$ one associates several interesting quantities in
the QFT$_{{\small 3D}}$ descending from the compactified M-theory on $X_{4}:$

$\bullet $ \emph{Gauge coupling regimes}\newline
Under compactification, one induces a basis set of abelian 1-form gauge
potentials \{$A^{\text{\textsc{a}}}$\}$_{1\leq \text{\textsc{a}}\leq \mathrm{%
\rho }}$. The 1-forms $A^{\text{\textsc{a}}}$ are given by the wrapping $%
\int_{C^{\text{\textsc{a}}}}\boldsymbol{C}_{3}$ of the 3-form potential $%
\boldsymbol{C}_{3}$ of the M-theory on $C^{\text{\textsc{a}}}$; the $A^{%
\text{\textsc{a}}}$'s couple to M2/$C^{\text{\textsc{a}}}$. Together with
the basis set of gauge potentials \{$A^{\text{\textsc{a}}}$\}, we have the
abelian gauge symmetry group $G=\prod\nolimits_{\text{\textsc{a}}}U(1)^{%
\text{\textsc{a}}}$ with gauge coupling matrix $\mathcal{G}_{\text{\textsc{ab%
}}}=\int_{X_{4}}J_{\text{\textsc{a}}}\wedge \ast J_{\text{\textsc{b}}}.$
This coupling matrix decomposes on $K3_{\bot }\times K3_{\Vert }$ in terms
of the Kahler 2-forms $J_{\text{\textsc{a}}}=(J_{a_{\Vert }},J_{i_{\bot }})$
and its Hodge dual $\ast J_{\text{\textsc{a}}}=(\ast J_{a_{\Vert }},\ast
J_{i_{\bot }})$ as%
\begin{equation}
\begin{tabular}{lllllll}
$\mathcal{G}_{a_{\Vert }b_{\Vert }}$ & $=$ & $\dint\nolimits_{X_{4}}J_{a_{%
\Vert }}\wedge \ast J_{b_{\Vert }}$ & ,\qquad & $\mathcal{G}_{a_{\Vert
}j_{\bot }}$ & $=$ & $\dint\nolimits_{X_{4}}J_{a_{\Vert }}\wedge \ast
J_{j_{\bot }}$ \\
$\mathcal{G}_{i_{\bot }j_{\bot }}$ & $=$ & $\dint\nolimits_{X_{4}}J_{i_{\bot
}}\wedge \ast J_{j_{\bot }}$ & ,\qquad & $\mathcal{G}_{i_{\bot }b_{\Vert }}$
& $=$ & $\dint\nolimits_{X_{4}}J_{i_{\bot }}\wedge \ast J_{b_{\Vert }}$%
\end{tabular}%
\end{equation}%
with leading dependence in the spectral parameter $\lambda $\ like%
\begin{equation}
\mathcal{G}_{\text{\textsc{ab}}}=\left(
\begin{array}{cc}
\mathcal{O}\left( \frac{1}{\lambda ^{2}}\right) & \mathcal{O}\left( 1\right)
\\
\mathcal{O}\left( 1\right) & \mathcal{O}\left( \lambda ^{2}\right)%
\end{array}%
\right) \qquad \underrightarrow{\text{ \ \ }\lambda \rightarrow \infty \text{
\ \ \ }}\qquad \left(
\begin{array}{cc}
0 & 1 \\
1 & \infty%
\end{array}%
\right)
\end{equation}%
Using results from the core of the paper, the inverse gauge metric $\mathcal{%
G}^{\text{\textsc{ab}}}$ denoted as follows
\begin{equation}
\mathcal{G}^{\text{\textsc{ab}}}=\left(
\begin{array}{cc}
\mathcal{G}^{a_{\Vert }b_{\Vert }} & \mathcal{G}^{a_{\Vert }j_{\bot }} \\
\mathcal{G}^{i_{\bot }b_{\Vert }} & \mathcal{G}^{i_{\bot }j_{\bot }}%
\end{array}%
\right)
\end{equation}%
behaves in the infinite distance limit as%
\begin{equation}
\left(
\begin{array}{cc}
\mathcal{O}\left( \lambda ^{2}\right) & \mathcal{O}\left( 1\right) \\
\mathcal{O}\left( 1\right) & \mathcal{O}\left( \frac{1}{\lambda ^{2}}\right)%
\end{array}%
\right) \qquad \rightarrow \qquad \left(
\begin{array}{cc}
\infty & 1 \\
1 & 0%
\end{array}%
\right)  \label{gab}
\end{equation}%
indicating the existence of two gauge regimes: weak and strong, and
suggesting the existence of holomorphic curves $\Upsilon _{\mathbf{q}}$ in
the CY4 where sit weak and gauge 3D effective field theories.

$\bullet $ \emph{Complex curves }$\Upsilon _{weak}$ and $\Upsilon _{strong}$%
\newline
Generic complex curves $\Upsilon _{\mathbf{q}}$ inside the $X_{4}$ are given
by the expansion $\sum_{\text{\textsc{a}}{\small =1}}^{\rho }q_{\text{%
\textsc{a}}}C^{\text{\textsc{a}}}$; they carry an integral charge vector $%
\mathbf{q}$ under the gauge symmetry $G$ with components $\left( q_{\text{%
\textsc{a}}}\right) $ splitting along the fiber and the base directions like
$\left( q_{a_{\Vert }},q_{i_{\bot }}\right) $. These generic curves support
the following objects: \newline
$\left( \mathbf{i}\right) $ the 1-form gauge potential $A_{\mathbf{q}%
}=\int_{\Upsilon _{\mathbf{q}}}\boldsymbol{C}_{3}$ which by substituting $%
\Upsilon _{\mathbf{q}}=$ $\sum_{\text{\textsc{a}}{\small =1}}^{\rho }q_{%
\text{\textsc{a}}}C^{\text{\textsc{a}}}$ expands in turn as a linear
combination of the individual gauge potential as follows $\sum_{\text{%
\textsc{a}}{\small =1}}^{\rho }q_{\text{\textsc{a}}}A^{\text{\textsc{a}}}$.
\newline
$\left( \mathbf{ii}\right) $ the abelian Maxwell-like gauge symmetry $%
U(1)_{\Upsilon _{\mathbf{q}}}$ is given by the linear combination $\sum_{%
\text{\textsc{a}}{\small =1}}^{\rho }q_{\text{\textsc{a}}}U(1)^{\text{%
\textsc{a}}}$. This expansion has two contribution like $U(1)_{\Upsilon _{%
\mathbf{q}_{\Vert }}}+U(1)_{\Upsilon _{\mathbf{q}_{\bot }}}$; the bloc $%
U(1)_{\Upsilon _{\mathbf{q}_{\Vert }}}$ originates from the base surface and
the $U(1)_{\Upsilon _{\mathbf{q}_{\bot }}}$ from the fiber,
\begin{equation}
U(1)_{\Upsilon _{\mathbf{q}_{\Vert }}}=\dsum\limits_{a_{\Vert }=1}^{\rho
_{\Vert }}q_{a_{\Vert }}U(1)^{a_{\Vert }}\qquad ,\qquad U(1)_{\Upsilon _{%
\mathbf{q}_{\bot }}}=\dsum\limits_{i_{\bot }=1}^{\rho _{\bot }}q_{i_{\bot
}}U(1)^{i_{\bot }}
\end{equation}%
$\left( \mathbf{iii}\right) $ An effective field theory \emph{EFT}$_{{\small %
3D}}$ located on a given complex curve $\Upsilon _{\mathbf{q}}$ has a gauge
symmetry $U(1)_{\Upsilon _{\mathbf{q}}}$. Its gauge coupling constant
denoted as \textrm{g}$_{_{\Upsilon _{\mathbf{q}}}}$ is a function of the 3D
gauge coupling $\mathrm{g}_{{\small YM}}$ and the integral charges $q_{\text{%
\textsc{a}}}.$ The squared \textrm{g}$_{_{\Upsilon _{\mathbf{q}}}}^{2}$ is
given the relation $\mathrm{g}_{{\small YM}}^{2}q_{\text{\textsc{a}}}%
\mathcal{G}^{\text{\textsc{ab}}}q_{\text{\textsc{b}}}$. Moreover, because of
the behaviour (\ref{gab}) in the spectral parameter $\lambda $, we
distinguish two interesting gauge regime limits: weak and strong given by%
\begin{equation}
\begin{tabular}{lllll}
{\small weak regime} & {\small :} & $\mathrm{g}_{_{weak}}^{2}$ & $\sim $ & $%
q_{i_{\bot }}\mathcal{G}^{i_{\bot }j_{\bot }}q_{j_{\bot }}$ \\
{\small strong regime} & {\small :} & $\mathrm{g}_{_{strong}}^{2}$ & $\sim $
& $q_{a_{\Vert }}\mathcal{G}^{a_{\Vert }b_{\Vert }}q_{b_{\Vert }}$%
\end{tabular}%
\end{equation}%
with%
\begin{equation}
\mathcal{G}^{i_{\bot }j_{\bot }}=\frac{1}{\mathcal{V}_{K3_{\Vert }}}\eta
^{i_{\bot }j_{\bot }},\qquad \mathcal{G}^{a_{\Vert }b_{\Vert }}=\frac{1}{%
\mathcal{V}_{K3_{\bot }}}\eta ^{a_{\Vert }b_{\Vert }}
\end{equation}%
where $\eta ^{i_{\bot }j_{\bot }}$ and $\eta ^{a_{\Vert }b_{\Vert }}$ are
respectively the metrics of the hyperbolic sublattices $\Lambda _{\bot
}^{\ast }$ and $\Lambda _{\Vert }^{\ast }$ and where $\mathcal{V}_{K3_{\bot
}}=\eta _{i_{\bot }j_{\bot }}\upsilon ^{i_{\bot }}\upsilon ^{j_{\bot }}/2$
and $\mathcal{V}_{K3_{\Vert }}=\eta _{a_{\Vert }b_{\Vert }}\upsilon
^{a_{\Vert }}\upsilon ^{b_{\Vert }}/2$ are the volumes of the two $K3$s.
These weak and strong gauge regimes are supported by the curves $\Upsilon
_{weak}$ and $\Upsilon _{strong}$ given by the expansions
\begin{equation}
\Upsilon _{weak}=\dsum\limits_{i_{\bot }=1}^{\rho _{\bot }}q_{i_{\bot }}%
\mathcal{C}^{i_{\bot }}\qquad ,\qquad \Upsilon
_{strong}=\dsum\limits_{a_{\Vert }=1}^{\rho _{\Vert }}q_{a_{\Vert }}\mathcal{%
C}^{a_{\Vert }}
\end{equation}%
Here the set \{$\mathcal{C}^{i_{\bot }}$\} generates a hyperbolic-like
sublattice $\Lambda _{\bot }^{\ast }$ with signature (1,$\rho _{\bot }-1$);
it is contained into the 22 dimensional K3$_{\bot }$ lattice $\Gamma _{\bot
}^{3,19}.$ Similarly, the set \{$\mathcal{C}^{a_{\Vert }}$\}\ generates also
a hyperbolic lattice $\Lambda _{\Vert }^{\ast }$ with signature ($1,\rho
_{\Vert }-1$); it is a sublattice of the 22 dimensional lattice $\Gamma
_{\Vert }^{3,19}.$\newline
Notice that if the fiber K3$_{\bot }$ (resp. the base K3$_{\Vert }$) does
not degenerate on the base K3$_{\Vert }$ (resp. the fiber K3$_{\bot }$), the
divisors $\mathcal{D}_{i_{\bot }}$ (resp. $\mathcal{D}_{\Vert }$) are given
\ by fibering the $C^{i_{\bot }}$ over K3$_{\Vert }$ (resp. $\mathcal{C}%
^{a_{\Vert }}$ over K3$_{\bot }$). Moreover, if $\mathfrak{i}:K3\rightarrow
X_{4}$ is an embedding in $X_{4}$; then the image of the Piccard group
Pic(K3) of generic K3 is given by $H^{2}(X_{4},\mathbb{Z})\cap
H^{1,1}(X_{4}).$

$\bullet $ \emph{BPS and anti-BPS curves }\newline
The the full charge lattice of the regular CY4 considered in our study is
given by $\Gamma _{\Vert }^{3,19}\oplus \Gamma _{\bot }^{3,19}$ with the 22
dimensional lattice $\Gamma ^{3,19}$ being the lattice H$^{2}\left( K3,%
\mathbb{Z}\right) $ decomposing into blocks like
\begin{equation}
\Gamma _{\Vert }^{3,19}=U_{\Vert }^{{\small \oplus }3}\oplus E_{8\Vert
}\oplus E_{8\Vert }\qquad ,\qquad U_{\bot }^{{\small \oplus 3}}\oplus
(E_{8\bot }\oplus E_{8\bot })
\end{equation}
with $U$ standing for the usual 2D hyperbolic lattice, and each copy $E_{8}$
representing the root lattice of exceptional $E_{8}$ Lie algebra. The 22+22
dimensional lattice $\Gamma _{\Vert }^{3,19}\oplus \Gamma _{\bot }^{3,19}$
contains the sublattice $\Lambda ^{\ast }{\small [\rho ]}=\Lambda _{\Vert
}^{\ast }{\small [\rho _{_{\Vert }}]}\oplus \Lambda _{\bot }^{\ast }{\small %
[\rho _{_{\bot }}]}$ with total dimension $\rho =\rho _{\bot }+\rho _{\Vert
} $ and partial $\rho _{\bot },\rho _{\Vert }$ $\leq 20.$ Furthermore, the
fibral sublattice $\Lambda _{\bot }^{\ast }$ decomposes into a self and
anti-self dual sublattices like $\Lambda _{\bot +}^{\ast }\oplus \Lambda
_{\bot -}^{\ast }$ in the sense that charge vectors $\mathbf{q}_{\bot
}=\sum_{i_{\bot }=1}^{\rho _{\bot }}q_{i_{\bot }}\mathbf{\varepsilon }%
^{i_{\bot }}$ splits also as
\begin{equation}
\mathbf{q}_{\bot }=\mathbf{q}_{\bot +}\oplus \mathbf{q}_{\bot -}
\end{equation}%
In this parameterisation, the metric $\eta ^{i_{\bot }j_{\bot }}=\mathbf{%
\varepsilon }^{i_{\bot }}.\mathbf{\varepsilon }^{j_{\bot }}$ has an
indefinite sign [type hyperbolic]; the number $\mathbf{q}_{\bot +}^{2}$ is
positive definite while the $\mathbf{q}_{\bot -}^{2}$ is negative definite.
\newline
Similarly, the base sublattice $\Lambda _{\Vert }^{\ast }$ decomposes also
into a self dual $\Lambda _{\Vert +}^{\ast }$ and anti-self dual $\Lambda
_{\Vert -}^{\ast }$ sublattices. The charge vector $\mathbf{q}_{\Vert
}=\sum_{a_{\Vert }=1}^{\rho _{\Vert }}q_{a_{\Vert }}\mathbf{\varepsilon }%
^{a_{\Vert }}$ with hyperbolic- like metric $\eta ^{a_{\Vert }b_{\Vert }}=%
\mathbf{\varepsilon }^{a_{\Vert }}.\mathbf{\varepsilon }^{b_{\Vert }}$
splits as well as
\begin{equation}
\mathbf{q}_{\Vert }=\mathbf{q}_{\Vert +}\oplus \mathbf{q}_{\Vert -}
\end{equation}
with $\mathbf{q}_{\Vert +}^{2}$ positive and $\mathbf{q}_{\Vert -}^{2}$
negative definite. Because of the property $\mathbf{q}^{2}=|\mathbf{q}%
_{+}|^{2}-|\mathbf{q}_{-}|^{2},$ we see that there are three region in the
charge lattice depending on the sign of $\mathbf{q}^{2}$. So, depending on
the region of $\Lambda _{\Vert }^{\ast }\oplus \Lambda _{\bot }^{\ast }$
where the curve $\Upsilon _{\mathbf{q}}$ is located, we distinguish
different types of particle states in the \emph{EFT}$_{{\small 3D}}.$ These $%
\Upsilon _{\mathbf{q}}$'s are given by movable curves inside K3 and non
movable ones as described below.

\subsection*{Movable curves in Mori cone of K3}

These are holomorphic curves $\Upsilon _{\mathbf{q}}=\sum_{\text{\textsc{a}}%
{\small =1}}^{\rho }q_{\text{\textsc{a}}}C^{\text{\textsc{a}}}$ that have
positive self intersection ($\Upsilon _{\mathbf{q}}^{2}>0$); that is charge
as $\mathbf{q}^{2}>0.$ This condition cane be $\left( i\right) $ either
\emph{simply} solved by charges $\mathbf{q=q}_{+}$ sitting in the self dual $%
\Lambda _{+}^{\ast },$ or $\left( ii\right) $ generally like $\mathbf{q}=%
\mathbf{q}_{+}+\mathbf{q}_{-}$ with the condition $|\mathbf{q}_{+}|^{2}>|%
\mathbf{q}_{-}|^{2}.$ They lie inside the movable cone of K3, and they
define rays $\Upsilon _{\mathbf{Q}_{n}}=n\Upsilon _{\mathbf{q}}$ with
positive integer n generating a tower of BPS particle states with charges $%
\mathbf{Q}_{n}=n\mathbf{q}$. These particle states, formally denoted as
\begin{equation}
M2/\Upsilon _{\mathbf{Q}_{n}}
\end{equation}%
are given by wrapping a M2 brane on $\Upsilon _{\mathbf{Q}_{n}}$ and have
interesting features; in particular: They are remarkably associated with a
non vanishing genus zero BPS invariants $\mathcal{N}_{n\Sigma }^{0}$ (often
termed as Gopakumar-Vafa invariant $\mathcal{N}_{n\Sigma }^{0}\neq 0$ with
integer $n\geq 1$). These BPS invariants are determined by the expansion
coefficients $c\left( m\right) $ of a meromorphic modular form $\vartheta
\left( q\right) $\ of weight $-2$ \textrm{\cite{30,31}} namely
\begin{equation}
\vartheta \left( q\right) =\sum_{m\geq -1}c\left( m\right) e^{i2m\pi \tau }%
\text{\qquad },\text{\qquad }c\left( m\right) \neq 0
\end{equation}%
with the relation%
\begin{equation}
\mathcal{N}_{\Upsilon _{\mathbf{Q}_{n}}}^{0}=c\left( \Upsilon _{\mathbf{Q}%
_{n}}^{2}/2\right) \text{\qquad }\Rightarrow \text{\qquad }\Upsilon _{%
\mathbf{Q}_{n}}^{2}\geq -2
\end{equation}%
The BPS states are also characterised by the so-called Gromov-Witten (GW)
invariants \textrm{\cite{32}} and by Noether-Leftshetz (NL) numbers \textrm{%
\cite{32,33}} for which all curve classes [$\Upsilon $] with the similar
intersection form and discriminant class in $\Lambda ^{\ast }$ have the same
invariant. \newline
The BPS particle states of this \emph{EFT}$_{{\small 3D}}$ have as well an
interesting realisation in terms of mode excitations of the \emph{heterotic}
string on $\mathbb{T}^{5}\times K3.$ This string comes by using two
dualities: first the duality between M-theory on $K3\times K3\times \mathbb{S%
}_{M}^{1}$ and Type IIA on $K3\times K3$; and second the duality between the
Type IIA string on $K3$ and the heterotic string on $\mathbb{T}^{4}$.
\newline
Furthermore, because here we have two K3 surfaces (the fiber $K3_{\bot }$
and the base $K3_{\Vert }$) in the Calabi-Yau compactification of M-theory
down to 3D, we can distinguish four sectors of K3 curves depending on their
location in $K3_{\bot }\times K3_{\Vert }$; these sectors are given by the
pairs%
\begin{equation}
\left( \Upsilon _{\bot +},\Upsilon _{\Vert +}\right) ,\qquad \left( \Upsilon
_{\bot +},\Upsilon _{\Vert -}\right) ,\qquad \left( \Upsilon _{\bot
-},\Upsilon _{\Vert +}\right) ,\qquad \left( \Upsilon _{\bot -},\Upsilon
_{\Vert -}\right)
\end{equation}%
This splitting goes with the existence of two kinds of dual \emph{heterotic}
strings for M-theory on $K3_{\bot }\times K3_{\Vert }$ due to the
automorphism symmetry $K3_{\bot }\leftrightarrow K3_{\Vert }$. The dual
strings are given by heterotic string on $\mathbb{S}_{M}^{1}\times \mathbb{T}%
^{4}\times K3_{\bot }$ and the heterotic on $K3_{\Vert }\times \mathbb{S}%
_{M}^{1}\times \mathbb{T}^{4}.$

\subsection*{Non movable curves in K3}

These are rigid curves $\Sigma $ living in the charge sublattice $\Lambda
_{\Vert }^{\ast }\oplus \Lambda _{\bot }^{\ast }$ of the Calabi-Yau fourfold
$K3_{\bot }\times K3_{\Vert }.$ They \textrm{lie} outside the movable cone
of the K3 surfaces and do not support BPS states. These holomorphic curves
have negative self intersection ($\Sigma ^{2}<0$), and they define a ray $%
\Sigma =\mathrm{k}\Sigma $ labeled by\ the integer \textrm{k} and the
generator $\Sigma $. This ray in the charge lattice defines a tower $%
\mathcal{T}_{\mathrm{k}}^{\text{\textsc{n-bps}}}$ of non-BPS particle
states. Examples of such curves $\Sigma $ are given by the hyperbolic family
\begin{equation}
\Sigma =l_{{\small U}}\mathcal{C}^{{\small U}}-l_{{\small T}}\mathcal{C}^{%
{\small T}}\qquad \leftrightarrow \qquad \Upsilon _{\mathbf{q}}=q_{{\small 1}%
}\mathcal{C}^{1}+q_{{\small 2}}\mathcal{C}^{2}
\end{equation}%
parameterised by two non zero integers $l_{{\small U}}$ and $l_{{\small T}}$
having the same sign (i.e: $l_{{\small U}}.l_{{\small T}}>0$). These curves
are generated by two 2-cycles $\mathcal{C}^{{\small U}}$ and $\mathcal{C}^{%
{\small T}}$ sitting inside K3 and realised in terms of a 2-torus $\mathbb{T}%
^{2}$ and a projective line $\mathbb{P}^{1}$ (2-sphere) as follows
\begin{equation}
\mathcal{C}^{{\small S}}=\mathbb{P}^{1},\qquad \mathcal{C}^{{\small T}}=%
\mathbb{T}^{2},\qquad \mathcal{C}^{{\small U}}=\mathbb{T}^{2}+\mathbb{P}^{1}
\end{equation}%
with the intersections
\begin{equation}
\mathcal{C}^{{\small T}}.\mathcal{C}^{{\small S}}=1,\qquad \ \mathcal{C}^{%
{\small T}}.\mathcal{C}^{{\small T}}=0,\qquad \mathcal{C}^{{\small S}}.%
\mathcal{C}^{{\small S}}=-2
\end{equation}%
From these relations, we learn the following features: First, the self
intersections $\mathcal{C}^{{\small U}}.\mathcal{C}^{{\small U}}=0$ and
\begin{equation}
\Sigma .\Sigma =-2l_{{\small U}}l_{{\small T}}<0\qquad \leftrightarrow
\qquad \Upsilon _{\mathbf{q}}.\Upsilon _{\mathbf{q}}.=\mathbf{q}^{2}<0
\end{equation}%
Second, the curves $\Sigma $ can be put in correspondence $\Upsilon _{%
\mathbf{q}}=\sum q_{\text{\textsc{a}}}\mathcal{C}^{\text{\textsc{a}}}$ with
self intersection $\Upsilon _{\mathbf{q}}.\Upsilon _{\mathbf{q}}.=\mathbf{q}%
^{2};$ this intersection factorises like $\mathbf{q}^{2}=2q_{{\small 1}}q_{%
{\small 2}}<0$ and reads in terms of the off diagonal\footnote{%
\ \ Hyperbolic diagonal metric is generated by taking a basis the two
2-cycles $\mathcal{C}_{+}=\mathcal{C}^{{\small U}}+\mathcal{C}^{{\small T}}$
and \ $\mathcal{C}_{-}=\mathcal{C}^{{\small U}}-\mathcal{C}^{{\small T}}$.
Their intersection matrix is $\mathcal{C}_{+}.\mathcal{C}_{+}=+2$ and $%
\mathcal{C}_{-}.\mathcal{C}_{-}=-2$ as well as $\mathcal{C}_{+}.\mathcal{C}%
_{-}=0.$} hyperbolic metric like $q_{\text{\textsc{a}}}\eta ^{\text{\textsc{%
ab}}}q_{\text{\textsc{b}}}$ with $\eta ^{\text{\textsc{aa}}}=0$ and $\eta
^{12}=1.$ In the 2-cycle basis \{$\mathbb{P}^{1},\mathbb{T}^{2}$\}, the
generic complex curve $\Sigma $ expands like $l_{{\small U}}\mathbb{P}%
^{1}+\left( l_{{\small U}}-l_{{\small T}}\right) \mathbb{T}^{2}$ and reduces
for the particular case where $l_{{\small U}}=l_{{\small T}}=l$ to the
multiple curve
\begin{equation}
\Sigma _{l}=l\mathbb{P}^{1}
\end{equation}%
with self intersection $\Sigma .\Sigma =-2l^{2}.$\newline
In the charge lattice $\Lambda ^{\ast }$ generated by \{$\mathbf{\varepsilon
}^{\text{\textsc{a}}}$\} and metric $\eta ^{\text{\textsc{ab}}}=\mathbf{%
\varepsilon }^{\text{\textsc{a}}}.\mathbf{\varepsilon }^{\text{\textsc{a}}},$
the particle states of the 3D effective gauge theory with gauge symmetry $%
U\left( 1\right) ^{2}$ are labeled by two integers $\left( q_{{\small 1}},q_{%
{\small 2}}\right) .$ It turns out that these integers can be interpreted in
terms of the left moving excitations $n_{L}$ in the dual heterotic string
description of M-theory on $K3_{\bot }\times K3_{\Vert }$. There, the
integer $l_{{\small U}}$ is seen as the string winding modes ($n_{\text{%
\textsc{w}}}=q_{{\small 1}}),$ and the integer $l_{{\small T}}$ just the KK
modes ($n_{\text{\textsc{KK}}}=q_{{\small 2}}$); and their product $q_{%
{\small 1}}q_{{\small 2}}$ is related to the moving excitation number as%
\begin{equation}
n_{L}=-q_{{\small 1}}q_{{\small 2}}\qquad ,\qquad n_{L}=-\frac{1}{2}\mathbf{q%
}^{2}  \label{left}
\end{equation}%
Recall that M-theory on $K3\times \mathbb{B}_{2}\times \mathbb{S}_{M}^{1}$
is dual to Type IIA string on $K3\times \mathbb{B}_{2}$, which in turn is
dual the heterotic string on $CY3_{\text{\textsc{het}}}\times \mathbb{T}%
_{M}^{3}$ with $\mathbb{T}_{M}^{3}=\mathbb{T}^{2}\times \mathbb{S}_{M}^{1}$
and elliptically fibered Calabi-Yau threefold $CY3_{\text{\textsc{het}}}.$
As such M-theory on $K3\times K3\times \mathbb{S}_{M}^{1}$, is dual to
heterotic string on $CY3_{\text{\textsc{het}}}\times \mathbb{T}_{M}^{3}$
with $CY3_{\text{\textsc{het}}}$ having both an elliptic and K3 fibrations as%
\begin{equation}
CY3_{\text{\textsc{het}}}=\mathbb{T}^{2}\times K3,\qquad \mathbb{T}_{M}^{3}=%
\mathbb{T}^{2}\times \mathbb{S}_{M}^{1}
\end{equation}%
Moreover, non-BPS particles states are given by wrapping M5 wrapping $%
K3\times \mathbb{S}_{M}^{1};$ the wrapping of M5 on $K3$ gives the so-called
Maldacena-Strominger-Witten (MSW) string. The wrapping of this solitonic MSW
string on $\mathbb{S}_{M}^{1}$ gives non-BPS particle states, which in the
dual string theory, is described by left moving excitations $n_{L}$ of a
heterotic string as in (\ref{left}).

\end{document}